# TITLE

Sequencing and characterisation of rearrangements in three *S. pastorianus* strains reveals the presence of chimeric genes and gives evidence of breakpoint reuse.

# AUTHORS

Sarah Kathryn Hewitt, Ian Donaldson, Simon C. Lovell and Daniela Delneri

# AFFILATIONS

Faculty of Life Sciences, University of Manchester, Manchester, United Kingdom

## ABSTRACT

Gross chromosomal rearrangements have the potential to be evolutionarily advantageous to an adapting organism. The generation of a hybrid species increases opportunity for recombination by bringing together two homologous genomes. We sought to define the location of genomic rearrangements in three strains of *Saccharomyces pastorianus,* a natural lager-brewing yeast hybrid of *Saccharomyces cerevisiae* and *Saccharomyces eubayanus*, using whole genome shotgun sequencing. Each strain of *S. pastorianus* has lost species-specific portions of its genome and has undergone extensive recombination, producing chimeric chromosomes. We predicted 30 breakpoints that we confirmed at the single nucleotide level by designing species-specific primers that flank each breakpoint, and then sequencing the PCR product. These rearrangements are the result of recombination between areas of homology between the two subgenomes, rather than repetitive elements such as transposons or tRNAs. Interestingly, 28/30 *S. cerevisiae- S. eubayanus* recombination breakpoints are located within genic regions, generating chimeric genes. Furthermore we show evidence for the reuse of two breakpoints, located in *HSP82* and *KEM1*, in strains of proposed independent origin.

## INTRODUCTION

Hybridisation in *Saccharomycetous* yeast occurs readily in natural and industrial environments [1,2,3,4,5,6,7], and may be a swift mechanism for evolutionary innovation. Investigating the genomics of successful natural hybrid species can provide valuable evolutionary insight into how the union of diverged genetic material can sculpt a genome more suited to its new environmental niche. These adaptations may include chromosomal rearrangements such as duplication, translocation, inversion and selective loss of genes or even whole chromosomes. The lager yeast *Saccharomyces pastorianus*, previously classified as *Saccharomyces carlsbergensis*, is a natural hybrid between *Saccharomyces cerevisiae* and a *Saccharomyces uvarum*-like species [4,8,9,10]. The *S. uvarum*-like species has most recently been identified as the Argentinean-isolate *Saccharomyces eubayanus*, which shows 99.5% identity to the non-*S. cerevisiae* portion of *S. pastorianus* [11]. *S. pastorianus* is thought to have arisen by spontaneous hybridisation in brewery conditions, maintained by human selection for colder brewing temperatures, a preference that is conferred by its *S. uvarum*-like parent [12].

So far, there is only one strain of lager yeast sequenced [9] and most of our knowledge of the genome composition of these natural hybrids derives from previous array-comparative genomic hybridisation studies (Array-CGH) performed on 17 strains of *S. pastorianus* [10]. This particular work identified two groups of lager yeasts: Group 1 strains contain roughly one haploid *S. cerevisiae* and one haploid *S. eubayanus* genome with significant loss of *S. cerevisiae* genes, whereas Group 2 strains contain one haploid *S. eubayanus* genome and a diploid *S. cerevisiae* genome. The differences between these two groups suggest that they may have had independent evolutionary origins, a theory given weight by both the aforementioned array-CGH analysis [10] and the differing distribution of transposons between the two groups [13]. Additionally, strains within each group are highly variable in their patterns of chromosomal loss, aneuploidy and gross chromosomal rearrangements. probably reflecting either evolutionary pressure from diverse brewery conditions or random genetic drift. [9,10],

Lager yeast chromosomes have been shown to have undergone recombination, generating chimeric chromosomes composed of genetic material from both parental species [10,14,15,16]. Typically, recombination between chromosomes within a non-hybrid yeast species is thought to be mediated primarily by transposons (Ty elements) [14,17,18,19], tRNAs [17,18], duplicated genes [18] or, as more recently proposed, origins of replication [20]. However, breakpoint formation in *S. pastorianus* is thought to be either Ty-mediated [10,14] or the result of recombination between homologous regions [15]. Studies have also demonstrated the role of high stress brewery conditions in promoting genomic rearrangements, such as localised areas of gene amplification and recombination [21]. Significantly, chromosomal rearrangements have been shown to confer adaptive traits in both wild and industrial yeasts including highly sulphite-resistant wine yeast [22], *flor* wine yeast [23] and wild copper-tolerant yeast [24]. Furthermore, rearrangements have been shown to contribute to speciation between species of yeast [25].

We sequenced three *S. pastorianus* strains to both characterise genomic breakpoints and shed further light on their formation and retention. We chose strains that have been used in a previous microarray study to provide a source of validation for our sequencing [10]. These strains have also been pre-classified into one of the two aforementioned groupings of *S. pastorianus*: two of the chosen strains of *S. pastorianus* are of Group 1 (DBVPG 6033 and DBVPG 6261) and one is of Group 2 (DBVPG 6257) [10]. These particular strains have the greatest level of differential gene loss and therefore the least amount of redundancy. The latter group is thought to have an independent evolutionary origin from the former group, allowing us to investigate similarities between non-related strains.

We have confirmed the location of each *S. cerevisiae- S. eubayanus* breakpoint at the single nucleotide level and identified both nearby repetitive elements and regions of homology. Significantly, we found that the majority of genomic breakpoints occurred within protein coding regions, generating chimeric genes. Furthermore, the presence of identical breakpoints in *KEM1* and *HSP82* is evidence of breakpoint reuse between strains of proposed independent origin.

## RESULTS AND DISCUSSION

### Genome assembly and analysis

The genomic DNA of three strains of *S. pastorianus*, DBVPG 6033 (*Saccharomyces carlsbergensis* type strain), DBVPG 6261 (*Saccharomyces monacensis* type strain) and DBVPG 6257 were sequenced using the SOLiD 4 Next Generation Sequencing platform and mapped to *S. cerevisiae* (sacCer2) and *S. uvarum* (sacBay MIT), which are representative of the *S. pastorianus* subgenomes. We used sacBay MIT as the reference genome for *S. eubayanus* due to its fully available sequence which is purportedly 7% diverged from *S. eubayanus* [11]. Visualisation of the *S. cerevisiae* chromosomes in the UCSC Genome Browser (http://genome.ucsc.edu/) is reported in Figure 1. *S. eubayanus* reads were viewed as contigs in the Integrative Genomics Viewer (http://www.broadinstitute.org/igv).

SOLiD sequencing allowed us to ascertain the approximate chromosomal copy number derived from relative read quantity. In total, DBVPG 6033, 6261 and 6257 are estimated to have between 32-38, 34-39 and 45-48 chromosomes respectively (Table 1). These chromosomes map to *S. cerevisiae*, *S. eubayanus* or a combination of both *S. cerevisiae* and *S. eubayanus* sequence (chimeric chromosomes). We estimated the *S. eubayanus* chromosome number by viewing multiple contigs (whole chromosome assembly was not available). When viewing these contigs, we accounted for the fact that *S. uvarum*-like chromosomes show reciprocal rearrangements between three sets of chromosomes: II-IV, VI-X and VIII-XV, comparative to *S. cerevisiae* [26].

*S. pastorianus* shows a high degree of aneuploidy and the chromosomal composition between strains is highly variable. Strains DBVPG 6033, 6261 and 6257 have eight, seven and sixteen complete *S. cerevisiae* chromosomes, respectively. They also have an estimated 16-21, 13-17 and 13-15 complete *S. eubayanus* chromosomes, and 11, 16 and 18 chimeric chromosomes, composed of both *S. cerevisiae* and *S. eubayanus* sequence (Table 1). The approximate number of total chromosomes in DBVPG 6033 and 6261 (Group 1) is 32-38 and 34-39 respectively, both roughly equal to a diploid with a few extra chromosomes (16x2= 32). The total number of chromosomes in DBVPG 6257 (Group 2) is 45-48, and roughly equates to a triploid (16x3= 48). These data support previous estimates of Group 1 strains generally being diploid-derived and Group 2 strains being triploid-derived [10]. Both Group 1 strains have lost their *S. cerevisiae* version of chromosomes VI and XII, i.e. there is no evidence of this sequence, even on a chimeric chromosome. DBVPG 6261 has additionally lost its *S.*

*cerevisiae* chromosome III and XIV sequence, whereas DBVPG 6033 has lost its *S. cerevisiae* chromosome XI sequence. In the Group 2 strain DBVPG 6257, there was no detection of *S. cerevisiae* chromosome V.

All three strains of *S. pastorianus* show evidence of homologous recombination between *S. cerevisiae* and *S. eubayanus* chromosomes IV, VII, XIII and XVI. Additionally, chromosome VIII and XV have recombined in both Group 1 strains. Chromosomes IX, X, and potentially XII, have recombined in Group 2 strain DBVPG 6257. Chromosomes I, VI and XVI remain largely stable, showing no evidence of *S. cerevisiae*- *S. eubayanus* recombination in any of the sequenced strains. Reciprocal recombination and inversion events could not be identified in this study since these rearrangements are copy-number neutral.

In agreement with previous analysis of these three strains of *S. pastorianus* [10], we did not detect any *S. cerevisiae* mitochondrial DNA. The restriction analysis of *COX2* in the three strains of *S. pastorianus* has indicated a *S. uvarum*-like mitochondrial sequence (data not shown), supporting the widely held notion that lager yeasts tend to inherit and/or retain only their *S. eubayanus* mitochondria [27]. The 2-micron plasmid maps to *S. cerevisiae* sequence in strains DBVPG 6033 and DBVPG 6257 but not DBVPG 6261. It is unknown if any 2-micron plasmids are *S. eubayanus*-derived.

**Chromosomal rearrangements**

We used the UCSC genome browser (http://genome.ucsc.edu/) to identify candidate breakpoints based on variations in *S. cerevisiae* read copy number across each chromosome (Figure 1). Using this technique, we were able to detect a total of 13 *S. cerevisiae*- *S. eubayanus* breakpoints in DBVPG 6033, 13 in DBVPG 6261 and 15 in DBVPG 6257. We used species-specific primers to confirm the presence of each breakpoint by PCR (Figure 2). Each successfully amplified PCR product was sequenced at GATC Biotech (Germany) (Figure S1). All the sequenced breakpoints were then aligned to the *S. cerevisiae* and the *S. uvarum* reference genomes (Figure S2). A total of 9/13 *S. cerevisiae*- *S. eubayanus* breakpoints were confirmed by PCR in DBVPG 6033, another 11/13 in DBVPG 6261 and 10/15 in DBVPG 6257 (Table 2). In DBVPG 6033, three of these breakpoints were located on chromosome XVI, two on VIII and the remaining four on chromosomes II, IV, VII and XIII. In DBVPG 6261, three were located on chromosome XV, two on chromosome XI, two on chromosome XIII and the remaining four on chromosomes IV, V, VII and XVI. In DBVPG 6257, three breakpoints were located on chromosomes XVI, two on XIII and the remaining five on chromosomes IV, VII, IX, X and XI. Breakpoints occurred most commonly but not exclusively at the subtelomeric regions of chromosomes.

The majority of the sequenced *S. cerevisiae*- *S. eubayanus* breakpoints occur within coding regions (Table 2), despite breakpoints in yeast usually being located in intergenic rather than intragenic regions [28]. A total of seven, eleven and ten intragenic breakpoints were located in DBVPG 6033, 6261 and 6257 respectively. In each strain there were a small number of candidate breakpoints which could not be amplified (Table S1). There are four unconfirmed breakpoints in DBVPG 6033 (one of which was within in a coding region), two in DBVPG 6261 (both of which were within in a coding region) and five in DBVPG 6257 (one of which was within a coding region). Notably, we detected a breakpoint at the MAT locus on chromosome III in strains DBVPG 6033 and 6257. Breakpoints to the right of the MAT locus in chromosome III have been noted previously in many strains of lager yeast [14,15,29]. There are two breakpoints that are in close proximity on chromosome X in strain DBVPG 6257, are in the vicinity of numerous Ty elements. A further three candidate breakpoints, one in each strain (at the far right of each chromosome X) could not be amplified, possibly due to their close proximity to both an AT-rich ARS element and the right telomeric region. The four remaining unsequenced candidate breakpoints are on chromosome XIII and XV in DBVPG 6033, chromosome VIII in DBVPG 6261 and chromosome XIII in DBVPG 6257.

The analysis of the chromosome copy number of *S. cerevisiae* and *S. eubayanus* (Table 1) showed that the majority of chimeric chromosomes were most likely formed as a result of either a single non-reciprocal recombination event from the *S. eubayanus* chromosome to the *S. cerevisiae* chromosome leaving both a chimeric copy and a full *S. eubayanus* copy of the sequence or a reciprocal event followed by a deletion. There are triplicate copies of seven breakpoint genes, four of which are in the purportedly diploid-derived Group 1 strains. These likely have formed through either gene conversion involving three chromosomes or duplication of subtelomeric regions, which are known to be sites of rapid gene duplication [30].

**Chimeric genes**

As a result of homologous recombination between chromosomes, several chimeric genes were formed (Table 2, see Figure S2 for full sequence alignments). Two of these genes, *KEM1*, a 5'-3' exonuclease, and *HSP82*, a molecular chaperone, are chimeric in all three strains. Interestingly, these two breakpoints occur in the same gene position in all three hybrids. The *KEM1 S. eubayanus> S. cerevisiae* breakpoint occurs 460(±4)/4587bp after the start of the gene and the *HSP82 S. eubayanus> S. cerevisiae* breakpoint occurs 1592(+15,-14)/2130bp after the start of the gene. Another chimeric gene, *UTP4*, is shared between both Group 1 strains, yet the position of the *S. eubayanus>S. cerevisiae* breakpoint differs between each strain: in DBVPG 6033, the breakpoint occurs 460(±16)/2259bp after the start of the gene, whereas in DBVPG 6261, the breakpoint occurs 868(+/-7)/2259bp of the gene. A previous whole genome sequencing of a strain of *S. pastorianus*, Weihenstephan 34/70 also identified breakpoints within these two shared genes [15]. Microarray data further indicated that both of these breakpoints are also shared between many of the other Group 1 and Group 2 strains [10]. Four additional breakpoints located in DBVPG 6257 were also found in identical gene positions in Weihenstephan 34/70: *KEM1*, *PRI2*, *GPH1* and *QCR2*. A further breakpoint gene, *PRP12*, identified in the study was also observed in our SOLiD sequencing data. However, although the sequencing of Weihenstephan 34/70 shows a reciprocal breakpoint within *TDH2*, we found that rearrangements at this site altered copy number. Either there has additionally been a chromosomal deletion or the event was non-reciprocal in DBVPG 6257. Furthermore, since reciprocal recombination events could not be detected in this study, it is possible that the breakpoint may have occurred twice in this location, once as a reciprocal event and once as a non-reciprocal event.

Interestingly, two of the chimeric genes are involved in ethanol metabolism, a key biochemical pathway in lager fermentation. *ALD2* is involved in the oxidation of ethanol and *TDH2* is a component of the tetramer glyceraldehyde-3-phosphate dehydrogenase, which is required for gluconeogenesis. Two further chimeric genes also play a role in energy metabolism. *GPH1* is involved in glycogen mobilisation and *GAL80* is a repressor of GAL genes in the absence of galactose. With the large genetic redundancy in all three strains of *S. pastorianus*, especially DBVPG 6257, which contains approximately two *S. cerevisiae* sets of chromosomes, chimeric gene copies may or may not significantly affect the hybrid organism. We investigated whether additional complete *S. cerevisiae* or *S. eubayanus* copies of each chimeric gene were present in the sequencing data. Seven of the sequenced chimeric genes, *HSP82* in all three strains, *CDH1*, *IRR1*, *KEM1* within DBVPG 6261, and *QCR2* are present in the chimeric form only since both full parental copies have been lost. *IRR1* is also an essential gene in *S. cerevisiae*, as well as being non-redundant. If it is presumed that *IRR1* is also essential in *S. pastorianus*, it can be deduced that *IRR1* is a functioning chimeric gene, given that this strain is viable. Only the chimeric form of *HSP82*, a molecular chaperone of protein complexes, is present in each strain of *S. pastorianus*, but its homologue *HSC82* has been retained in both parental forms. The remaining 21 chimeric genes have one or more non-chimeric homologues. Two chimeric genes, *KEM1* in DBVPG 6033 and *TDH2* have additional homologous copies of both *S. cerevisiae* and *S. eubayanus* genes present in the genome. A further two chimeric genes, *FKS3* and *KEM1* in DBVPG 6257 have at least one additional *S. cerevisiae* homologue remaining in the genome. The final 17 chimeric genes are complemented by at least one additional complete *S. eubayanus* homologue. The identification of chimeric genes within *S. pastorianus* is of importance in phylogenetic analysis, since they have the potential to weaken phylogenetic signal and contribute to incongruence [31,32].

Previous studies on the functionality and fitness of chimeric genes show mixed results. The chimeric gene *GPH1* in the lager strain CMBS-33 contains a disruptive base insertion within its initial *S. eubayanus* sequence, and the resultant gene is not expressed [33]. However, a recent study located a recurrent in-frame breakpoint within *MEP2*, an ammonium permease, in clones of lab-created hybrids of *S. cerevisiae* and *S. uvarum* that were evolved under nitrogen-limiting conditions [34]. The experimentally evolved strains bearing the rearrangement were fitter than the non evolved ones in nitrogen-limiting competition experiments. A number of studies have also examined the fitness effects of rearrangements involving non-homologous genes. The *LG-FLO1*, involved in flocculation appears to have been inactivated in non-flocculent lager yeast by a non-reciprocal translocation of *S. cerevisiae* YIL169c into its C-terminal region in various non-flocculent strains [16]. However, a sulphite-resistant gene found in wine yeast, *SSU1*-R was generated by recombination between the promoter regions of *SSU1* and *ECM34* genes and has been found to grant increased sulphite resistance compared to the wild type allele [22,35].

**The reuse of breakpoints**

The reuse of breakpoints is a relatively new hypothesis that challenges the long held random breakage model of chromosomal rearrangements [36] and is gaining momentum in studies of mammalian and fly genomes. Studies of mammalian genomes have indicated that breakpoint regions may be reused throughout evolution at a rate between 7.7% and 20% [37,38,39]. The term breakpoint reuse, first coined by Pevzner and Tesler [40], applies to regions of overlapping breakpoints and is not limited to breakpoints shared at the nucleotide level. Although it is unclear whether this overlap in usage is random or non-random, there is increasing evidence of association of these evolutionary breakpoint regions with fragile sites (heritable regions prone to breakage and reorganization) [40,41], with telomeric and centromeric regions [37], with segmental duplications [37,38], and gene dense regions [37,39]. Moreover, fragile sites prone to breakage, rather than functional constraints on genes, are thought to have been instrumental in shaping gene organisation, at least in Drosophila [42]. In our study, we see the reuse of two breakpoints, *HSP82* and *KEM1* in all three sequenced strains of *S. pastorianus* (Table 2). Interestingly, mutagenized lager yeast strains selected under heat stress and high osmotic stress showed a rearrangement in *YGL172w*, which is adjacent to *KEM1* [21], a further indication of breakpoint reuse in this region of the genome. Furthermore, the aforementioned study detected four other rearrangements in or around breakpoint genes that were identified in our study: *TDH2*, *GPH1* and in ORFs adjacent to *PRP8* and *CDH1*. Interestingly, a fully sequenced *S. cerevisiae- S. eubayanus* breakpoint in *GPH1*, in the lager strain CMBS-33, differs in location from a breakpoint found in DBVPG 6257 [33]. The breakpoint identified in our study occurs after the first 1496(+6,-5)bp of the gene, whereas the breakpoint in CMBS-33 occurs after the first 330-360bp of the gene. Finally, a breakpoint present in both Group 1 strains in *UTP4* was resolved into two distinct breakpoints at the nucleotide level, and are in close proximity to a breakpoint approximately 30kb upstream of this site in DBVPG 6257 (Table 2, Figure 2). This region on chromosome IV is potentially a site of independent evolutionary breakpoint reuse.

**Mechanisms of breakpoint formation**

Chromosomal translocation requires the induction of double-stranded DNA breaks followed by incorrect repair of these breaks using an erroneous homologous and repetitive sequence [43]. A recent study found that the potential of a double stranded break in the genome to cause changes in genome copy number increases when the breakage occurs within non-repetitive DNA rather than repetitive DNA [44]. This effect was far more pronounced in hybrid diploids comparative to non-hybrid diploids. This would suggest that any breakpoint that was to randomly occur within a coding region may be more likely to promote a genomic rearrangement in *S. pastorianus* than if the break was to occur within a repetitive element.

We looked for the presence or absence of Ty elements, their flanking LTRs, tRNAs and origins of replication in proximity to each identified breakpoint. Using the sequence data mapped to each *S. cerevisiae* SacCer2 chromosome in UCSC genome browser (http://genome.ucsc.edu/), we manually recorded the nearest repetitive genomic feature to each breakpoint (Table S2). The majority of breakpoints, having occurred within coding regions, were not immediately flanked by repetitive elements, although Dunn and Sherlock [10] have observed clustering of breakpoints near repetitive features in the genomes of lager yeast. Breakpoint proximity to a repetitive element ranges between 0.6kb and 39kb with a mean of 11.4kb. Five sequenced breakpoints were less than 5kb from an element; eleven were between 5 and 10kb away, eight between 10kb and 20kb away and four that are father than 20kb from a repetitive element. Despite the lack of proximity of repetitive elements to breakpoints as a trend, one breakpoint gene, *TDH2* on chromosome XII in DBVPG 6257 is situated adjacent to an ARS, a feature known to promote chromosomal translocation [20]. Eight breakpoints genes, *HSP82* and *KEM1*, in all three strains; *CHD1* in DBVPG 6261 and *QCR2* in DBVPG 6257 are located adjacent to subtelomeric *S. cerevisiae* regions on chromosomes XVI, VII, V and XVI respectively. These regions appear to have undergone significant duplication.

Large areas of homology are known to induce recombination in yeast, and this mechanism is utilised widely for yeast gene deletion in the laboratory [45]. More recently however, very small areas of microhomology have also been indicated in the formation of chromosomal breakpoints in wine yeast [3,22]. Since the reference parental species of *S. pastorianus* are closely related, with an average of 80% nucleotide identity in coding regions [26] we view the induction of recombination via homologous regions in lager yeast a likely hypothesis. Furthermore recombination has occurred more frequently in these coding regions than in non-coding regions, which have an average nucleotide identity of only 62% [26]. We identified areas of microhomology between the two subgenomes (Table S2).

Whatever the underlying sequence that facilitates breakpoint formation, it is likely that one of three events is potentiating breakpoint formation: the unstable nature of newly formed hybrids, an increase in the occurrence of double stranded breaks or an evolutionary pressure for recombination. The double stranded breaks required to initiate recombination are induced at a higher frequency in stressful brewery conditions. A previous study investigating the formation of breakpoints in lager yeast under high stress conditions isolated a number of heat stress and high-gravity stress tolerant mutants generated by EMS-mutagenesis of an existing lager strain, CMBS-33 [21]. The mutants were found to contain breakpoints that are at or near those observed in other natural strains *of S. pastorianus*. Breakpoints arose in a gene directly adjacent to *KEM1*, and within *PRP8, CHD1, TDH2* and *GPH1*, all of which have been identified as existing breakpoints in our sequenced strains of *S. pastorianus*. These findings promote the existence of breakpoint hotspots in the *S. pastorianus* genome and give evidence for the role of stress in promoting and maintaining genomic breakpoints.

**Conclusion**

Our whole genome sequencing of three strains of *S. pastorianus* allowed the identification of *S. cerevisiae*- *S. eubayanus* chromosomal breakpoints at a single nucleotide resolution. The majority of *S. cerevisiae*- *S. eubayanus* breakpoints are located within coding regions and were most likely formed as a result of homology and microhomology between the two parental subgenomes, rather than via repetitive elements in the genome. PCR sequencing of breakpoints enabled the further characterisation of these recombination-generated chimeric genes. The greater resolution granted by PCR sequencing allowed us to verify that the breakpoints within *HSP82* and *KEM1* have occurred at an identical genomic location in all three strains. We determined that two different breakpoints have occurred within *UTP4* in the two Group 1 strains. Although the breakpoints are in different positions, this will still be regarded as an example of breakpoint reuse. Interestingly, we note the presence of a chimeric gene *IRR1* in DBVPG 6257 of *S. pastorianus* has lost both parental homologues. Since *IRR1* is also an essential gene, this indicates that the chimeric gene is efficiently utilised by the hybrid. The presence of chimeric genes in the genome also has the potential to weaken the phylogenetic signal of these genes, which could promote incongruence in phylogenetic analyses [32]. Future studies on the function and fitness of chimeric genes may reveal their evolutionary role in facilitating the adaption of *S. pastorianus* to high stress brewery conditions.

## MATERIALS AND METHODS

**Strains and media**

*Saccharomyces pastorianus* strains DBVPG 6033 (GSY129), DBVPG 6261 (GSY134) and DBVPG 6257 (GSY132) were obtained from DBVPG Industrial Yeasts Collection, University of Perugia, Italy. Yeast was grown at 25$^{o}$C, 200rpm for 20 hours in YPD (1% yeast extract, 2% peptone, 2% glucose) and genomic DNA extracted using Wizard Genomic DNA Purification Kit (Promega).

Genome sequencing reference strain for *Saccharomyces cerevisiae,* sacCer2 was obtained via the UCSC Genome Browser (http://genome.ucsc.edu/). Genome sequencing reference strain for *Saccharomyces uvarum*, sacBay MIT was obtained from the *Saccharomyces* Genome Database (SGD, http://www.yeastgenome.org).

**SOLiD sequencing**

The genomic DNA of three strains of *S. pastorianus* was sequenced using Next Generation Sequencing Applied Biosystems SOLiD 4 platform. Using BFAST (http://sourceforge.net/projects/bfast/files/), the reads were mapped to the *S. cerevisiae* reference genome “sacCer2” obtained from UCSC (http://genome.ucsc.edu/), which includes 16 chromosomes, the mitochondrial genome and the 2 micron plasmid. The ‘-a 3’ flag of the post-process step was used to obtain unique best scoring alignments. The *S. cerevisiae* ORFs were used to find *S. eubayanus* consensus ORFs in the *S. uvarum* reference strain “sacBay MIT” obtained from SGD (http://www.yeastgenome.org/). BFAST files were filtered to retrieve sets of reads with 0, 0 to 1 or 0-5 mismatches to each reference genome. Generally, 0 to 1 mismatches was found to be the most suitable cut-off value, having the best agreement to previous microarray data by Dunn and Sherlock (2008).

**Data deposition**

Raw reads from this study have been deposited at the European Nucleotide Archive under the accession number PRJEB4654.

**Chromosomal analysis and breakpoint identification**

We used the *S. pastorianus* SOLiD sequence mapped to the pre-annotated *S. cerevisiae* genome sequence via the UCSC Genome Browser (http://genome.ucsc.edu/) to identify both *S. cerevisiae* chromosome copy number and potential chimeric chromosomes comprising both *S. cerevisiae* and *S. eubayanus* sequence. These candidate breakpoint regions were identified visually by their abrupt and sustained reduction in *S. cerevisiae* reads along a chromosome. Due to difficulties in mapping and analysing repetitive regions, telomeres were excluded. Similarly, changes in read number due to the presence of a yeast transposon (Ty) or other repetitive element were noted.

Copy number of *S. eubayanus* chromosomes was estimated using the Integrative Genomics Viewer (http://www.broadinstitute.org/igv/) to view multiple *S. eubayanus* contigs. Naturally existing rearrangements between chromosomes within the *S. uvarum* genome (chromosomes II-IV, VI-X and VIII-XV) were taken into account when estimating copies of *S. eubayanus* chromosomes.

## Primer design and PCR

Species-specific primers were designed to flank each predicted breakpoint area. The *S. cerevisiae* primers were designed using *S. cerevisiae* sequence obtained directly from the UCSC Genome Browser (http://genome.ucsc.edu/). The *S. eubayanus* primers were designed by finding the *S. uvarum* orthologue of the nearest *S. cerevisiae* gene using the SGD Synteny Viewer (http://www.yeastgenome.org). This orthologue was then mapped to *S. pastorianus* to find consensus sequences for the *S. eubayanus* portion of the *S. pastorianus* genome. The *S. eubayanus*-specific primer was then designed within this consensus sequence. Candidate primers were generated for both *S. cerevisiae* and *S. eubayanus* sequences using Primer 3 (http://frodo.wi.mit.edu). To circumvent the potential for non-specific binding between the two closely related subgenomes, these primers were then carefully selected for species-specificity using the Fungal BLAST tool in SGD (http://www.yeastgenome.org). For ease of amplification, primers were designed to anneal no more than a few thousand base pairs apart but with sufficient sequence either side of the breakpoint for clear identification of each subgenome. Primer sequences are available in Table S3.

PCR conditions were optimised for each breakpoint to obtain pure homogeneous chimeric sequence. The PCR product was separated by electrophoresis on 1% (w/v) agarose gel. PCR products were purified prior to sequencing using QIAquick PCR Purification Kit (Qiagen, UK).

## Sequencing of breakpoint PCR product

Purified PCR products were sequenced at GATC Biotech (Germany). The resulting sequences were analysed using the Fungal BLAST tool in SGD (http://www.yeastgenome.org) to identify DNA with high identity to that of either *S. uvarum* or *S. cerevisiae* either side of the predicted breakpoint.

**Analysis of sequence identity**

Percentage nucleotide identity between each subgenome and *S. pastorianus* was calculated using *S. cerevisiae* and *S. uvarum* sequences obtained from SGD (http://www.yeastgenome.org/) and Clustal Omega (http://www.ebi.ac.uk/Tools/msa/clustalo/) Amino acid identity was calculated similarly but chimeric nucleotide sequences were first converted to protein sequences using Expasy Translate Tool (http://web.expasy.org/translate/).

## ACKNOWLEDGEMENTS

The authors wish to thank Dr Casey Bergman for help in designing the bioinformatics approach and suggesting analysis of breakpoint reuse.

## FIGURE LEGENDS

**Figure 1. Representation of *S. pastorianus* reads mapped to *S. cerevisiae* chromosomes in the UCSC Genome Browser.** Full set of *S. cerevisiae* chromosomes is displayed for each strain using the UCSC Genome Browser. Depth of track corresponds to read number so potential *S. cerevisiae- S. eubayanus* breakpoints can be located and ploidy can be deduced. An abrupt and sustained reduction in read number is indicative of a *S. cerevisiae- S. eubayanus* breakpoint. Track depth window is set to accommodate three copies of a chromosome, since *S. cerevisiae* chromosome copy number in these strains generally varies between zero and three.

**Figure 2. Diagram showing different breakpoint positions in *S. pastorianus* chromosome IV.** To locate the breakpoint at the nucleotide level, species-specific primers are designed on the Watson strand around putative breakpoints (SC F: *S. cerevisiae* forward primer, SU F: *S. uvarum* forward primer. SC R: *S. cerevisiae* reverse primer. SU R: *S. uvarum* reverse primer). Strains are labelled Group 1 or Group 2 according to their previous assignment [10].

## SUPPORTING INFORMATION

**Figure S1. Full set of sequences across each breakpoint in all three strains of *S. pastorianus*.** The region covering each breakpoint was amplified using species-specific primers. The PCR products were sequenced on the Watson strand at GATC (Germany).

**Figure S2. Multiple sequence alignment of each *S. pastorianus* breakpoint sequence to parental species.** *S. cerevisiae (*Scer, *Saccharomyces* Genome Database), *S. uvarum* (MIT_Sbay or WashU_Sbay, *Saccharomyces* Genome Database) and the region sequenced over the breakpoints in each *S. pastorianus* strain were aligned using Clustal Omega. Note that ORF sequences obtained from the *Saccharomyces* Genome Database are taken from Cliften *et al*.[46] and Kellis *et al*.[26]. Low quality ends of breakpoint sequences were trimmed before alignment. Breakpoint area is demarcated in grey with central position underlined. All gene sequences are 5’-3’.

**Table S1. Breakpoints which were not successfully amplified by PCR.**

**Table S2. Analysis of the breakpoint region in each strain of *S. pastorianus*.**

**Table S3. Primer sequences for amplification of breakpoint regions.**

## TABLES

### Table 1. Estimation of chromosome copy number in *S. pastorianus*

| **DBVPG strain** | **6033** | **6261** | **6257** | **6033** | **6261** | **6257** | **DBVPG strain** | **6033** | **6261** | **6257** |
|---|---|---|---|---|---|---|---|---|---|---|
| ***S. cerevisiae* chromosome** | ***S. cerevisiae* copies** | | | **Chimeric copies** | | | ***S. eubayanus* chromosome**[a] | ***S. eubayanus* copies**[c] | | |
| I | 1 | 2 | 1 | 0 | 0 | 0 | I | 1 | 1 | 2 |
| II | 0 | 1 | 2 | 1 | 0 | 0 | II-IV | 1 | 1 | 1 |
| III | 2 | 0 | 2 | 1 | 0 | 2 | III | 0 | 2-3 | 0 |
| IV | 0 | 0 | 0 | 1 | 1 | 2 | IV-II | 2 | 1 | 1 |
| V | 1 | 0 | 0 | 0 | 3 | 0 | V | 1-2 | 0 | 2-3 |
| VI | 0 | 0 | 1 | 0 | 0 | 0 | X-VI | 1-2 | 1-2 | 1 |
| VII | 1 | 0 | 2 | 1 | 3 | 1 | VII | 1 | 0 | 0 |
| VIII | 0 | 1 | 2 | 1 | 1 | 0 | VIII-XV | 0-1 | 0-1 | 1 |
| IX | 1 | 2 | 0 | 0 | 0 | 3 | IX | 1-2 | 1 | 0 |
| X | 1 | 1 | 1 | 0 | 0 | 1 | VI-X | 2 | 2 | 1 |
| XI | 0 | 0 | 0 | 0 | 1 | 2 | XI | 2 | 1 | 1 |
| XII | 0 | 0 | 1 | 0 | 0 | 1 | XII | 1-2 | 1-2 | 1-2 |
| XIII | 0 | 0 | 0 | 1 | 2 | 2 | XIII | 1 | 0 | 0 |
| XIV | 1 | 0 | 2 | 0 | 0 | 0 | XIV | 1 | 2 | 1 |
| XV | 0 | 0 | 2 | 1 | 1 | 0 | XV-VIII | 1 | 0 | 1 |
| XVI | 0 | 0 | 0 | 1-2 | 2-3 | 2-3 | XVI | 0 | 0 | 0 |
| **TOTAL** | 8 | 7 | 16 | 8-9 | 14-15 | 16-17 | **TOTAL** | 16-21 | 13-17 | 13-15 |
| | | | | | | | **TOTAL CHROMOSOMES** | **32-38** | **34-39** | **45-48** |

An estimation of chromosome number based on reads mapped to *S. cerevisiae* chromosomes and *S. uvarum* contigs.
[a]*S. uvarum* chromosomes are known to have undergone reciprocal recombination [26]

## Table 2. Genomic location of *S. cerevisiae*- *S. eubayanus* breakpoints

| DBVPG strain | Genome position[a] | Systematic name | Standard name | Breakpoint location from start codon in each chimeric gene[b] |
|---|---|---|---|---|
| 6033 | chrII:780898-780906 | YBR289w | SNF5 | *S. cerevisiae> S. eubayanus* at 1240(±4)/2694bp |
| | chrIV:1116214-1116246 | YDR324c | UTP4 | *S. eubayanus> S. cerevisiae* at 460(±16)/2259bp |
| | chrVII:179656-179664 | YGL173c | KEM1 | *S. eubayanus> S. cerevisiae* at 460(±4)/4587bp |
| | chrVIII:433730-433738 | YHR165c | PRP8 | *S. eubayanus> S. cerevisiae* at 3226(±4)/7251bp |
| | chrVIII:451231-451276 | Intergenic | Intergenic | Intergenic |
| | chrXIII:843626-843635 | YMR287c | MSU1 | *S. eubayanus>S. cerevisiae* at 1714(+5-4)/2910bp |
| | chrXVI:97019-97048 | YPL240c | HSP82 | S. *eubayanus> S. cerevisiae* transition occurs at 1592(+15-14)/2130bp |
| | chrXVI:482992-483014 | YPL036w | PMA2 | *S. eubayanus> S. cerevisiae* at 124(±11)/2805bp |
| | chrXVI:906846-906880 | Intergenic | Intergenic | Intergenic |
| 6261 | chrIV:1115815-1115829 | YDR324c | UTP4 | *S. eubayanus> S. cerevisiae* at 868(±7)/2259bp |
| | chrV:507240-507245 | YER164w | CHD1 | *S. eubayanus> S. cerevisiae* at 1856(+3-2)/4401bp |
| | chrVII:179656-179664 | YGL173c | KEM1 | *S. eubayanus> S. cerevisiae* at 460(±4)/4587bp |
| | chrXI:60183-60193 | YKL203c | TOR2 | *S. cerevisiae> S.* eubayanus at 3173(±5)/7425bp |
| | chrXI:285492-285506 | YKL080w | VMA5 | *S. cerevisiae> S. eubayanus* at 826(±7)/1179bp |
| | chrXIII:172148-172153 | YML051w | GAL80 | *S. eubayanus> S. cerevisiae* at 557(+3-2)/1308bp |
| | chrXIII:882708-882716 | YMR306w | FKS3 | *S. eubayanus> S. cerevisiae* at 1555(±4)/ 5358bp |
| | chrXV:496862-496866 | YOR092w | ECM3 | *S. cerevisiae> S. eubayanus* at 1737(±2)/1842bp |
| | chrXV:526436-526442 | YOR109w | INP53 | *S. eubayanus> S. cerevisiae* at 1161(±3)/3327bp |
| | chrXV:561420-561424 | YOR127w | RGA1 | *S. cerevisiae> S. eubayanus* at 252(±2)/3024bp |
| | chrXVI:97019-97048 | YPL240c | HSP82 | S. *eubayanus>S. cerevisiae* transition occurs at 1592(+15-14)/2130bp |
| 6257 | chrIV:1148746-1148747 | YDR338c | Putative protein of unknown function | *S. eubayanus> S. cerevisiae* at 715(+1-0)/2088bp |
| | chrVII:179656-179664 | YGL173c | KEM1 | *S. eubayanus> S. cerevisiae* at 460(±4)/4587bp |
| | chrIX:306349-306368 | YIL026c | IRR1 | *S. eubayanus> S. cerevisiae* at 1560(+10-9)/3447bp |
| | chrX:453941-453961 | YJR009c | TDH2 | *S. eubayanus> S. cerevisiae* at 724(±10)/999bp |
| | chrXI:354021-354023 | YKL045w | PRI2 | *S. cerevisiae> S. eubayanus* at 886(±1)/1537bp |
| | chrXIII:602993-602998 | YMR170c | ALD2 | *S. eubayanus> S. cerevisiae* at 86 (+3-2)/1521bp |
| | chrXIII:657849-657854 | YMR196w | Putative protein of unknown function | *S. eubayanus> S. cerevisiae* at 2807(+3-2)/3297bp |
| | chrXVI:97019-97048 | YPL240c | HSP82 | S. *eubayanus>S. cerevisiae* transition occurs at 1592(+15-14)/2130bp |
| | chrXVI:862792-862803 | YPR160w | GPH1 | *S. cerevisiae> S. eubayanus* at 1496(+6-5)/2709bp |
| | chrXVI:919950-919951 | YPR191w | QCR2 | *S. eubayanus> S. cerevisiae* at 574(+1-0)/1107bp |

[a]Breakpoint position in genome based on *S. cerevisiae* sequence UCSC SacCer2 June 2008
[b]Breakpoint position within hybrid gene. Brackets indicate extent of *S. cerevisiae*/*S. eubayanus* sequence overlap or ambiguity region.

**Figure 1**

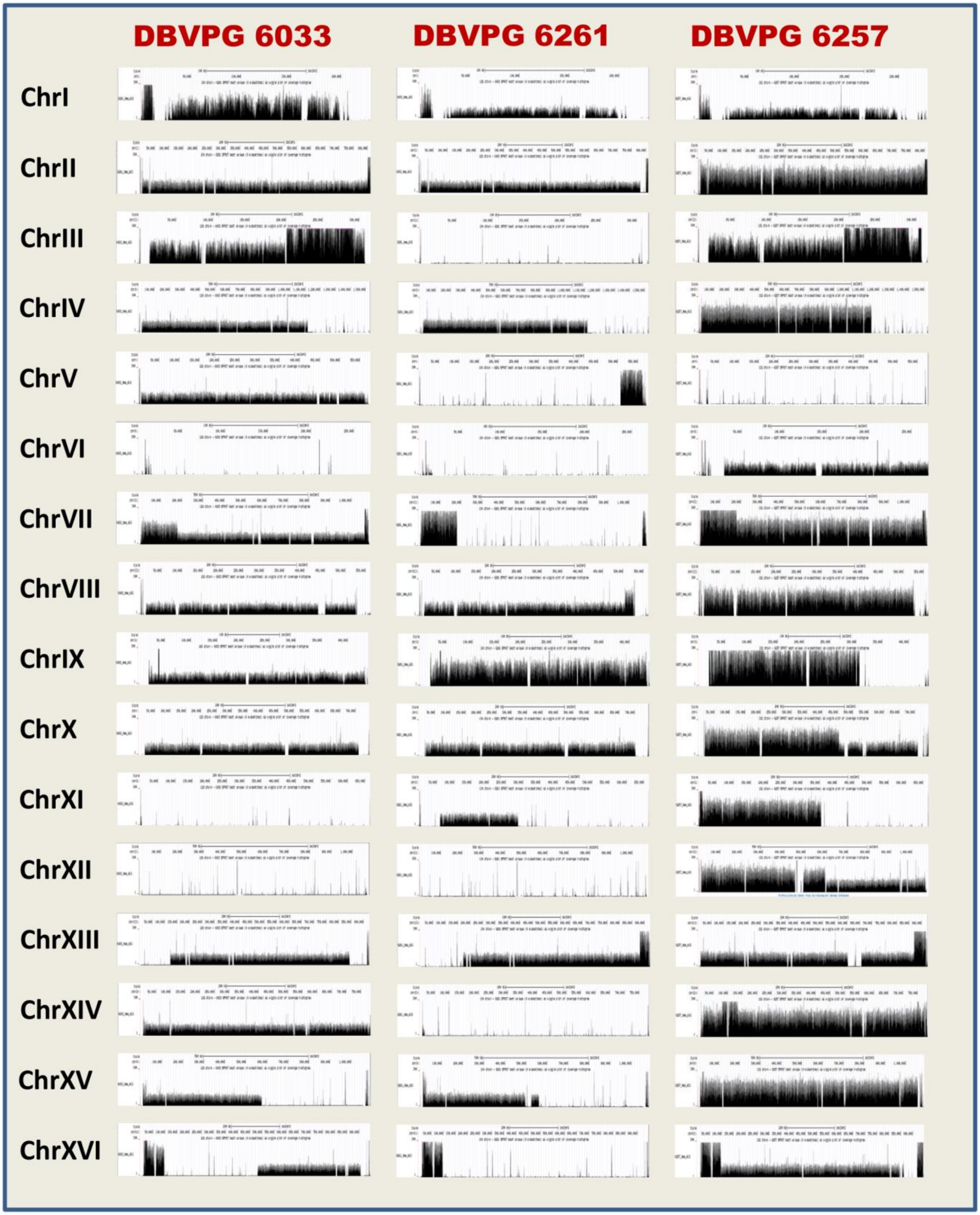

**Figure 2**

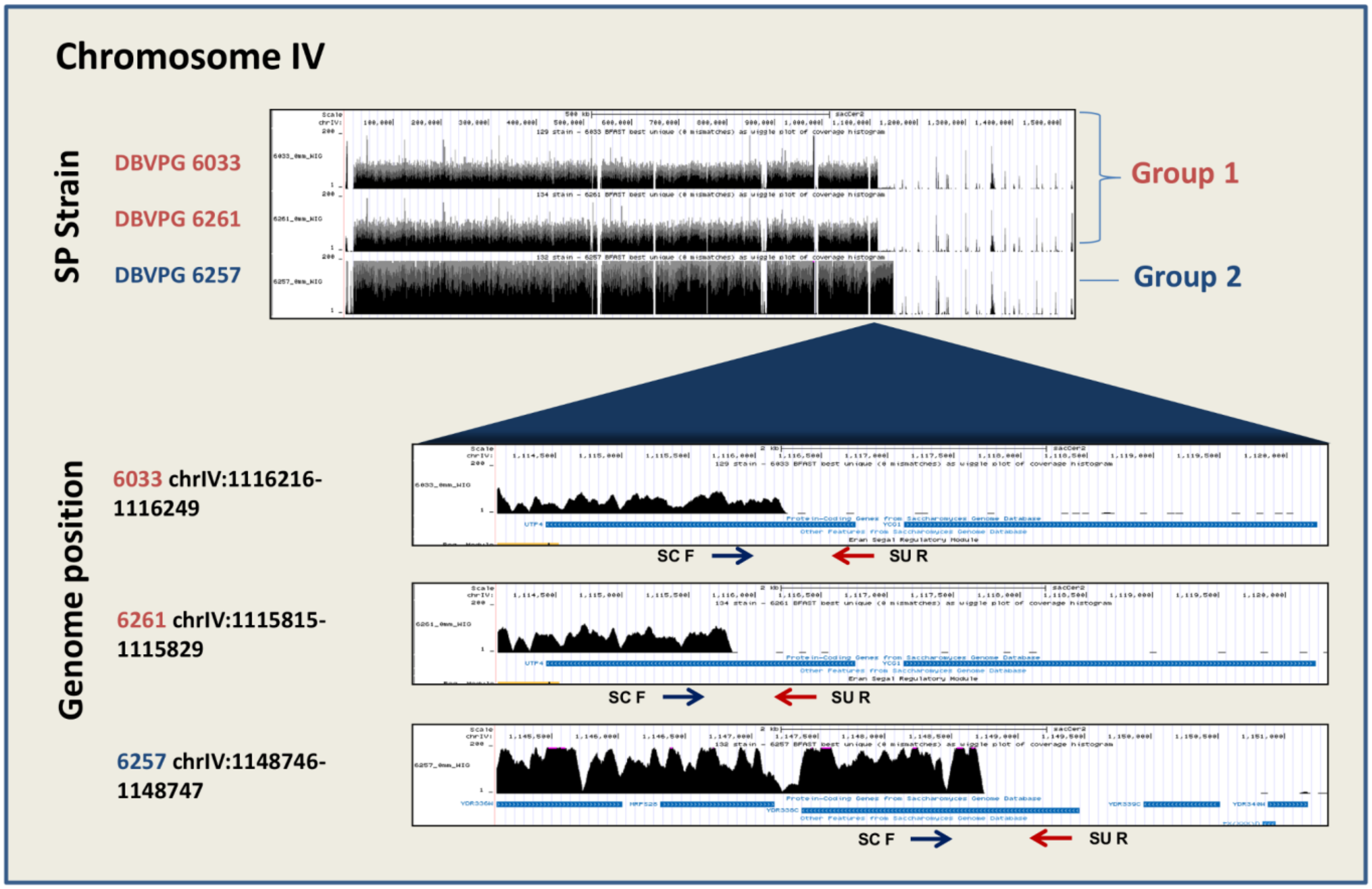

## Supplementary materials

## Table S1

| DBVPG strain | Genome position* | Systematic name | Standard name | Notes |
|---|---|---|---|---|
| 6033 | chrIII:201,200 | Intergenic | Intergenic | Encompasses MATALPHA |
| | chrX:710,200 | Intergenic | Intergenic | ARS 1022 |
| | chrXIII:123,200 | Intergenic | Intergenic | Apparent *S. bayanus* transposon |
| | chrXV:575,150 | YOR133w | EFT1 | - |
| 6261 | chrVIII:505,000 | YHR203c | RPS4B | - |
| | chrX:702,580 | YJR145c | RPS4A | - |
| 6257 | chrIII:201,200 | Intergenic | Intergenic | Encompasses MATALPHA |
| | chrX:710,200 | Intergenic | Intergenic | ARS |
| | chrX:531,500 | Intergenic | Intergenic | LTR |
| | chrXII:599,000 | Intergenic | Intergenic | TY1 LTR |
| | chrX:543,330 | Intergenic | Intergenic | LTR |
| | chrXIII:870,900 | YMR302c | PRP12 | - |

* Approximation based on reads mapped to *S. cerevisiae* sequence UCSC SacCer2 June 2008.

## Table S2. Analysis of the breakpoint region in each strain of *S. pastorianus*.

| DBVPG Strain | Genome position[a] | Systematic name | Standard name | Gene sc:su nt identity[b] | Microhomology [c] | Local homology[d] | Nearest genomic feature[e] |
|---|---|---|---|---|---|---|---|
| 6033 | chrII:780898-780906 | YBR289w | SNF5 | 79% | 8bp in overlap | 46bp at 2011-2057bp | 11.3kb to ARS |
| | chrIV:1116214-1116246 | YDR324c | UTP4 | 85% avail seq | - | 32bp at site of breakpoint, in overlap | 6kb from ARS |
| | chrVII:179656-179664 | YGL173c | KEM1 | 83% | 8bp in overlap 11bp in *S. cerevisiae* | 34bp at 499-533bp and 58bp at 832- 890bp | 6kb from ARS |
| | chrVIII:433730-433738 | YHR165c | PRP8 | 83% | 8bp in overlap, 14bp in *S. bayanus* region | No | 14kb FROM ARS |
| | chrVIII:451231-451276 | - | - | N/A | | | 3.4kb from ARS |
| | chrXIII:843626-843635 | YMR287c | MSU1 | 78% | 9bp in overlap | No | 5.9kb from Ty1 LTR |
| | chrXVI:97019-97048 | YPL240c | HSP82 | 87% 8bp indel | 19bp overlap | No | 19.4kb from ARS |
| | chrXVI:482992-483014 | YPL036w | PMA2 | 83% | 15bp in overlap | 53bp at 1575-2027bp | 28.9kb from ARS |
| | chrXVI:906846-906880 | - | - | N/A | | | 25.6kb from Ty4 |
| 6261 | chrIV:1115815-1115829 | YDR324c | UTP4 | 85% avail seq | 14bp in overlap | No | 5.6kb from ARS |
| | chrV:507240-507245 | YER164w | CHD1 | 83% | 5bp in overlap | No | 7.5kb from ARS |
| | chrVII:179656-179664 | YGL173c | KEM1 | 83% | 8bp in overlap 11bp in *S. cerevisiae* | 34bp at 499-533bp and 58bp at 832- 890bp | 6kb from ARS |
| | chrXI:60183-60193 | YKL203c | TOR2 | 82% | 10bp in overlap | No | 4.3kb from ARS |
| | chrXI:285492-285506 | YKL080w | VMA5 | 88% | 14bp in overlap | 32bp at 900-932bp | 16.5kb from TY1 LTR |
| | chrXIII:172148-172153 | YML051w | GAL80 | 84% | 5bp, 11bp in *S. bayanus* | 38bp at 346-383bp and 41bp at 907-947bp | 3.3kb from t-RNA-Tyr |
| | chrXIII:882708-882716 | YMR306w | FKS3 | 83% | 8bp in overlap | no | 15kb from ARS |
| | chrXV:496862-496866 | YOR092w | ECM3 | 83% | 4bpin overlap, 8bp in S.bayanus | 35bp at 1774bp-1808bp | 9.3kb from tRNA |
| | chrXV:526436-526442 | YOR109w | INP53 | 82% | 6bp in overlap | 39bp at 1261- 1298bp | 39kb from tRNA |
| | chrXV:561420-561424 | YOR127w | RGA1 | 76% | No | No | 5kb from ARS |
| | chrXVI:97019-97048 | YPL240c | HSP82 | 87% 8bp indel | 19bp overlap | No | 19.4kb from ARS |
| 6257 | chrIV:1148746-1148747 | YDR338c | - | 83% | No | No | 2.3kb from tRNA and TY1 LTR |
| | chrVII:179656-179664 | YGL173c | KEM1 | 83% | 8bp in overlap 11bp in *S. cerevisiae* | 34bp at 499-533bp and 58bp at 832- 890bp | 6kb from ARS |
| | chrIX:306349-306368 | YIL026c | IRR1 | 80% | 19bp in overlap region | No | 5.4kb from TY1 LTR and tRNA |
| | chrX:453941-453961 | YJR009c | TDH2 | 96% | 20bp in overlap | 77bp at 637-713bp | 0.6kb from ARS |
| | chrXI:354021-354023 | YKL045w | PRI2 | 82% | 8bp in *S. cerevisiae* region | No | 24.5kb from ARS |
| | chrXIII:602993-602998 | YMR170c | ALD2 | 81% | 5bp in overlap | No | 8.7kb from ARS |
| | chrXIII:657849-657854 | YMR196w | - | 84% | 5bp in overlap | No | 8.4kb from ARS |
| | chrXVI:97019-97048 | YPL240c | HSP82 | 87% 8bp indel | 19bp overlap | No | 19.4kb from ARS |
| | chrXVI:862792-862803 | YPR160w | GPH1 | 85% | 11bp in overlap | 41bp at 1449-1503bp | 2.3kb from tRNA |
| | chrXVI:919950-919951 | YPR191w | QCR2 | 82% | 9bp in adjacent bayanus seq | No | 13.2kb from ARS and Ty4 |
| | | | **Mean** | **83.5%** | | | **11.41kb** |
| | | | **Stdev** | **3.64%** | | | **9.2kb** |

[a]Based on *S. cerevisiae* sequence UCSC SacCer2 June 2008; [b]Nucleotide identity between *S. cerevisiae* and *S. uvarum* homologues (http://www.yeastgenome.org). ; [c]Regions of microhomology between 5-30bp observed between *S. cerevisiae* and *S. uvarum* homologues at the breakpoint site using SGD Fungal Alignment (http://www.yeastgenome.org). ; [d]Regions of local homology over 30bp observed between *S. cerevisiae* and *S. uvarum* homologues at or within a few hundred bp of the breakpoint site using SGD Fungal Alignment (http://www.yeastgenome.org); [e]Nearest repetitive feature to the site of breakpoint formation annotated in the *S. cerevisiae* genome. Feature distance obtained using the UCSC Genome Browser (http://genome.ucsc.edu/).

## Table S3. Primer sequences for amplification of breakpoint regions.

| DBVPG strain | Breakpoint location | Primers[a] | Primer sequence (5'-3') | Tm (°C) | Product size (kb) |
|---|---|---|---|---|---|
| 6033 | chrII:780898-780906 | 6033_SNF5_F (Sc) | GATCCATGAGTGGACAACCTC | 59.8 | 1.23 |
| | | 6033_SNF5_R (Se) | GTACCAGTCAACGCAAGCATA | 57.9 | |
| | chrIV:1116214-1116246 | IVUTP4_6261_F (Sc) | AGGGCCTCCCTACGACTAAA | 59.4 | 5.46 |
| | | IVUTP4_6261_R (Se) | CACCGTAAGGGCTCGAATAG | 59.4 | |
| | chrVII:179656-179664 | VII6261_172a_F (Sc) | AATGTGGCCCATGAGTAGACA | 57.9 | 1.95 |
| | | VII6261_172a_R (Sb) | CCGAACAGTCCTCCTGATGT | 59.4 | |
| | chrVIII:433730-433738 | 6033_PRP8_F (Sc) | GTTAACGAAGTTGGCACCCTA | 60.3 | 1.37 |
| | | 6033_PRP8_R (Se) | ACAACACTTGAGTGAAGCTTGG | 58.4 | |
| | chrVIII:451231-451276 | 6033_ENO2_F (Se) | AGACCACTTTTGTCGACCTTTTC | 58.9 | 2.10 |
| | | 6033_ENO2_R (Sc) | ATGCATATCACTGTGGTCGTG | 57.9 | |
| | chrXIII:843626-843635 | 6033_MSU1_F (Sc) | TTGGGTGTACCGAGTATCCTG | 59.8 | 0.92 |
| | | 6033_MSU1_R (Se) | GCTGCTCTCCCCTACATCAGT | 61.8 | |
| | chrXVI:97019-97048 | Hsp82_6261_F (Sc) | AAGTTGGTTCGTCCAAACTGA | 55.9 | 2.89 |
| | | Hsp82_6261_R (Se) | TATCGCGTCTTGGTCTTCTTG | 57.9 | |
| | chrXVI:482992-483014 | 6033_PMA2_F (Se) | ATAGTGACGGCGTGCATCTT | 57.3 | 1.67 |
| | | 6033_PMA2_R (Sc) | AAATCCGACAGATGCGTTTAAT | 54.7 | |
| | chrXVI:906846-906880 | 6033_GDB1_F (Sc) | CAAGCCTGGAGTTCTGCTTGT | 59.8 | 1.8 |
| | | 6033_GDB1_R (Se) | AGGATTTCTACTCGCGCTTTG | 57.9 | |
| 6261 | chrIV:1115815-1115829 | IVUTP4_6261_F (Sc) | AGGGCCTCCCTACGACTAAA | 59.4 | 5.46 |
| | | IVUTP4_6261_R (Se) | CACCGTAAGGGCTCGAATAG | 57.3 | |
| | chrV:507240-507245 | VCHD1_6261_F (Sc) | GCCAAGTACGTGGTCTGAAAA | 57.9 | 2.03 |
| | | VCHD1_6261_R (Se) | TGCGACTCTCCTAAGTCTGGA | 59.8 | |
| | chrVII:179656-179664 | VII6261_172a_F (Sc) | AATGTGGCCCATGAGTAGACA | 57.9 | 1.95 |
| | | VII6261_172a_R (Se) | CCGAACAGTCCTCCTGATGT | 59.4 | |
| | chrXI:60183-60193 | XITOR2_6261_F (Se) | CACGCTTATCGGAGTTTTGC | 57.3 | 2.0 |
| | | XITOR2_6261_R (Sc) | GCTAATGCAAGGGGTATCTCC | 59.8 | |
| | chrXI:285492-285506 | XI_VMA5_6261_F (Sc) | GACCTTAATCGGGGGTAGGG | 61.4 | 0.85 |
| | | XI_VMA5_6261_R (Se) | CCATAACGCAACACGGATTC | 57.3 | |
| | chrXIII:172148-172153 | 3rdGAL80_6261_F (Se) | CTACTGGGTACAGGGCAACAC | 61.8 | 2.05 |
| | | NEWGAL80_6261_R (Sc) | GTCGTTTGCTCTAGTTCCACTGTA | 61.0 | |
| | chrXIII:882708-882716 | FKS3_6261_F (Se) | CTAAGCTGTCGCTTTCAGACG | 59.8 | 2.19 |
| | | FKS3_6261_R (Sc) | AATGGGGTACCAAAACGGTAA | 55.9 | |
| | chrXV:496862-496866 | XV6261_093C_F (Sc) | TCTTTTAGGCGCTACTCTTGGT | 58.4 | 1.23 |
| | | XV6261_093C_R (Sb) | TAAGGCAGGCGGATTGTTAG | 57.3 | |
| | chrXV:526436-526442 | XV6261_109W_F (Se) | AGGCACCTCTCTGATCAACC | 59.4 | 1.18 |
| | | XV6261_109W_R (Sc) | TATTCCCAGCCATACCCCTA | 57.3 | |
| | chrXV:561420-561424 | XV6261_127a_F (Sc) | ACACTTTGGGGACCTGCTG | 58.8 | 1.8 |
| | | XV6261_127a_R (Se) | GCCCCTCCGAATCTTTCATA | 57.3 | |
| | chrXVI:97019-97048 | Hsp82_6261_F (Sc) | AAGTTGGTTCGTCCAAACTGA | 55.9 | 2.89 |
| | | Hsp82_6261_R (Se) | TATCGCGTCTTGGTCTTCTTG | 57.9 | |
| 6257 | chrIV:1148746-1148747 | 6257_338_F (Sc) | TAAAAGCATGGCACTTCCAATAC | 63.8 | 1.74 |
| | | 6257_338_R (Se) | ACAGTCGATTTCGGAAGTAACG | 51.2 | |
| | chrVII:179656-179664 | VII6261_172a_F (Se) | AATGTGGCCCATGAGTAGACA | 57.9 | 1.95 |
| | | VII6261_172a_R (Sc) | CCGAACAGTCCTCCTGATGT | 59.4 | |
| | chrIX:306349-306368 | 6257_IRR1_F (Sc) | ACAGACCGTGGAATACGTTCA | 57.3 | 1.19 |
| | | 6257_IRR1_R (Se) | AATAGTCACTGGCAACCTGGTC | 60.3 | |
| | chrX:453941-453961 | 6257_TDH2_F (Sc) | GGTCCTTCAGGCATCAAAT | 54.5 | 3.42 |
| | | 6257_TDH2_R (Se) | CGTAAGCCTGGGTAGTCATAAT | 58.4 | |
| | chrXI:354021-354023 | 6257_PRI2_F (Sc) | GCTAACCTCCACCGATCAAA | 57.3 | 2.13 |

| | | | | |
|---|---|---|---|---|
| | 6257_PRI2_R (Se) | GCTGGACGAAAGGAAGATTG | 57.3 | |
| chrXIII:602993-602998 | 6257_ALD2_F (Sc) | TAACCGTGTTACCGGCTGCT | 54.4 | 1.86 |
| | 6257_ALD2_R (Se) | AGTGGGGGTACGCAAAATCA | 57.4 | |
| chrXIII:657849-657854 | 6257_196W_F (Se) | AGGTCGAGTGTCCGGTAGGT | 61.4 | 2.46 |
| | 6257_196W_R (Sc) | GTTCCCTGGCATGAATACGA | 57.3 | |
| chrXVI:97019-97048 | Hsp82_6261_F (Sc) | AAGTTGGTTCGTCCAAACTGA | 55.9 | 2.89 |
| | Hsp82_6261_R (Se) | TATCGCGTCTTGGTCTTCTTG | 57.9 | |
| chrXVI:862792-862803 | 6257_GPH1_F (Sc) | CAACAACGCGCAGAGTCTAT | 57.3 | 2.5 |
| | 6257_GPH1_R (Se) | CTGGGACCCCAATCAAGATA | 57.3 | |
| chrXVI:919950-919951 | 6257_QCR2_F (Se) | ACGCTCCCAGTAAAGTGTCC | 59.4 | 0.925 |
| | 6257_QCR2_R (Sc) | GATTCATTTTGGACGGCATT | 53.2 | |

[a]Species specificity of each primer is stated in parentheses. Sc: *S. cerevisiae*. Se: *S. eubayanus*.

**Figure S1**

## DBVPG 6033

### SNF5/YBR289w, chrII:780898-780906

AATTAAACTTACCCAAGTACCAAACTATTCAATACGATCCACCAGAAACCAAGCTACCATATCCAACCTATTGGT
CAGACAAAAAAGCAGATACGGATACTTTGTTGTACGAACAAATTATCCAGCGTGATAAAATTAACAAATATTCGC
TAATAAGAGAAACCAATGGTTACGATCCGTTTAGCATTTATGGATTTAGTAATAAAGAGTATATTAGTAGACTGT
GGCATACACTGAAGTATTATCAAGATTTGAAGAACACTAGAATGAAATCTATCACAAGCACTTCTCAGAAGATTC
CTTCGGCAAGTATTTGGGGAAATGGTTACTCAGGATACGGGAACGGAATCACAAATACAACCACAAGAGTTATAC
CTCAAGTTGAAGTTGATAATAGAAAACACTACCTTGAAGATAAGTTGAGAGTTTATAAACAGGCAATGAGTGAAA
CAACAGAGGAATTGGTTCCGATAAGATTGGAATTCGACCAAGATCGTGACAAATTCTTCCTAAGAGATACATTGC
TATGGAACAAAAATGATGAACTTATCAAGATTGAAGAATTTGTGGATGACATGATGCGAGATTATCGATTTGAAG
ATTCGACCAGGGAGCAACACATAGATACCATATGTCAGTCCATACAAGAACAAATCCAAGAATTTCAAGGAAATC
CATATTTAGAATTGAATCAGGATCGCTTGGGTGGTGATGATTTACGAATTAGAATCAAGCTGGATATTGTGGTGG
GACAAAACCAACTTATCGATCAATTTGAATGGGATATCTCTAATANTGATAATTGTCCCGAAGAATTTGCAGAAT
CCNTGTGCCCAGAACTGGAATTGCCCGGGTGAGTTCNNTTACNTCCATTGCACACTCGATAAGAGAACAGGTTCA
CATGTAC

### UTP4/YDR324c, chrIV:1116214-1116246

TCTGTaCTaGAATCGGTTCCCATTGTCCATATCTTAACAGTAGATTCGCTCCATGAAACAACAAGGCGTTGCTCT
TTGTTAACTAAAACATTCTTTGAATAAGGTTCGACAGTTGGCATCTTCCTGtAGTTTCCATTAGAAAAAGAAGTA
AGTGAGTTGATGACTAGTGTTTTTTCAACACCTCCTGAAACTAGAAAATCTGCACCTTTAGATTGGTATGCACAT
ATTGCTCTAATGTCGTTTCCATGAAGCAACCTATTAGAAGAATTTACCCATCTGTTGTTCTTTTGAGATTTGTTA
GTGTTTTGAGAAAATTGAAAGATTTTTCTGTCCACACCAGCACTAAAAACATAATTATTATCAGTATCGGTAGTT
AGACACAGTACGTCTGCATCGTGCGCCTTAAATGACTGGTTTAGCGTGGCAAACTGGAAATCCCAGAATTTAATG
GAGCCTGTAGAATCACCAGAGGCAATCTGATCAGTTCTTGGTAAATATATAACTGACCAAACTAGAGTTGATTCT
TTTTTGGCCTTGTCGACCTTCATAGTGTGTAATAGACGACCCATGTTTTCGTCATTTTTTTGTGCAGACCAAATC
CTTATTCTACCATCAGAACAACCACCAATCACGAAGTCATCCTTTTTCCAAGCCAAAGTCAATACTCTGGCTTCT
TGTCTCATCAnAATAGTATCGTGTTCCAAGACACCCGGCCCCcCAGanaTATCTataAGTACCACAGTCCCATTA
TCGCAGCCCACTGACAGCTTATCTTGGGaTTCATTTATGGCAATAGACCAGATCACGCCTGCATTAcAGTCATAA
TTTCTTAATGGTAAACCCGTAgCCAAgTCCCATT

### KEM1(XRN1)/YGL173c, chrVII:179656-179664

TTGAAATCCTTTTGGGATTTTAAATGCCTTATAAAGTTCATGATCTTGTGTTCACCTTCACCTGGAACTTCATGG
CCAGAAAATATGATTTGCACTTCCCTCCATTTGGAATCGTTAGAAAT*T*TTGTCATGAATGAAATACTGTAAATTT
TTTGTCAGTTTAGCCATAAATTCAGTACCTGGAGTAATACAGTTTGAGTCGAACGGCTCACCTTTAGGAATTTCG
TCACCATTTTCAATAGCTTTTTTCATAGCTTTTTCGGCGTCCATGGCAGTTCTAAATCTACGAGCTCTCTGTTGA
TTCATCTTCGCGCGGGGGGCCACACCATCGATAGCCATGTAAAAAATTTGCTTGGGCTTGATAGTTTGAAAAAGA
TGATCAATATACGTGCAGATCTTTGCAAAAACCTCTTCTTCAGTTAATCGCTTCGTCACGTCATCGTCGTTACCA
TGTGTACAGTTATGTAAAATCGAATTCATATCCAGATATAGGTTATCAAACTCAGGAATCTGAGTCCCTTCAATA
AGTTGTAGAATCATGGGCCATCTTTCTGAGATGTACCTGAAAAATTTCGGAATACCCATGCTGTACTAGTATATG
TTGGTTACTGTTTTTACAGTTGCTGCTTTTTTCTTTATAGTACACGAATCGTCTCACGATTGGTCAATTGCTTAT
TGCAGCCTAAATATAA

### PRP8/YHR165c, chrVIII:433730-433738

TACGTCCTgaCGTATTAGCCTCATGCGCGAATCTTTGGGCCAACATTTTctaTTATTgtATCCGATGCTGTTTTC
AAAGTTTGGATCTGGATTCTCTGCCAAGTATTCATCAGTCAATTCCTCTCCCTCATCTTCTTCAAAGTGAAAAAG
CATATATATACGATCTAAATATCTGGTGTAAAGTCTGATCGGATGTGCCTTTTCTACTTCTTTGCTCTTGAATTG
CATAAATTCATTTGGATTGTTAGCTGGACCAGCCAAATCTGTAGCCCTTTCCTGACCCAATAGTAAAAGATCTAT
AACTAGTCCGTAATATTGGAATATGAAAGAAGCGAACTTCAACCCGCGTATTAAGCCATATTTGTTGACGTGACT
CATATCTTTAAAGTTGATAACAACATTATTTTTTGCAGTAATATAGTCAGCAATATTAGGATCTACAATCAGACG
AAGTAATCTATT*T*AACAAAGTAAAATCAATCTTTTCAGCCATTTCCCTTAACGTGGTCTCGAGCAAAACTGTGGA
TTGACCTTTAGAGACATCCCAAATGTCAGATAGATTGTTTATGCCTTGAGACCACTTGTATACCAGAAGAGGTGG
TATCTCTGAATCACTTGGTTTTATCCAGTTGGGGAAAAGCTTCCTTTGATCGGCTTCGTACCACAAATACTGATC
AAGATATGCATCGGTGATTTTTTCCAAGGGATCGACTGCATAAACAGGagaTATATTTTGATAGTTTTCCATCAT
AGTCAAATCAACAGGCTTAAAAACACGTTGCGTCAACAAGTATTTTTTTATTCTATTCAAAGTATCGTGGGGGTT
ATCATAAGCTTCCTCTATTAACGCCAGTTCTTCCCTTTCAgaTGCATTTAAACGTACTTTTGATGCATAAACGTC
CTTTagaTTTTCTAATGCAanaACCAAnATTTTGGTATCATTTTTgTAAGTT

**Intergenic, chrVIII:451231-451276**

TTGGTACCAGGTCAGCATACACACTACTAGGGTAGTTTCTTTGGCTGTATTGATCNTTTGGTTCAACGTGGTCCT
ACATTTTTTTTTTTCCATTGCATTCGGGCTTTCAACTTATTGTCTTTTGGATCTCTGTCGAAGGTCGGGGAAGTG
TATTGTGAGGAATTGGCACGAATATTTATAACGAAATGCTTTATCTTTTTTTTTAGGTTGATTGTTAAAAACGTT
TTTATAGCTAAGGCAATTGTAATTAATTCTCATTTGTATCTTCCCTTCTTTTATTTTCTTATATTTTCTTATTTA
TTTTTCTTTTATCTTATTTTCTTTCATAACCCCAAGCAACTAATACTATAACATACAATAATAAGGGCTGTCTCT
AAAGTTTACCCTAAATCCGTCTACNACTCCCGGGGNAACCCAACCGTCAAANTCAAATTANCCCCCGAAAAGGGG
GTTTTCAAATCCNTTGTTCCNTCNGGNGCCTCCACCGGTGTCCACAAANCTTTGGAAATGAAANATGAAAACAAA
TCCAAGTGNATGGGTAAGGGTGTTATGAACGCTGTCAACAACGTCAACAACGTCTTTGCTGCTGCTTTCGTCAAG
GCCAACCTANATGTTAAGGACCAAAAGGCCGTCAATGACTTCTTGTTGTCTTTGGATGGTACCGCCAACAAGTCC
AAGTTGGGGTGCTAACGCTATCTTGGGGTGTCTCCATGGCCGCTGCTAGAGCCGCTGCTGCTNAAAAAGAACGTC
CCATTGTACCAACNTTTGGCTGACTTGTCTAAGTCCAAGACCTNCTCCATACGTTTTGCCAGTTCCATTCTTGGA
ACGTTTTNNAACGGGNGGTNCCCACGCTGGTGGNNGNTTNGNCTTNCANGAANTCATGATTGCTCNNCTGGGT

**MSU1 (DSS1)/YMR287c, chrXIII:843626-843635**

AnTGCCTTTCCGGATTttaAACTGTCATACATAATTTCAATTGATTTTCCACTGCATTTAGAAGTGATTTTAACA
TCAACAGAGAAGGATATAGTTTTCGTCCTTTGTCCTTGTTTCCCCAAATCTGATAAGTGACAAATAGATTGAGGT
AACATAGGAACAACCGTATCTGGTAAATATGTGGTAAATGATCTTTTCAAGGCAACATTTAGGATATCTGTACTT
ATACCCTCAATATCAACATTAGTGCTTTCTGGAAACAAAGAAGTAGGATCTGCAATATGGATATGTAAAGTGTAC
AATCCGTCCTTTTTGTGGTTTTCTATCGAGATTCCATCGTCAATCTCATGCGCAGTCTCCGAATCTATACAAAAA
ACCTTAAGATCACCAAAATCATATCTATCACCATTAATATTGGAATTACTTGCTTGCAGGTTTTCTATGTCTGTC
AACTCATAAAGCTTTTGCTGTGATTGCCCCAATTTCGAAGATGCAGGCAAAGCTAAATCCATATTCAGTAACAAC
GGATTGATCATTTTATTGGGTAAAATTTCATTGACCAGATCTTGACATATATCTCTTGTTATATCAAAACTCTTA
TAGAGTTCTATCTTTCTAAATATTTTGGAGATTAAGGTGACTATAACTCCATTATTGTGGAAGTTTCCCGCGGCA
AAATCTTTCAGCAGCTGAATGACAGAAGGGTATAAAGATGAGATATCTTTATATTTTCTTTCGTTCACCAACTTT
ACAAATCGGTCGATGTCCTTGTAGTTGTCAGCTTCTAGCTTCTCTATTACCTGTGCGTAGTATAAGTGTTGGGAT
TTCAAAGGGATGATTGTAACTGATGTag

**HSP82/YPL240c, chrXVI:97019-97048**

GTCCTTgACAGTCTTGTCTTGAGCACCACCTTCGTCAACTCTCTTTTTCAATTCCTTGATAATTGGAGATTTTGG
AGAAATTTCGAAAGTCTTCTTGGAAGACATGTAGGAGGACATGGAAGAGTCTCTCAAGGCTTGAGCCTTCATGAT
TCTTTCCATGTTAGCAGACCAACCAAATTGACCAGTTCTGATAGCAGCTGGGGCATCCAACAATTTGTAAGAAAC
AACAACTTTCTCCACTTGGTCACCCAAAATTTCTTTCAAGGCCTTGGTCAATGGTTCATATTCTTTGATCTCCTT
CTCTCTTTCAGCTTTTTCTTCGTCAGTTTCTTCCAGTTCGAAATCCTTGGTGATATCGACCAAAGTCTTACCTTC
GAATTCCTTCAATTGGGTGAAAGCGTATTCATCGATTGGATCAGTCAAGAACAAAACTTCAAAGTTCTTAGCCTT
CAAAGCGTCCAAGAATGGGGATTTTTCAACGGCCTTTAGAGATTCACCTGTGATGTAGTAAATGTTCTTTTGGTG
TTCTGGCATTCTGGTGATGTAATCGGTCAAAGAAGTTAATTCGTCGACGGACTTGGTGGAGTTGTAACGTAGTAA
CTTAGCCAAGGCAGCTCTGTTTTGAGTGTCTTCGTGGACACCCAATTTGATGTTCTTAGCGAAGGCAGAGTAGAA
CTTTTCGAATTGTTCAGAGTCTTCAGCAATTTCGTTGAAAGCTTCAATTACCTTCTTGACGATGTTCTTTCTGAT
AACCTTCATGATCTTGTTTTGTTGTAGCATTTCTCTGGACAGGTTCAaTGGTAAGTCTTCAGAGTCAACAACACC
CTTGACAAAGGACATCCATTCTGGGATCAAGTCTTCAGCTTCGTCGGTGangAaAACACGACGAACATACAACTT
GATgTTgTtcttctTCTTCTTACTTTCgAATAantCAAATGGannTCTCTTTGGaATGTACAagaTGGCTCTGAA
TTCTaACTGACCTTcnaCGGaaaagtGCTTAAcataTAGtGGgtctTCCCa

**PMA2/YPL036w, chrXVI:482992-483014**

aTnCnaCcaaTCAgacaACCGGCAGCCAAAATAGCAGCTGCCTCCATAACGAATTGAATCGGCCCTACGAAGAAC
ATCAAAAACTTCACAATCAACGATTCATTCTCCTCAGCCATTTGATTTAACCCATATTTCTTTCTTCTCCTGGCG
ACTTCATCCGAAGTCAAACCATACGCAGGgtCCGTAGAAAGGTCCTTTTCAGGAACAACCCTTTGGCCAGCGTGC
ACCCCATCAGTACGTATTTCTTCTTCACCAGATTCATCACCCTCACCGTAGTTAGATTGTAGTTCATCAATCAAT
TGGTCTATATCTTCATCTGAGTCAGAATCATCAGCGGCAGAAGCGGCTGCGGCCTTACGTGGAGCGGCTGCAGGC
GATGGCGAGGCCGATGTAGAAGAAGACGAAGACGAAGACGAGGACGAAGCACCATTGTTCGCTGCAGGCTTCTCT
CGGTATGGCTTTGCACCACTAGAAGACATAATACAAAAATGTTTTCTCTTTCTTGAATCGCAATGCTACACAGTT
AAAAATGAAAAAGAAAGCACAGTAGCTTGAGCAGCATACGCCATACAGAAAGAACAAGAGAGATGGACAGGAAAG
GAGACACGATACGCCTTTATATAAGCGCAAGTACAGTAAGCGAACCATCACCACTGCTACCAATCACAGTTTAGG
GGACGTTGACAAGAGTGTGTTCATTGCTATGTGTTTGTCCATTGATAACCATTTCCCCCTTCAACATTCGTTCGT
CGACGTTCATCCTCCCTTTTTCAGTGCAGGCTGCCCCGagaATTGACTTTGCAATTCCCTTAAACAaganATTCT

TtTTGTTCTCCCTCGACGATTTTTGGAGTAAAGGAgACCTTACGCCGGCTTTCCTTTTCCTTGCAGCAATGCTTT
CTTATTCTTGCATGGGTTTCTTTTTGTcacTGCCtTCcCTAATCcgCaCTCGTGCcnnatt

### Intergenic, chrXVI:906846-906880

ctCCGTaGAAgattGGTATCGGCTcnnnnctGTAGAACATATTAGCAATGTTGCTTTCAAGAAAAAACTATCTAT
CAATTGAAGAACCTCTTTCTCGTTATTCAGATTGGCAACCATGATTCATACTACTCGATCGAGTGGTATGTCTAT
TATTTTGTACAACACCTCTTTTCGAAAGCGTAAATGGGATGCACTTTATTCCCGTAAAGTTGGCTCTATAACTAT
GTGACAGAATGGGAATCTCCACACCTTTGTTGTGCTTTTTTTCTCTATCTTTACCTGCCGCTGTTTACTTGGCTA
AAGATCTAAATCCAAGCGAACTAAACAAAAAATAAGGATACTAACTGCCTGTATTACACAAATGGCTCTCTATCC
TCTTCTATTACGTACGTTACAGCCCTTTTGCCACTCGTTCCGACGTGCCTTGCCCCTTTGGTAACTTTTGTCTCT
CACTTTCACGTGTGTTCCCTTTATTTCTTTAATTt

## DBVPG 6261

### UTP4/YDR324c, chrIV:1115815-1115829

gnncCCATTGTCcaTATcttaACAGTAgaTTctctcCATGAAACAACAAGGCGTTGCTCTTTGTTAACTAAAACA
TTCTTTGAATAaggTTcgACAGTTGGCATCTTCCTGTAGTTTCCATTagaAAAAGAAGTAAGTGAGTTGATGACT
AGTGTTTTTTCAACACCTCCTGAAACTAGAAAATCTGCACCTTTAGATTGGTATGCACATATTGCTCTAATGTCG
TTTCCATGAAGCAACCTATTAGAAGAAttTacCCATCTGTTGTTCTTTTGAGATTTGTTGCTATTTTGAGAAAAT
TGGAAGATCTTTctATCCAcaccatcACTAaacacGTAATTGTTAtCAATATCagTAGTTAggcaTAagacgtCT
GCATCATGTGCtttaaAagac

### CHD1/YER164w, chrV:507240-507245

AanatcaTccaGTTAAGATCCTCCAATTTTTTCTGGTTGTCTGTGGCAGTGAACATaTTGCCTGCACCGAACTTC
AGGATTGCCGATAATTCACCGGCATTTGGTTCATTCTTCTTAGTGTATTTGTTACCATCTGTCACTCCAAGAGAA
ATAATAGCATATTCCAGAATCATTTTCTTCCGTGCTCTTTCTAATACCTCTTCCTCTACTGTGTCTTTTGAAACC
AACCTATACACCATAACGTGATTTTTTTGGCCAATACGATGAGCTCTAGCCATTGCTTGTAAATCGGCTTGCGGA
TTCCAATCGGAATCAAAAATCACAACGGTATCGGCGGTCATTAAGTTGATACCCAAACCACCAGCACGAGTAGAA
AGTAAAAATACGAAGTCGTTTGAATCCGGAGAGTTGAAATGATCAATAGATATTCTTCTTTGAGCAGATGGTACT
GTACCATCTAACCTTTGGAAGTTAATACCTTTAATGGATAAATAGTCGCCTAAAATGTCAAGCATTCTGACCATT
TGTGAAAAAATCAACACGCGGTGCCCATCTTTCTTCAATCTGGTCAATAATTGGTCTAAAAGAACCATCTTACCC
GAAGACATGATCAAACCTCTTAGTACGTTTTCTCGAGTCATTTTACCATCCCCAAATTTCTGTAAGACGCGCTCT
TCAGCATTATCGAAGagaTATGGATGGTTCGATGCCttTTTc

### KEM1/YGL173c, chrVII:179656-179664

TanaTGtCTngTatTCTGGTTGAaTCCTTTtGGgatTTTAAATGCCTTATAAAGTTCATGATCTTGTGTTCACCT
TCACCTGGAACTTCATGGCCAGAAAATATGATTTGCACTTCCCTCCATTTGGAATCGTTAGAAATTTTGTCATGA
ATGAAATACTGTAAATTTTTTGTCAGTTTAGCCATAAATTCAGTACCTGGAGTAATACAGTTTGAGTCGAACGGC
TCACCTTTAGGAATTTCGTCACCATTTTCAATAGCTTTTTTCATAGCTTTTTCGGCGTCCATGGCAGTTCTAAAT
CTACGAGCTCTCTGTTGATTCATCTTCGCGCGGGGGGCCACACCATCGATAGCCATGTAAAAAATTTGCTTGGGC
TTGATAGTTTGAAAAAGATGATCAATATACGTGCAGATCTTTGCAAAAACCTCTTCTTCAGTTAATCGCTTCGTC
ACGTCATCGTCGTTACCATGTGTACAGTTATGTAAAATCGAATTCATATCCAGATATAGGTTATCAAACTCAGGA
ATCTGAGTCCCTTCAATAAGTTGTAGAATCATGGGCCATCTTTCTGAGATGTACCTGAAAAATTTCGGAATACCC
ATGCTGTACTAGTATATGTTGGTTACTGTTTTTACAGTTGCTGCTTTTTTCTTTATAGTACACGAATCGTCTCAC
GATTGgTCAattGctTATTGCAGCCtaAATAtaaCGCAGtgnangGATATAacgC

### TOR2/YKL203c, chrXI:60183-60193

gaagtCGGTACCCAATTGCAaaAGAAGCAAACTCAATGTATTCATGGTtGctTTTGTTAGTTCTCTGTCCCCgtT
ATTCAAGATTCGCACCAGTGCCTGAACAATCCTTGATGACATTTCTGAGAGGTTGATATTCTTTGCTAATCTCCC
CAACGTAATGATTGCAAtTTTTTTTAAACTTCCAGCGGAGTATTCAGTCATTCTGACGACTACGGGCATAATTAA
ATGGGAGTAATCCTCTAAATTGGATCCAAAAGTCACTAAGGACTTCAGTATACGGATAGAAACAATTTTTTTATT
AGACTGGTCGTTCTCAAGAATATCAAGGAAAAAGGTTAGGGTCTCGGGAACAAATCTTTTAAACTCACCTTCCAG
AGCCTTAGATATCGATTCTATGACAGAAATAATTGTGATTTGTAGTTTAATGATCGGGAAAAACTCCCTGATCAC
ACCATAAATTTTCTCGACATGGGGCCTAATATGTTGCTTGACAATTGAGATGAGAGATCCCAGTTGCTGAAAATA
AAAGTCAAGTTGGGACAGCGGGCACGAACGCATGACTAAAATGATACCTGGAATAATTTGATCCAAAAAGGAGAC
ACATCGTAAAGCAAGGTTTTGAAAAATATGCATAATAGCTTGaAtngcaggnngTGTGATGGATTGACAaCGATG
GATCATTCAATATCTtCATCAGATTnngAtT

**VMA5/YKL080w, chrXI:285492-285506**

```
atcAAAGAtTtgaTAACaTTGATTTCtAATGaAtCTTCTCAATTAGACGCCGACGTCAGAGCTACTTATGCAAAT
TACAACAGCGCTAAAACTAACTTGGCTGCTGCTGAGAGAAAGAAGACGGGTGACCTTTCTGTCAGATCCTTGCAT
GATATTGTCAAGCCCGAAGACTTCGTTCTTAATTCTGAACATTTAACTACTGTTCTAGTAGCAGTTCCCAAAAGT
TTAAAATCCGATTTCGAAAAATCGTACGAAACTTTATCCAAGAACGTTGTACCAGCATCTGCCAGCGTGATTGCA
GAGGATGCTGAGTATGTTTTGTTCAATGTTCATTTGTTCAAGAAAAACGTTCAAGAATTCACAACAGCTGCTAGA
GAGAAGAAATTCATTCCTCGTGAATTTAACTACTCGGAGGAATTAATTGACCAGTTGAAAAAAGAGCATGACTCT
GCTGCAAGTTTAGAGCAATCTTTacgTGTCCAACTAGTTAGATTGGCCAAGACTGCTTATGTTGATGTCt
```

**GAL80/YML051w, chrXIII:172148-172153**

```
angctCatCTaTCnGCTCTTGCTCTGGaaTATTATTGAAAACCATTGCATTTATCCTGGAAAAGTACGAACTTGT
CATGTATTGTAAAATATCGATTGTGTGACCAAATGTTGTGGTTACCAGATCTACACCGTTCCCGATTTCATAGAT
GTATTTTGGTGATTTAACAGGCCTTTCGTAGCCATACCAACCGCCGTTTCCTGCAATTTCTATGGAGTTGATGTC
GCCAATGTAACCTTCAGAAATTAGTTCTTTGGCTCTCAAGATGTATGGTGATTTACGGCCTTGTAAAGAAATAAT
AGTTTGTAGTGCACGTTCGGCAGCTGCCTTATAAATCGATTCTGCTTGATCCAGAGAACATGCAAGGGCCCATTC
TACGAAAAGATACTTGAGATTCGGATTATTTTGGGAGTATTTCAACAGGGGCATTAGAACCTCATAATGGCTGGC
CACCTGGATGGTTATCACTATCATGTCCACGGTGGCAGAAGATGCAAATGATTCTAAAGTGGGGAAGGCCGTGGC
ATTGCTAAGTTTCAGTTGTTGGATAGTGGCAATGGAGGTCTCTATCTTTGGGTTGTATAAGGCAGTAATTTGAAA
TTGGGACGATAGCTGCAGTATGGCGGGATAATGAGTTTTGATCGCCCACCCTTTGGTAGCGTTGAGGCCAATGAA
TCCGACTCTTATGGGGGCTGCATTGGGTACTGTAGAGACGGAAGATCTCTTGTTATAGTCCATGACTGGAAGGCA
CGAGAAAACTAGAGGTCGAACTTGTATACGGAAGTGGAGTAGTCTGCAGTGGCCAACAATGATGTAAGGTATGTG
TTTACGTGCAAAAAGAAAAnAAAAntgTGATTTTTTGGATaTTAcncatatacCAagCAaacatTTTcgcAGTTT
TTCGGGCAanagtGctcCGGCAGTnnaAAAAATGtaggnnntTATTATTatTAAnATTatACTATTatattg
```

**FKS3/YMR306w, chrXIII:882708-882716**

```
ccTTGttCgAGtggGgTTTTnnnncTcgAGAATGGCCTGGCGCTCaacATtnntcACGAAGAATGATTGGCCTCC
TTGTTTGtntCgTAATCAATTTAGGACCATCCATTTATGTTTTGGGGTTTTTCGAGTGGGATGTCCATtcgAAAT
CAGCATATATCGTGtCGATCATTCAATTAATAATTGCACTTCTAACCACCCTTTTTTTtGCTATCAGGCCCTTGG
GCGGCCTATTTCGTCCATATCTGAATAAAGACAAAAGGCATCGAAGATACGTCTCATCTCAGACTTTTACCGCTT
CGTTTCCCAAGCTGGCAGGAAGAAGCAAATGGTtctcTTACGGGCTATGGGTATTCGTATTTTTGGCAAAATACA
TTGAGTCCTAtTTTTTTTTGACCTTGTCCCTCAGGGACCCCATCAGGGTCCTCTCTATTATGGATTTGtCCAGAT
GTCAAGGTGAATATTTGTTGGGTCCTATTCTATGTAAATGGCAAGCCAAAATTACATTAGTTCTCATGCTGCTTT
CTGACTTGGGCCTGTTTTTTCTCGACACTTACCTTTGGTACATTATTTGCAACTGTATTTTTTCCATTGTACTGT
CATTTTCCCTTGGTACTTCAATTCTCACGCCATGGAAGAATGTATACTCTAGATTGCCAAAAAGGATATATTCCA
AAATCCTTGCTACTTCAGAGATGGATGTAAAATTTAAAGCAAAAATACTGATATCGCAGGTTTGGAATGCCATTG
TTATATCAATGTATAGGGAACATCTTCTCTCCATTGAGCATTTACAAAGACTCTTGTTTCAGCAAGTTGACTCCT
TAATGGGAGACACAAGAACCCTGAAATCGCCTACATTTTTCGTTGCACAAGATGATTCCACGTTCAAGTCAATGG
AATTTTTTCCATCAAATTcagAGGCAAAAAGAAGGATAtCCTTTTTGcCCAatcCCtGgCGACCCCCAtT
```

**ECM3/YOR092w, chrXV:496862-496866**

```
tggaatcCGCAGTGGTATTAGTCTTTCTCAGACAATGTATCATGccnnnctTTGGTGTCTTGTGGTGTGACCGTC
TAGTGAAGGCAGGATGGCTAAATTGGGAAAACGACAAGATGTTATTGTTTGTTACCGCTATTACTTGGAACTTAC
CAACAATGACCACCTTAATCTACTTCACTGCAAGTTATACCCCTGAGGACGAAACTGAACCCGTTCAGATGGAAT
GCACTTCTTTCTTCTTGATGCTTCAGTATCCTCTGATGGTCGTTAGTTTACCATTTTTGGTGTCTTATTTCATAA
AGGTACAAATGAAACTATGATTTTCTCTAGAAATCTAAAACATTAAAAAACAAAAGATCCACCCCATGCATTCCC
CTTCTCATTTTTATTTATCATAACCGAAGCAAGTCACAATTGTTGTTTTATTGATCTTTTACATCTGCTATATAA
AAAGTACTACTTCCCTTGCTTTTGTTATCGAAGTAGAATTcGCTTGATTCAAACTTTATAGAGTCAAAAAGTATA
TAAACATAAAAAAATaatTaTAAAATCCAcg
```

**YOR109w/INP53, chrXV:526436-526442**

```
tnagcCagtgTTTGACAAGCACATTATGAAGTCAGTAGAAAAATACGGCCCCgtTCACGTTGTCAACCTGtTATC
AACAAAATCTTCAGAAATAGAACTTTCAAAGCGATACAAAGAGCATCTAACGCACTCAAAAAAGTTGAATTTCAA
TAAAGATGTATTCTTAACAGAGTTCGATTTTCACAAAGAGACTTCTCAGGAAGGGTTTTCCGGTGTCAGAAAACT
TATTCCATTAATATTGGACTCTCTTTTATCTTCTGGCTATTATTCTTACGATGTTAGAGAAAAAAAGAACATATC
TGAACAACATGGCATATTTAGGACCAACTGTTTAGATTGTTTGGATAGAACAAATTTAGCTCAGCAAATTATTTC
TTTGGCTGCTTTTAGAACTTTTCTCGAAGATTTCCGATTGATTGGTTCAAATTCGTTCATCGACGATGATGATTT
```

CGTTTCTAAACATAACACCCTGTGGGCTGATCACGGTGATCAAATATCCCAAATATATACTGGTACTAATGCTTT
GAAGTCCTCCTTTTCAAGAAAAGGTAAAATGTCACTTGCTGGGGCATTATCAGACGCCACAAAATCGGTCAGCAG
AATATATATTAACAATTTCATGGATAAAGAAAAGCAACAAAATATCGATACTTTGTTGGGAAGGTTACCGTATCA
GAAAGCAGTGCAaCTTTATGATCCCGTAnACGAATACGTAAGTACGAnATTACAAAGCATGTCTGATAAGTTCAC
ATCAACCTCCAACATTAACTTGTTAGTAGGATCATTCAATGTTAATGGAGcnnCCAAgnnaGTTGATTTATCnna
GTGgttATTTCCAATc

**RGA1/YOR127w, chrXV:561420-561424**

tattatagcTTTTTGnnnaannnaAgGATAGCTGAttcAGGTACTAGTGGtGGAnngngcgGCAtaTTAAAATgg
cATCAACTgnnCCCAATGaacaaTTTCcntCCTGCGTACGAtgcAAaGAATTtATTACCACGGGGCATGCATaTG
AGTTGGGTTGTGATAGATGGCACAcacATTGTTTCGCTTGTTACAAATGTGAGAAACCATTAAGCTGCGAATCTG
AtTTTTTAGTCCTTGGAACAGGTGCTTTGATCTGCTTTGATTGTTCCGATTCTTGTAAAAATTGCGGTAAAAAGA
TTGATGATTTGGCCATAATACTGTCTTCGTCAAATGAAGCTTATTGTTCAGATTGTTTCAAATGTTGCAAGTGTG
GTGATAATATTGCTGATTTGCGATACGCAAAGACCAAGCGGGGCTTATTCTGTTTAAATTGCCACGAGAAGCTAT
TAGCCAAAAGGAAATATTACGAGGAGAAAAAAAGACGACTTaAAAAAAATTTGCCCAGTCTTCCTACTCCCGTGC
TTGACAATGATTCTATTGATGTAACTTCAATTACGGCAGTTGCACCCAAAAGGTCATCTAGTAGACCTGTATCAC
CGGTTAAGAAAATATCCTTAGAGTCCGAATCTATGAAAGATATAGCAATAGAAACCAACTCGAGCGATATCATTC
CGCATTTCATCACTGGGTATGATGATAGCGACGATAATTCTGGAAGTTCGAAATTCGGTTCTAATATTTCAATAG
ACATTATAGAACCACAGCAAGATAGCGCGGAGCATGCAAAAGATGAGAAAGTAGAAGAGGTGAAGGTGCATTCTA
GAAACGCATCGCTTGACATAGCATTGGATGCTACTCCAAGTCATAAGGTCTTGTTGGGCAACACGGAACctCCGA
GCCGCTCTAAGATTTTATTAAACAAAACACCATTAAGGAATTCATCCGGACAATATGTCGCAAAATCTCCAAGTT
CTTATAGACAGGGTATAGTTGTTAATGATAGCTTtGAGGAAAaTAACCAGGTCGAACctCCAAAtGACGgtCCCG
AACCGCAAnnag

**HSP82/YPL240c, chrXVI:97019-97037**

GTCCTTGaCaGTCTTGTCTTGAgcACcACCTTCGTCAACTCTCTTTTTCAATTCCTTGATAATTGGAGATTTTGG
AGAAATTTCGAAAGTCTTCTTGGAAGACATGTAGGAGGACATGGAAGAGTCTCTCAAGGCTTGAGCCTTCATGAT
TCTTTCCATGTTAGCAGACCAACCAAATTGACCAGTTCTGATAGCAGCTGGGGCATCCAACAATTTGTAAGAAAC
AACAACTTTCTCCACTTGGTCACCCAAAATTTCTTTCAAGGCCTTGGTCAATGGTTCATATTCTTTGATCTCCTT
CTCTCTTTCAGCTTTTTCTTCGTCAGTTTCTTCCAGTTCGAAATCCTTGGTGATATCGACCAAAGTCTTACCTTC
GAATTCCTTCAATTGGGTGAAAGCGTATTCATCGATTGGATCAGTCAAGAACAAAACTTCAAAGTTCTTAGCCTT
CAAAGCGTCCAAGAATGGGGATTTTTCAACGGCCTTTAGAGATTCACCTGTGATGTAGTAAATGTTCTTTTGGTG
TTCTGGCATTCTGGTGATGTAATCGGTCAAAGAAGTTAATTCGTCGACGGACTTGGTGGAGTTGTAACGTAGTAA
CTTAGCCAAGGCAGCTCTGTTTTGAGTGTCTTCGTGGACACCCAATTTGATGTTCTTAGCGAAGGCAGAGTAGAA
CTTTTCGAATTGTTCAGAGTCTTCAGCAATTTCGTTGAAAGCTTCAATTACCTTCTTGACGATGTTCTTTCTGAT
AACCTTCATGATCTTGTTTTGTTGTAGCATTTCTCTGGACAGGTTCAATGGTAAGTCTTCAGAGTCAACAACACC
CTTGACAAAGGACATCCATTCTGGGATCAAGTCTTCAGCTTCGTCGGTGATGAAAACACGACGAACATAcaACTT
GATGTTgntcntcntCTTCTTACTTTCGAAtaaGTCAAATGGancTCTCTTTGGAATgTACAagaTGGCTCTGAA
TTCTAACTGACCTT

## DBVPG 6257

**YDR338c, chrIV:1148746-1148747**

aCaTGACAGCAAAAggtataTATATGAccnnngaAAAAGCAATGCAACGTTGAaggtGAActnctaCACTGTAGA
annnncCAGAGCCATACGCTTGAGGGCATAGAGTATctAGACTAGTGGCAATACcctcgAATAtcgCTagtGtTA
TATTAGAAGTCATGGatgCTAAGGACACAGctgCTAGTTCATTTTTGcCTaagTGGCCCACAGTTAATGAACATA
CCATAGGGAAAAtctGTTCCAATAAGAATGTGAAAATTAAAGGGAAAGAGTAAGATGCTAATACTTTTGCTTCTG
ATTTGAAGGTCACATTACTATGAGCCAATGCAGCTAGTGCTTCTGGATCATTCTCATCTAGGTCTTGATATATCG
TGGGCACCGAACCTCGACTGTTTACAGATACGAATGATGGCCTTCTACGATGTGAGATACCTTTTGCATCCATTG
TTGCATTTTGGTTTTGTTTCTGTTTCTGGGCTTGAAGCCTGTTCATAAATTCGTCATATTCCAGTTCTCTTCCGG
GATATCCTTGTGTTCTATCGGGTATGTTGCTTTGACACTCCTCCTCATTATCAGAGTTATACACTGACGAGTATC
GTCGTTCATGTTCATGTAGTATCCAGTCAACGTCTTGTGGAGAAAGTTCAGGAGTCTCTGACACTCTTGAAGGTA
AAGAAATGGTTCTACTAAGCTCGCCATCCTCCTCGTCGTGGTAATATAAATCAGg

**KEM1/YGL173c, chrVII:179656-179664**

GTAanaCnaTGTCTCGTATTCTGGTTGAAATCCTTTTGGGATTTTAAAtGcCtTATAAAGTTCATGATCTTGTGT
TCACCTTCACCTGGAACTTCATGGCCAGAAAATATGATTTGCACTTCCCTCCATTTGGAATCGTTAGAAATTTTG
TCATGAATGAAATACTGTAAATTTTTTGTCAGTTTAGCCATAAATTCAGTACCTGGAGTAATACAGTTTGAGTCG
AACGGCTCACCTTTAGGAATTTCGTCACCATTTTCAATAGCTTTTTTCATAGCTTTTTCGGCGTCCATGGCAGTT
CTAAATCTACGAGCTCTCTGTTGATTCATCTTCGCGCGGGGGGCCACACCATCGATAGCCATGTAAAAAATTTGC
TTGGGCTTGATAGTTTGAAAAAGATGATCAATATACGTGCAGATCTTTGCAAAAACCTCTTCTTCAGTTAATCGC
TTCGTCACGTCATCGTCGTTACCATGTGTACAGTTATGTAAAATCGAATTCATATCCAGATATAGGTTATCAAAC
TCAGGAATCTGAGTCCCTTCAATAAGTTGTAGAATCATGGGCCATCTTTCTGAGATGTACCTGAAAAATTTCGGA
ATACCCATGCTGTACTAGTATATGTTGGTTACTGTTTTTACAGTTGCTGCTTTTTTCTTTATAGTACACGAATCG
TCTCACGATTGGTCAATTGCTTATTGCAGCCTAAATATAACGCAGTGAAGGATATAAGGCCGTTAnAnAaTAGAC
TATTAACAGGAAGAAACGGAAGTAGAAGGCAGAGAGTAAATAATAnAACAACTGGCTCACGTGTAGAAGGCCAAA
AGAATCTTATTGAAATGGAGATCAGcnnaGGGAAAAAaganAGGCTGCTGTAGTannGTGCGTAATATAACTTtT
ACAACAACTAantgGGGAGTTg

**IRR1/YIL026c, chrIX:306349-306368**

GGGAaCaTGGagTgatGTccAaTTCCTCATCGTCAACATCATCGTCATCGCTGTTATTATTGGCCGAGTTTTGTA
ACTTTTGGATCAGTTCATTTGATTCAGTGTCAGAGATCAGCAGATTGCAAATAGTTTTCAAGTGAGTGGAAATGT
AAGGAGACAAAAACTCTGCTGCTTGTGTTAGCATACGGATCTTTGTCCTTGAATCAACTTCAGCGCAATCCTTCA
AGTGATAAATTAAAGAGTCATTCAGAATCTTGATAAATATACCGACTTGAACAACGGGTCCAACTTCTAACCCGT
CGACTTCTTTGGGCAAGTCTTCATGCGTCTTGATGAATTCGTCAAATTTCTCTTTTATTACTCTTGCTAAGAACT
TGGCTACAGTAGATAAAAATTTGGACCTTTTATTGAATGAGGATGTTTTGAAAGGGTCGAATTCTTCATCAAACA
TTAAGCTAGAAATCATTAGAATCTCAAAATCATCCAAATAGCCCAGTGATGACGCTTCAGTTAGAATTTGAACAC
TATGAATTCTGACATCAAGATTGACGTCATGGATGGCGACCTCCAGGATCTTAGATTTAAACCTTTCAAAGACTT
GGCGGATCGCGGAATTATCACTGGATTTGCTGTTATGATTTTGGATGATCAAATGTGGAAGAATCTTGGCAACTT
GTAACCTGACAGAGACGGAATTATCGCTGAGTAGCCAGCCAAAATATTTCAnAnATGTGACCTTGAGGAAATATT
CTGGATAGTTtTTAATCCAGATCGAtaAGTGCAACATGGACTCAGAACGAATTGAATCGGanaTGTCctTGTACC
TGTGCACGAAACACAACTTCAcaaTATTATCAATAACACCTTCAATGACAACTTTACTACCTTGaGTTTCAGCAA
TgGTGCTtTCgaGCTTTTCTAAAGTTTTCTTatTGGGccTTTTTTTctTTTctTCTAGTGACaACTGTCTAGTCA
ATTTGGCTA

**TDH2/YJR009c, chrX:453941-453961**

GATGATAGTAGTGTCAGTTATACTTCAGGTTATGTTAcnnnnGGATAATGATCACGGctAAAACGGTCGAATGTA
AGCATATATCTTTCGATTgTATAATTGTTCCCAAATACTACAGCATCTCAAGGaAAAAAAACAAAAACTTCCAAA
AAAATCGAATCCCTGAGGAATCTTTAATACATTTTCAATCTATTTAAGTTTTATAAACGTGTATATGAGATGTCA
TGAGCATGAATTATTAATAATAAAAACTAAATCATTAAAGTAACTTAAGGAGTTAAATTTAAGCCTTGGCAACGT
GTTCAACCAAGTCGACAACTCTGGTAGAGTAACCGTATTCGTTGTCGTACCAGGAAACCAACTTGACGAACTTTG
GAGACAATTGGATACCAGCGGCAGCATCGAAGATGGAAGAGTTAGAGTCACCCAAGAAGTCAGAGGAGACAACAG
CGTCTTCAGTGTAACCCAAGACACCCTTCAACTTACCTTCAGCGGCAGCCTTGACAACCTTCTTGATTTCATCGT
AGGTGGTTTCCTTGTTCAACTTGACAGTCAAGTCAACAACGGAGACATCGACGGTTGGGACTCTGAAAGCCATAC
CGGTCAACTTACCTTGTAATTCAGGCAAGACCTTACCGACAGCCTTGGCGGCACCGGTGGAGGATGGGATGATGT
TACCGGAAGCGGTTCTACCACCTCTCCAGTCCTTGTGGGATGGACCATCGACAGTCTTTTGAGTGGCGGTCATGG
AGTGGACAGTGGTCATCAAACCTTCTTCAATACCGAAAGCATCGTTGATAACCTTGGCCAATGGAGCcaAACAGT
TAGTGGTACAGGAAGCGTTGGAGaCAATCTTCAAGTCAGAAGTGTATTTGTCTTCGTTAACACcCATAACGAACA
TTGGggCGGTGGAAGATGGAGCAGTGATGACAACCTTCTTGGCacCAGCGTCAATGTGCttnTGAGCA

**PRI2/YKL045w, chrXI:354021-354023**

CntGGaCntaCCTTaCTacaGTTtATCtcaATGAAGAAAAGGCGGAATTATCTCATCAGTTGTATccnncagtTT
CCGCGTCTCTACAGTTCCAATTGAATTTAAACGAGGAACATCAAAGAAAACAGTATTTTCAACAGGAAAAATTCA
TAAAATTACCTTTCGAAAATGTGATAGAACTGGTAGGAAACCGTTTAGTGTTTTTGAAGGACGGGTACGCATATT
TACCACAATTCCAGCAATTGAATTTACTTTCTAATGAGTTTGCTAGCAAATTAAACCAGGaGTTAATAAAAACGT
ACCAGTACTTACCAAGACTGAATGAGGATGACCGGTTGTTACCAATTCTGAATCATCTTTCGTCAGGATACACAA
TTGCAGACTTTAACCAGCAAAAGGCAAACCAATTCGGTGAAAATGTAGACGATGAAATAAATGCACAAAGTGTGT
GGTCCGAGgAgATCAGCTCGAATTATCCATTGAGTATCAAAAACTTAATGGAGGGTTTGAAGAAAAACCATCATT
TGAGATATTACGGTAGGCAACAACTAAGCCTGTTTCTAAAGGGAATTGGGTTGAGCGCCGACGAAGCTTTGAAAT
TTTGGTCAGAAGCTTTCACAAGAAACGGCAACATGACGATGGAAAAGTTCAATAAAGAGTACCGCTACAGTTTTA
GACACAACTACGGCCTCGAAGGTAACAGAATCAACTACAAACCGTGGGACTGTCACACCATTCTATCTAAGCCCA
GACCCGGCCGCGGCGATTATCACGGATGCcCATTCCGTGACTGGAGCCACGACAGACTATCTGCAGAACTGCGTT
CCATGAAACTCACcCAAGCACAAATCATAAGTGTCCTagaTTCGTGCCAAAnagGCGagTACACAATCGCTTGCA
CCAAAGTATTCGAAATAACACACAATTCCgCCTCAGCAgaCTTGGanaTCGGCGAACAAACTCATATCGcaCATC

CTAaccTctaTTTTGAAAGGTCaAGGCAACTGCAAAagaAACAACaGAAACCGgaAAAGGAAAAACTTTCCAAta
gtGCCAataaCCAGT

### ALD2/YMR170c, chrXIII:602993-602998

GcnatGgcTaGAGgnaATTccaTGGAACGATTTGAGCAACAACACCAAAAGGAACTTTcnnagtATATGCAAACt
tGtTAAAAGTCAATGGTATGGTTGCACCCttgTCAAACTtATCAGCGGACCCAGCAAAATATCTGGTAAGCTgtA
AAATTTGTGCCAAATCACCTTTGGCATTTGAATGATAAGGCTTTCCAGCGTCTAAAGTCTCTAACGCGGCAAGTG
TGTCtTgCTCCTCCTCAATAAGTTTTAATAAGTTTGAAAGATAAATACCACGttgCTCAGAagATGTCTTCGACC
AAACGTTATCAAAAGCAGCCCTGGCAGCTTTCACAGCTTTGTCTACATCCTTTTCGTTAGCTGCTTGGAAGGATG
TTATCGGTTCGCCAGTAGCTGGGTTCACAGTTTCGATGGTTTTTCCATCTGATGATGGACAAAATTCGTTGTTGA
TGAACAACCCAAGTGGTTGTTTAACAGAGATTTTCAATTGTGGGATTTTGAGGTCTGTGTATAAATCAGGCATGG
CGCTTTAGTTTTGTTTGTGGATTTTGTTGATAAAAGTGAGGCGAAAGAAAGGAGAGTAATTTGAAACTATTTTCA
CTGGCCTCGCGCTTTTATATAAGTCCTATTGACTGTGCTTCAATTACGTGAGCATTGATCATTAAGCATAATCAC
GTCCCCAACCCCTAAACGACTTGGAGATAGCACATTGGaGAGGGGTGTATAGTAnAGGGTTTTAGATTGTATATT
ATTAGTATCTAAAGGCCGCCGTGCTTCAAAGGGCGTGGAGaAGTAagaCGCAAAGCACGGCGATAGCGATCTGCT
TGCCTCTGGCGGCAAATTagaTAAGGGTCGGGTCACTCGaAACAAAAGCGagaAAAAAATGATAGATGATAGATg
aanaagataaGCTATTTTCCAAACTAAAACACTTCATATCAaGATTTCCtaTTAannTTTGTGGCTAGGCATTAA
CGGAAGcaCTTcgaATCTCTACAACGTTtTCTCAtTctTcncAAAAttgaACTAGGtannnGGCAAtGcGgAAAa
CTTgaaTAAAnagAannCAAGTtgaagataanaAAnAAATTTTACGGgccaACCagTtgtCCTT

### YMR196w, chrXIII:657849-657854

TTGCCGANACTTGGGTATCGTATGATTCACTTGTTTGTACCAGATGAAAATGNNNAGCGTGCCGTTCATTATGGT
GATCACTCTAAATTCTTATCCACTGATCCATATTTCAAGGATTATGTACCATTCTTTGAATACTTCGATGGTGAC
TCAGGAAGAGGGCTTGGTGCTTCACACCAATGTGGTTGGACTGCTCTTGTGGCCAAATGGATAAGTGATGTAGGT
ATATCCTGTGTAAGACTACCTCGTACGCCAAGATCATCTGTGGCAACGACCGCTTCAACAGAGAGCTCTGAGCAA
GGTCCCAAAATGAAGAGAATGGCAAGACGTAAGAGTGCAAAGTCTTTGGTAAACTACACTGCCACCATTTTGGAC
TTAACCGAAGAAGAAAAGCGCCATCATAGGATAGGGGGCACCCATTCTGGGTTGACACCACAAAGCAGCAATTCA
AGTGACAAGGCTAGACATTTGATGGAGGAAATGAATGAAGAGGAAGGTATTCACGAAACTGTGGTACCTGAAGAT
CGTCACAACTTTGAAACCAAGCTTATAGGCAAGCTAAAAGATAAGGTGAAAAATATGAAAGTAACTGACAAGGCT
AAAGATGAGGCTATAGACCCAATGGACCCGATGAGTCCGTTTGAATAANGATGTGTCTTGATTAACTGCGGTGAA
CCTTTCAATATC

### HSP82/YPL240c, chrXVI:97019-97048

ataActtaGtcagtCCTTGaCaGTCTTGTCTTGAGCACCACCTTCGtCaACTCTCTTTTTcaATTCCTTGATAAT
TGGAGATTTTGGAGAAATTTCGAAAGTCTTCTTGGAAGACATGTAGGAGGACATGGAAGAGTCTCTCAAGGCTTG
AGCCTTCATGATTCTTTCCAtgTTAGCAGACCAACCAAATTGACCAGTTCTGATAGCAGCTGGGGCATCCAACAA
TTTGTAAGAAACAACAACTTTCTCCACTTGGTCACCCAAAATTTCTTTCAAGGCCTTGGTCAATGGTTCATATTC
TTTGATCTCCTTCTCTCTTTCAGCTTTTTCTTCGTCAGTTTCTTCCAGTTCGAAATCCTTGGTGATATCGACCAA
AGTCTTACCTTCGAATTCCTTCAATTGGGTGAAAGCGTATTCATCGATTGGATCAGTCAAGAACAAAACTTCAAA
GTTCTTAGCCTTCAAAGCGTCCAAGAATGGGGATTTTTCAACGGCCTTTAGAGATTCACCTGTGATGTAGTAAAT
GTTCTTTTGGgTGTTCTGGCATTCTGGTGATGt

### GPH1/YPR160w, chrXVI:862792-862803

aCgataCTTtGCtCAGgtaAGgaGTTGAGGTTGAAACAGCAGTACTTCTGGTGTGCTGCATcCttACACGACATC
TTAAGAAGATTCAAAAAATCCAAGAGGCCATGGACTGAATTTCCTGACCAAGTGGCTATTCAGTTGAATGATACT
CATCCAACTTTAGCCATCGTTGAATTACAGAGAGTTTTGGTCGATCTAGAAAAACTAGATTGGCACGAGGCTTGG
GACATCGTGACCAAGACTTTTGCTTATACTAACCACACTGTTATGCAAGAGGCCCTGGAAAAATGGCCCGTCGGC
CTCTTTGGCCATTTGCTACCCAGACATTTGGAAATTATATATGATATCAACTGGTTCTTCTTGCAAGATGTGGCC
AAAAAATTCCCCAAGGATGTTGATCTTTTGTCTCGTATATCCATCATCGAAGAAAACTCTCCAGAGAGACAGATC
AGAATGGCCTTTTTGGCTATTGTTGGTTCTCATAAAGTCAACGGTGTTGCGGAATTGCACTCTGAATTAATTAAG
ACCACCATCTTCAAAGATTTCGTCAAATTCTACGGTGCATCAAAGTTTGTCAACGTTACTAACGGTATCACACCA
AGAAGATGGTTGAAGCAAGCTAACCCTAACTTGGCTAGATTGATTAGCAAAACTCTTAACGATCCTACAGAGGAC
TATCTACTAGACATGACAAAGTTAACTCAACTGGCAAAGCACCTTGAGGATAAGAAGTTTTTGAAAGAGTGGAAT
CAAGTCAAACTCAATAATAAGATCAGATTGGTGGACCTAATCAAAAAaGAAAATGGTGGTGAAGACATCATTAAC
AGAGAGTATCTAGACGATACTtTGTTTGATATGCAAGTTAAACGTATTCACGAGTATAAACGTCAACAACTAAAC
GTCTTTGGTATTATTTACCGTTACTTAGCAATGAAAAATATGCTAgagaACGGTGCTTCCATCGAGGAAGTGGCC
AanaAATATCCACGTAAGGTTTCTATCTTCGgtGGTAanantGCaCCCGGTTACTACATGGCTAAgtTGatcATC
AAaCTGGTCAACTCTGTcncTgaAATTGTtaaCAACGACGAATCAATcGACGaCTTATtgaAagttgtcttCATT
GCTGatTanaat

**QCR2/YPR191w, chrXVI:919950-919951**

GCaccagnacAGgtgtAGCCcATCTTTTGAACAGGTTCAACTTCCaGaACACAAATGCTAGGTCTGCGTTGAGAT
TGGTCAGAGAATCCGAATTATTAGGGGGAAATTTTAAGTCCACTTTGGATAGGGAATACATCACTCTAAAAGCTA
CATTCTTGAGGGACGACCTTCCTTACTACGTCAATGCCTTGGCAGATGTGTTGTATAAGACTGCCTTCAAACCCC
ACGAGCTGTCTGAATCTGTTTTGCCTGCTGCCAGATACGATTATGCGGTTGCTGAGCAGTGCCCCGTAAAAAGCG
CAGAAGAACAGTTATTCGCTATTACATTTAGAAAGGGCTTGGGAAATCCATTGTATTACGACGGTGTGGAAAAAG
TCAGCTTGCAAGATATCAAGGATTACGCTGaCAAAGTCTACACTAAAGAGAATCTTGAAATTACAGGTGAAAATA
TTGTTGAGGCCGATTTGAAAAGATTTGTTGACGAGTCACTGTTAAGCACTTTGCCTGCAGGTAAGTCATTGGTGA
GTAAATCCGAACCAAAATCCTTTTTGGGTGAAGAAAACAGGGTAAGGTTTATCGGTGACTCCGTTGCCGCCATTG
GTATCCCGGTAAACAAAGCCTCCCTAGCTCAATATGAAGTATTGGCCAACTATTTGACCTCTGCCCTATCCGACC
TTTCCGGCTTAGTCAGCTCGGCTAAACTTGATAAATTCACTGACGGCGGCCTGTTTACTCTGTTTGTAAGAgACC
AAGACAGCGCCGTGGTATCTTCCAACATCAAGAAAATTGTTGCGGATTTGAAgaagGGCAAGGACTTATCCCCTG
CaatAATTACACAAAgtTAAanaTGct

**Figure S2**

## DBVPG 6033

### SNF5/YBR289w, chrII:780898-780906

```
SGD_Scer_SNF5/YBR289W      CCAACTATTGGCCAACTTCCTCAACTTCCAAAATTAAACTTACCCAAGTACCAAACTATT    936
MIT_Sbay_c596_2303         CCAACTGTAGGACAACTTCCTCAATTACCTAAATTAAACTTACCTAAGTTTCAAACGATT    927
6033_SNF5                  ------------------------------AATTAAACTTACCCAAGTACCAAACTATT    29
                                                          ************* ****: ***** ***

SGD_Scer_SNF5/YBR289W      CAATACGATCCACCAGAAACCAAGCTACCATATCCAACCTATTGGTCAGACAAAAAAGCA    996
MIT_Sbay_c596_2303         CAGTACGACCCACCAGAATCCAAGTTACCATACCCAACTTATTGGTCAGATAAAGGAGCA    987
6033_SNF5                  CAATACGATCCACCAGAAACCAAGCTACCATATCCAACCTATTGGTCAGACAAAAAAGCA    89
                           **.***** *********:***** ******* ***** *********** ***..****

SGD_Scer_SNF5/YBR289W      GATACGGATACTTTGTTGTACGAACAAATTATCCAGCGTGATAAAATTAACAAATATTCG    1056
MIT_Sbay_c596_2303         GACACAGATACTTTATTGTATGAACAAATAATCCAGCGCGATAAAATCAATAAATTTTCC    1047
6033_SNF5                  GATACGGATACTTTGTTGTACGAACAAATTATCCAGCGTGATAAAATTAACAAATATTCG    149
                           ** **.********.***** ********:******** ******** ** ****:***

SGD_Scer_SNF5/YBR289W      CTAATAAGAGAAACCAATGGTTACGATCCGTTTAGCATTTATGGATTTAGTAATAAAGAG    1116
MIT_Sbay_c596_2303         TTAGTGAGGGAAACTAACGGTTACGATCCGTTCAGTATTTATGGATTCAGTAATAAAGAA    1107
6033_SNF5                  CTAATAAGAGAAACCAATGGTTACGATCCGTTTAGCATTTATGGATTTAGTAATAAAGAG    209
                            **.*.**.***** ** ************** ** *********** ***********.

SGD_Scer_SNF5/YBR289W      TATATTAGTAGACTGTGGCATACACTGAAGTATTATCAAGATTTGAAGAACACTAGAATG    1176
MIT_Sbay_c596_2303         TATATAAGTAGACTGTGGCATACATTGAAATATTATCAGGATCTGAAAAATACTAGGATG    1167
6033_SNF5                  TATATTAGTAGACTGTGGCATACACTGAAGTATTATCAAGATTTGAAGAACACTAGAATG    269
                           *****:****************** ****.********.*** ****.** *****.***

SGD_Scer_SNF5/YBR289W      AAATCTATCACAAGCACTTCTCAGAAGATTCCTTCGGCAAGTATTTGGGGAAATGGTTAC    1236
MIT_Sbay_c596_2303         AAATCTATAACTAACACCTCACAGAAGATTTCATCCGCAAGTATTTGGGGAAATGGTTAT    1227
6033_SNF5                  AAATCTATCACAAGCACTTCTCAGAAGATTCCTTCGGCAAGTATTTGGGGAAATGGTTAC    329
                           ********.**:*.*** **:********* *:** **********************

SGD_Scer_SNF5/YBR289W      TCAGGGTATGGTAATGGGATTACGAATACAACTACCAGAGTTATTCCACAAGTAGAAGTT    1296
MIT_Sbay_c596_2303         TCAGGGTACGGGAACGGAATCACAAATACAACCACAAGAGTTATACCTCAAGTTGAAGTT    1287
6033_SNF5                  TCAGGATACGGGAACGGAATCACAAATACAACCACAAGAGTTATACCTCAAGTTGAAGTT    389
                           *****.** ** ** ** ** **.******** **.********:**:*****:******

SGD_Scer_SNF5/YBR289W      GGAAATAGGAAGCATTACCTAGAGGATAAATTAAAAGTCTATAAACAGGCCATGAATGAG    1356
MIT_Sbay_c596_2303         GATAATAGAAAACATTACCTTGAAGATAAGTTGAGAGTTTATAAACAGGCAATGAGTGAA    1347
6033_SNF5                  GATAATAGAAAACACTACCTTGAAGATAAGTTGAGAGTTTATAAACAGGCAATGAGTGAA    449
                           *.:*****.**.** ****:**.*****.**.*.*** ***********.****.***.

SGD_Scer_SNF5/YBR289W      ACATCGGAACAGTTAGTTCCCATAAGATTGGAGTTCGATCAAGATCGTGACAGATTCTTC    1416
MIT_Sbay_c596_2303         ACGTCAGAGGAATTGGTTCCGATAAGATTGGAGTTCGACCAAGATCGTGACAAATTCTTC    1407
6033_SNF5                  ACAACAGAGGAATTGGTTCCGATAAGATTGGAATTCGACCAAGATCGTGACAAATTCTTC    509
                           **.:*.**. *.**.***** *********** .***** ************* .*******

SGD_Scer_SNF5/YBR289W      CTCAGGGACACTTTGTTATGGAACAAAAATGACAAGCTTATTAAAATTGAAGACTTTGTG    1476
MIT_Sbay_c596_2303         CTAAGAGATACATTGTTATGGAATAAAAATGATGAACTAATCAAGATTGAAGAGTTTGTG    1467
6033_SNF5                  CTAAGAGATACATTGCTATGGAACAAAAATGATGAACTTATCAAGATTGAAGAATTTGTG    569
                           **.**.** **:*** ******* ******** .*.**:** **.********* ******

SGD_Scer_SNF5/YBR289W      GACGACATGTTGCGAGATTACCGATTTGAGGACGCTACGAGAGAGCAACACATTGATACT    1536
MIT_Sbay_c596_2303         GATGACATGATGCGAGATTACCGATTTGAAGATTCAACCAGGGAGCAACACATTGACGCC    1527
6033_SNF5                  GATGACATGATGCGAGATTATCGATTTGAAGATTCGACCAGGGAGCAACACATAGATACC    629
                           ** ******:********** ********.**  * ** **.***********:** .*

SGD_Scer_SNF5/YBR289W      ATTTGTCAATCTATACAAGAGCAGATTCAGGAGTTTCAAGGAAATCCATATATAGAGTTG    1596
MIT_Sbay_c596_2303         ATATGCCAGTCCATACAAGAACAAATCCAAGAATTCCAAGGAAATCCATATTTAGAATTA    1587
6033_SNF5                  ATATGTCAGTCCATACAAGAACAAATCCAAGAATTTCAAGGAAATCCATATTTAGAATTG    689
                           **:** **.** ********.**.** **.**.** ***************:****.**.

SGD_Scer_SNF5/YBR289W      AATCAGGACCGTCTAGGCGGTGATGACTTGAGAATTAGAATCAAGCTGGATATTGTCGTG    1656
MIT_Sbay_c596_2303         AACCAGGATCGCTTGGGTGGTGATGATTTACGAATTAGGATCAAGTTGGATATTGTGGTG    1647
6033_SNF5                  AATCAGGATCGCTTGGGTGGTGATGATTTACGAATTAGAATCAAGCTGGATATTGTGGTG    749
                           ** ***** **  *.** ******** **..*******.****** ********** ***
```

### UTP4/YDR324c, chrIV:1116214-1116246

```
SGD_Scer_UTP4/YDR324C      ACTGTGGTAACAGAATGGGATTTAGCAACAGGTTTACCATTAAGAAACTATGATTGCAAT      420
MIT_Sbay_c442_4963         ACTGTCGTTACAGAATGGGACTTGGCCACAGGTTTACCGTTAAGAAATTACGACTGTAAT      348
UTP4                       -------------AATGGGAcTTGGcTACGGGTTTACCATTAAGAAATTATGACTgTAAT      47
                                        ******* **.*  **.********.******** ** ** *  ***

SGD_Scer_UTP4/YDR324C      TCAGGTGTGATATGGTCTATTTCCATCAACGATTCACAAGACAAGCTGTCTGTGGGTTGC      480
MIT_Sbay_c442_4963         GCAGGTGTGATCTGGTCTATTGCCATCAACGATTCCCAAGATAAACTGTCTGTGGGTTGC      408
6033_UTP4                  GCAGGCGTGATCTGGTCTATTGCCATAAATGAAtCCCAAGATAAGCTGTCAGTGGGCTGC      107
                            **** *****.********* ****.** **: *.***** **.*****:***** ***

SGD_Scer_UTP4/YDR324C      GATAATGGGACGGTGGTCCTCATAGATATCTCTGGCGGGCCTGGTGTCTTGGAACACGAT      540
MIT_Sbay_c442_4963         GATAATGGGACTGTTGTACTTATAGATATTTCTGGGGGGCCAGGTGTCTTGGAACACGAT      468
6033_UTP4                  GATAATGGGACTGTGGTACTtatAGATAtntCTGgGGGGCCGGGTGTCTTGGAACACGAT      167
                           *********** ** **.**   *****   ***  ***** ******************

SGD_Scer_UTP4/YDR324C      ACTATTTTGATGAGACAAGAAGCCAGAGTATTGACTTTGGCTTGGAAAAAGGATGACTTC      600
MIT_Sbay_c442_4963         ACTATTCTAATGAGACAAGAAGCCAGAGTATTGACTTTAGCTTGGAAGAAGGATGACTTT      528
6033_UTP4                  ACTATTnTGATGAGACAAGAAGCCAGAGTATTGACTTTGGCTTGGAAAAAGGATGACTTC      227
                           ****** *.*****************************.********.***********

SGD_Scer_UTP4/YDR324C      GTGATTGGTGGTTGTTCTGATGGTAGAATAAGGATTTGGTCTGCACAAAAAAATGACGAA      660
MIT_Sbay_c442_4963         GTTATTGGTGGTTGTTCTGATGGTAGAATAAGGATTTGGTCGGCTCAAAAGGGTGAAGAA      588
6033_UTP4                  GTGATTGGTGGTTGTTCTGATGGTAGAATAAGGATTTGGTCTGCACAAAAAAATGACGAA      287
                           ** ************************************** **:*****...***.***

SGD_Scer_UTP4/YDR324C      AACATGGGTCGTCTATTACACACTATGAAGGTCGACAAGGCCAAAAAAGAATCAACTCTA      720
MIT_Sbay_c442_4963         AACATGGGCCGTCTATTGCATACCATGAAAGTCGACAAAGCCAAGAGAGAATCAACTTTA      648
6033_UTP4                  AACATGGGTCGTCTATTACACACTATGAAGGTCGACAAGGCCAAAAAAGAATCAACTCTA      347
                           ******** ********.** ** *****.********.*****.*.********** **

SGD_Scer_UTP4/YDR324C      GTTTGGTCAGTTATATATTTACCAAGAACTGATCAGATTGCCTCTGGTGATTCTACAGGC      780
MIT_Sbay_c442_4963         GTTTGGTCCGTTATCTATTTGCCAAATACTGATCAAATCGCATCTGGTGATTCTACCGGC      708
6033_UTP4                  GTTTGGTCAGTTATATATTTACCAAGAACTGATCAGATTGCCTCTGGTGATTCTACAGGC      407
                           ********.*****.*****.****.:********.** **.**************.***

SGD_Scer_UTP4/YDR324C      TCCATTAAATTCTGGGATTTCCAGTTTGCCACGCTAAACCAGTCATTTAAGGCGCACGAT      840
MIT_Sbay_c442_4963         TCCATAAAGTTCTGGGATTTCCAATTCGCTACATTGAACCAGTCTTTTAAAGCACATGAT      768
6033_UTP4                  TCCATTAAATTCTGGGATTTCCAGTTTGCCACGCTAAACCAGTCATTTAAGGCGCACGAT      467
                           *****:**.**************.** ** **. *.********:*****.**.** ***

SGD_Scer_UTP4/YDR324C      GCAGACGTACTGTGTCTAACTACCGATACTGATAATAATTATGTTTTTAGTGCTGGTGTG      900
MIT_Sbay_c442_4963         GCAGACGTCTTATGCCTAACTACTGATATTGATAACAATTACGTGTTCAGTGCTGGTGTG      828
6033_UTP4                  GCAGACGTACTGTGTCTAACTACCGATACTGATAATAATTATGTTTTTAGTGCTGGTGTG      527
                           ********. *.** ******** **** ****** ***** ** ** ************

SGD_Scer_UTP4/YDR324C      GACAGAAAAATCTTTCAATTTTCTCAAAACACTAACAAATCTCAAAAGAACAACAGATGG      960
MIT_Sbay_c442_4963         GATAGAAAAATCTTCCAATTCTCCCAAAATAGCAACAAATCTCAAAAAAACAACAGATGG      888
6033_UTP4                  GACAGAAAAATCTTTCAATTTTCTCAAAACACTAACAAATCTCAAAAGAACAACAGATGG      587
                           ** *********** ***** ** ***** *  ***************.************

SGD_Scer_UTP4/YDR324C      GTAAATTCTTCTAATAGGTTGCTTCATGGAAACGACATTAGAGCAATATGTGCATACCAA      1020
MIT_Sbay_c442_4963         GTGAACTCTTCCAATAGGTTACTTCATGGTAACGAA------------------------      924
6033_UTP4                  GTAAATTCTTCTAATAGGTTGCTTCATGGAAACGACATTAGAGCAATATGTGCATACCAA      647
                           **.** ***** ********.********:*****.

SGD_Scer_UTP4/YDR324C      TCTAAAGGTGCAGATTTTCTAGTTTCAGGAGGTGTTGAAAAAACACTAGTCATCAACTCA      1080
MIT_Sbay_c442_4963         ------------------------------------------------------------      924
6033_UTP4                  TCTAAAGGTGCAGATTTTCTAGTTTCAGGAGGTGTTGAAAAAACACTAGTCATCAACTCA      707

SGD_Scer_UTP4/YDR324C      CTTACTTCTTTTTCTAATGGAAACTACAGGAAGATGCCAACTGTCGAACCTTATTCAAAG      1140
MIT_Sbay_c442_4963         ------------------------------------------------------------      924
6033_UTP4                  CTTACTTCTTTTTCTAATGGAAACTaCAGGAAGATGCCAACTGTCGAACCTTATTCAAAG      767

SGD_Scer_UTP4/YDR324C      AATGTTTTAGTTAACAAAGAGCAACGCCTTGTTGTTTCATGGAGCGAATCTACTGTTAAG      1200
MIT_Sbay_c442_4963         ------------------------------------------------------------      924
6033_UTP4                  AATGTTTTAGTTAACAAAGAGCAACGCCTTGTTGTTTCATGGAGCGAATCTACTGTTAAG      827

SGD_Scer_UTP4/YDR324C      ATATGGACAATGGGAACCGATTCTAGTACAGAACAGAATTATAAGCTAGTTTGCAAGTTA      1260
MIT_Sbay_c442_4963         ------------------------------------------------------------      924
6033_UTP4                  ATATGGACAATGGGAACCGATTCtAGtACAGA----------------------------      859
```

## KEM1(XRN1)/YGL173c, chrVII:179656-179664

```
SGD_Scer_XRN1/YGL173C        ------------------------------------------------------------     0
MIT_Sbay_c403_8392           ------------------------------------------------------------     0
6033_KEM1                    TTATATTTAGGCTGCAATAAGCAATTGACCAATCGTGAGACGATTCGTGTACTATAAAGA     60


SGD_Scer_XRN1/YGL173C        -----------------------------------------------ATGGGTATTCCAA     13
MIT_Sbay_c403_8392           -----------------------------------------------ATGGGTATTCCAA     13
6033_KEM1                    AAAAAGCAGCAACTGTAAAAACAGTAACCAACATATACTAGTACAGCATGGGTATTCCGA     120
                                                                            ***********.*

SGD_Scer_XRN1/YGL173C        AATTTTTCAGGTACATCTCAGAAAGATGGCCCATGATTTTACAGCTTATTGAGGGAACAC     73
MIT_Sbay_c403_8392           AATTTTTCAGATACATCTCAGAGAGATGGCCTATGATTCTACAACTTATTGAAGGGACTC     73
6033_KEM1                    AATTTTTCAGGTACATCTCAGAAAGATGGCCCATGATTCTACAACTTATTGAAGGGACTC     180
                             **********.***********.******** ****** ****.********.**.**:*

SGD_Scer_XRN1/YGL173C        AGATTCCTGAGTTTGATAACTTATACCTGGATATGAATTCGATTTTACATAATTGTACGC     133
MIT_Sbay_c403_8392           AGATTCCCGAGTTTGATAACCTGTATCTGGACATGAATTCGATTTTACATAACTGTACGC     133
6033_KEM1                    AGATTCCTGAGTTTGATAACCTATATCTGGATATGAATTCGATTTTACATAACTGTACAC     240
                             ******* ************ *.** ***** ******************** *****.*

SGD_Scer_XRN1/YGL173C        ATGGTAACGACGATGATGTAACCAAGCGATTAACTGAAGAAGAGGTTTTTGCAAAAATCT     193
MIT_Sbay_c403_8392           ATGGTAACGACGATGACGTGACGAAGCGACTAACTGAAGAAGAGGTTTTTGCCAAGATCT     193
6033_KEM1                    ATGGTAACGACGATGACGTGACGAAGCGATTAACTGAAGAAGAGGTTTTTGCAAAGATCT     300
                             **************** **.** ****** **********************.**.****

SGD_Scer_XRN1/YGL173C        GTACGTATATCGATCACCTTTTTCAAACAATCAAGCCCAAGAAGATTTTCTACATGGCTA     253
MIT_Sbay_c403_8392           GCACGTATATCGACCATCTTTTTCAAACCATCAAGCCCAAGCAAATTTTTTACATGGCTA     253
6033_KEM1                    GCACGTATATTGATCATCTTTTTCAAACTATCAAGCCCAAGCAAATTTTTTACATGGCTA     360
                             * ******** ** ** *********** ************.*.***** **********

SGD_Scer_XRN1/YGL173C        TTGATGGTGTGGCCCCTCGTGCCAAGATGAATCAACAAAGAGCTCGTAGATTCAGAACCG     313
MIT_Sbay_c403_8392           TCGATGGTGTGGCTCCCCGTGCGAAGATGAATCAACAGAGAGCTCGTAGGTTTAGAACTG     313
6033_KEM1                    TCGATGGTGTGGCCCCCCGCGCGAAGATGAATCAACAGAGAGCTCGTAGATTTAGAACTG     420
                             * *********** ** ** ** **************.***********.** ***** *

SGD_Scer_XRN1/YGL173C        CTATGGATGCAGAAAAAGCCTTGAAGAAGGCTATTGAGAATGGTGACGAGATTCCTAAGG     373
MIT_Sbay_c403_8392           CCATGGACGCCGAAAAAGCTATGAAAAAAGCTATTGAAAATGGTGACGAGATTCCTAAAG     373
6033_KEM1                    CCATGGACGCCGAAAAAGCTATGAAAAAAGCTATTGAAAATGGTGACGAAATTCCTAAAG     480
                             * ***** **.******** :****.**.********.***********.********.*

SGD_Scer_XRN1/YGL173C        GTGAGCCATTTGATTCGAATTCTATTACTCCAGGTACGGAGTTTATGGCCAAATTGACCA     433
MIT_Sbay_c403_8392           GTGAACCGTTCGACTCAAATTGTATTACTCCAGGTACTGAATTCATGGCCAAACTGACAA     433
6033_KEM1                    GTGAGCCGTTCGACTCAAACTGTATTACTCCAGGTACTGAATTTATGGCTAAACTGACAA     540
                             ****.**.** ** **.** * *************** **.** ***** *** ****.*

SGD_Scer_XRN1/YGL173C        AAAACTTACAATATTTTATTCACGACAAGATTTCTAACGATTCCAAATGGAGGGAAGTGC     493
MIT_Sbay_c403_8392           AAAATTTACAGTATTTCATTCATGACAAGATCTCTAACGATTCTAAGTGGAGAGAAGTAC     493
6033_KEM1                    AAAATTTACAGTATTTCATTCATGACAAAATTTCTAACGATTCCAAATGGAGGGAAGTGC     600
                             **** *****.***** ***** *****.** *********** **.*****.*****.*

SGD_Scer_XRN1/YGL173C        AAATCATATTTTCTGGCCATGAAGTTCCAGGTGAAGGTGAACACAAGATCATGAACTTTA     553
MIT_Sbay_c403_8392           AAATTATATTTTCTGGCCATGAAGTTCCAGGTGAAGGTGAGCACAAAATCATGAATTTCA     553
6033_KEM1                    AAATCATATTTTCTGGCCATGAAGTTCCAGGTGAAGGTGAACACAAGATCATGAACTTTA     660
                             **** ***********************************.*****.******** ** *

SGD_Scer_XRN1/YGL173C        TAAGGCATTTAAAATCCCAAAAGGATTTCAACCAGAATACGAGACATTGTATTTACGGTC     613
MIT_Sbay_c403_8392           TAAGGCATTTAAAATCACAGAAGGATTTTAACCAAAATACGAGACATTGTATTTATGGTC     613
6033_KEM1                    TAAGGCATTTAAAATCCCAAAAGGATTTCAA-----------------------------     691
                             ****************.**.******** **
```

## PRP8/YHR165c, chrVIII:433730-433738

```
SGD_Scer_PRP8/YHR165C        TCACCAATACCATTTCCACCTTTAACTTACAAAAATGATACTAAGATATTAGTTCTTGCC     2745
WashU_Sbay_Contig599.17      TCACCAATCCCATTTCCACCTCTAACTTACAAAAATGATACCAAAATTCTGGTGCTTGCA     2754
6033_PRP8                    -----------------------AACTTAcAAAAATGATACCAAAATnTTGGTtntTGCA     37
                                                    ****** *********** **.**  *.**   ***.

SGD_Scer_PRP8/YHR165C        CTCGAAGACCTAAAAGATGTCTACGCATCAAAAGTACGTTTAAATGCATCTGAAAGGGAA     2805
WashU_Sbay_Contig599.17      TTAGAAAATCTAAAGGACGTTTATGCATCAAAAGTACGTTTGAATGCATCTGAAAGGGAA     2814
```

```
6033_PRP8                     TTAGAAAAtctAAAGGACGTTTATGCATCAAAAGTACGTTTAAATGCAtcTGAAAGGGAA     97
                               *.***.*   ***.** ** ** ****************.******  **********

SGD_Scer_PRP8/YHR165C         GAACTTGCGTTGATAGAGGAGGCTTATGATAATCCTCACGATACTTTGAACAGAATCAAA     2865
WashU_Sbay_Contig599.17       GAACTAGCATTAATAGAGGAAGCCTATGATAATCCTCACGATACTTTGAATAGAATTAAA     2874
6033_PRP8                     GAACTGGCGTTAATAGAGGAAGCTTATGATAACCCCCACGATACTTTGAATAGAATAAAA     157
                              ***** **.**.********.** ******** ** ************** ***** ***

SGD_Scer_PRP8/YHR165C         AAGTACTTGTTGACCCAGCGAGTTTTTAAGCCTGTTGATATAACCATGATGGAAAACTAT     2925
WashU_Sbay_Contig599.17       AAATATTTGTTGACGCAACGTGTCTTTAAGCCCGTTGATTTGACTATGATGGAAAACTAT     2934
6033_PRP8                     AAATACTTGTTGACGCAACGTGTTTTTAAGCCTGTTGATTTGACTATGATGGAAAACTAT     217
                              **.** ******** **.**:** ******** ******:*.** ***************

SGD_Scer_PRP8/YHR165C         CAAAACATTTCCCCTGTTTATTCAGTTGATCCCTTGGAAAAGATTACCGATGCATATCTT     2985
WashU_Sbay_Contig599.17       CAAAATATCTCTCCTGTTTATGCAGTCGATCCCTTGGAAAAAATTACCGATGCATATCTT     2994
6033_PRP8                     CAAAATATAtctCCTGTTTATGCAGTCGATCCCTTGGAAAAAATCACCGATGCATATCTT     277
                              ***** **    ********* **** **************.** ***************

SGD_Scer_PRP8/YHR165C         GATCAGTATTTATGGTACGAAGCTGACCAACGAAAGCTTTTCCCAAACTGGATTAAACCA     3045
WashU_Sbay_Contig599.17       GACCAGTATTTGTGGTACGAAGCCGATCAAAGAAAGCTTTTCCCCAACTGGATAAAACCA     3054
6033_PRP8                     GATCAGTATTTGTGGTACGAAGCCGATCAAAGGAAGCTTTTCCCCAACTGGATAAAACCA     337
                              ** ********.*********** ** ***.*.***********.********:******

SGD_Scer_PRP8/YHR165C         AGTGATTCAGAGATACCACCTCTTCTGGTATACAAGTGGACTCAGGGTATAAACAACCTA     3105
WashU_Sbay_Contig599.17       AGTGATTCCGAGATCCCGCCCCTTCTGGTATACAAGTGGTCTCAAGGCATAAACAATCTA     3114
6033_PRP8                     AGTGATTCAGAGATACCACCTCTTCTGGTATACAAGTGGTCTCAAGGCATAAACAATCTA     397
                              ********.*****.**.** *****************:****.** ******** ***

SGD_Scer_PRP8/YHR165C         TCTGAAATTTGGGATGTATCCAGAGGCCAATCTGCAGTTTTGCTTGAGACTACTTTGGGT     3165
WashU_Sbay_Contig599.17       TCTGACATCTGGGATGTATCTAAAGGTCAATCCACAGTCTTGCTTGAAACCACGTTAGGG     3174
6033_PRP8                     TCTGACATTTGGGATGTCTCTAAAGGTCAATCCACAGTTTTGCTCGAGACCACGTTAAGG     457
                              *****.** ********.** *.*** *****  .**** ***** **.** ** **..*

SGD_Scer_PRP8/YHR165C         GAAATGGCCGAAAAAATTGACTTTACTTTACTTAATAGATTACTTCGCCTGATTGTAGAT     3225
WashU_Sbay_Contig599.17       GAAATGGCTGAAAAGATTGATTTTACTTTGTTAAATAGATTACTTCGTCTGATTGTGGAC     3234
6033_PRP8                     GAAATGGCTGAAAAGATTGATTTTACTTTGTTAAATAGATTACTTCGTCTGATTGTAGAT     517
                              ******** *****.***** ********. *:************** ********.**

SGD_Scer_PRP8/YHR165C         CCTAATATTGCTGACTATATCACTGCAAAAAATAATGTTGTTATCAACTTTAAAGATATG     3285
WashU_Sbay_Contig599.17       CCAAATATTGCCGACTACATCACTGCAAAGAATAATGTTATCATTAACTTTAAGGATATG     3294
6033_PRP8                     CCTAATATTGCTGACTATATTACTGCAAAAAATAATGTTGTTATCAACTTTAAAGATATG     577
                              **:******** ***** ** ********.*********.* ** ********.******

SGD_Scer_PRP8/YHR165C         AGTCACGTCAACAAATATGGCTTAATACGCGGGTTGAAGTTCGCTTCTTTCATATTCCAA     3345
WashU_Sbay_Contig599.17       AGTCACGTTAACAAATACGGTTTAGTGCGCGGATTACAGTTTGCCTCTTTCATATTCCAA     3354
6033_PRP8                     AGTCACGTCAACAAATATGGCTTAATACGCGGGTTGAAGTTCGCTTCTTTCATATTCCAA     637
                              ******** ******** ** ***.*.*****.**..**** ** ***************

SGD_Scer_PRP8/YHR165C         TATTACGGACTAGTTATAGATCTTTTACTATTGGGTCAGGAAAGGGCTACAGATTTGGCT     3405
WashU_Sbay_Contig599.17       TATTATGGATTAGTTATTGATCTCTTATTATTAGGCCAAGAAAGGGCTGCTGATTTGGCT     3414
6033_PRP8                     TATTACGGACTAGTTATAGATCTTTTACTATTGGGTCAGGAAAGGGCTACAGATTTGGCT     697
                              ***** *** *******:***** *** ****.** **.*********.*:*********

SGD_Scer_PRP8/YHR165C         GGTCCAGCTAACAATCCAAATGAATTTATGCAATTCAAGAGCAAAGAAGTAGAAAAGGCA     3465
WashU_Sbay_Contig599.17       GGCCCTGCCAACAATCCAAATGATTTTATGCAATTCAAAAGCAGAGAAACAGAAAAGACA     3474
6033_PRP8                     GGTCCAGCTAACAATCCAAATGAATTTATGCAATTCAAGAGCAAAGAAGTAGAAAAGGCA     757
                              ** **:** **************:**************.****.****. *******.**

SGD_Scer_PRP8/YHR165C         CATCCGATCAGACTTTACACCAGATATTTAGATCGTATATATATGCTTTTTCACTTTGAA     3525
WashU_Sbay_Contig599.17       CATCCGATCAGGCTCTATACTAGATATTTGGATCGCATTTATATGGTCTTTCACTTTGAA     3534
6033_PRP8                     CATCCGATCAGACTTTACACCAGATATTTAGATCGTATATATATGCTTTTTCACTTTGAA     817
                              ***********.** ** ** ********.***** **:****** * ************

SGD_Scer_PRP8/YHR165C         GAAGATGAGGGAGAGGAATTGACTGATGAATACTTGGCAGAGAATCCAGATCCAAACTTT     3585
WashU_Sbay_Contig599.17       GAAGGCGAGGCAAATGAACTGTCTGATGAATATCTAGCGGAAAACCCAGATCCAAACTTT     3594
6033_PRP8                     GAAGATGAGGGAGAGGAATTGACTGATGAATACTTGGCAGAGAATCCAGATCCAAACTTT     877
                              ****. **** *.* *** **:**********  *.**.**.** ***************
```

## MSU1 (DSS1)/YMR287c, chrXIII:843626-843635

```
SGD_Scer_DSS1/YMR287C       GAAATCCATCTGAATACAGCGTTGCTGTCGCCTATATCAGTAACGATTATTCCCTTAAAA     1140
MIT_Sbay_c753_19436         AAAATACACTTGAATACTGCGCTGCTCTCCCCTACATCAGTAACAATCATCCCTTTAAAA     1140
6033_MSU1                   ---------------------------------ACATCAGTTACAATCATCCCTTTGAAA     27
                                                             * ******:**.** ** ** **.***

SGD_Scer_DSS1/YMR287C       TCACAACATTTATATTATGCGCAGGTGATAGAGAAGCTAGAAGCCAACAGTTATAGAGAG     1200
MIT_Sbay_c753_19436         TCCCGACACTTATACTACGCACAGGTAATAGAGAAGCTAGAAGCCGACAACTACAAAGAC     1200
```

```
6033_MSU1                   TCCCAACACTTATACTACGCACAGGTAATAGAGAAGCTAGAAGCTGACAACTACAAGGAC    87
                            **.*.*** ***** ** **.*****.***************** .***. ** *..**

SGD_Scer_DSS1/YMR287C       GTCAATAAGTTTGTAAAATTGGTGAATGAAAGGAAATATCGCGATATATCAGCTTTATAT    1260
MIT_Sbay_c753_19436         ATCGACCGATTTGTAAAGTTGGTGAATGAAAGGAAATATAGAGATATCTCGTCTTTGTAC    1260
6033_MSU1                   ATCGACCGATTTGTAAAGTTGGTGAACGAAAGAAAATATAAAGATATCTCATCTTTATAC    147
                            .**.* ...********.******** *****.******...*****.**. ****.**

SGD_Scer_DSS1/YMR287C       CCTTCTGTGATCCAATTATTGAAAGATTTTGCTGCTGGGAATTTTCACAATAACGGAATT    1320
MIT_Sbay_c753_19436         CCTTCCGTTATTCAGCTGCTGAAAGATTTTGCCGCCGGAAATTTCCACAACAATGGAATC    1320
6033_MSU1                   CCTTCTGTCATTCAGCTGCTGAAAGATTTTGCCGCGGGAAACTTCCACAATAATGGAGTT    207
                            ***** ** ** **. *. ************* ** **.** ** ***** ** ***.*

SGD_Scer_DSS1/YMR287C       ATAGTAGCTTTGATCTCAAAAATATTCAGAAAGATAGAACGCTATAAGGATTGTGACATC    1380
MIT_Sbay_c753_19436         ATAGTCACTTTAATCTCTAAAATATTTAGAAAAATAGAACTCTATAAGGGTTCCGATATA    1380
6033_MSU1                   ATAGTCACCTTAATCTCCAAAATATTTAGAAAGATAGAACTCTATAAGAGTTTTGATATA    267
                            *****..* **.***** ******** *****.******* *******..**  ** **.

SGD_Scer_DSS1/YMR287C       ACAAGAGATATATGTCAAGATTTAATCAATGAGATAACACCCAACTCAATACCCAATCCA    1440
MIT_Sbay_c753_19436         ACAAGAGATATATGTCAAGATTTGGTCAATGAAATCTTACCCAATAAAATGATCAATCCG    1440
6033_MSU1                   ACAAGAGATATATGTCAAGATCTGGTCAATGAAATTTTACCCAATAAAATGATCAATCCG    327
                            ********************* *..*******.** : ****** :.***.. ******.

SGD_Scer_DSS1/YMR287C       TTGTTACTTAACATGGATTTAGCATTACCTGCTTCCTCGAAGTTAGTGAAATGGCAACAA    1500
MIT_Sbay_c753_19436         TTGTTACTGAATATGGATTTAGCTTTGCCTGCTTCTTCGAAATTGGGGCAATCGCAGCAA    1500
6033_MSU1                   TTGTTACTGAATATGGATTTAGCTTTGCCTGCATCTTCGAAATTGGGGCAATCACAGCAA    387
                            ******** ** ***********:**.*****:** *****.**.* *.*** .**.***

SGD_Scer_DSS1/YMR287C       AAACTCTACGACTTAACTAATATAGAAGAATTGCAATGGAAAAAATCCGGCACTGATGAC    1560
MIT_Sbay_c753_19436         AAGCTCTATGAGTTGACAAATATAGAAAACATGCAAAGAAAAAAACTCTAATATTAATGGT    1560
6033_MSU1                   AAGCTTTATGAGTTGACAGACATAGAAAACCTGCAAGCAAGTAATTCCAATATTAATGGT    447
                            **.** ** ** **.**:.* ******.*. *****  .*.:** ** .. * *.***.

SGD_Scer_DSS1/YMR287C       GATAGGTACGATTTTGGCGATCTTAGGGTTTTTTGTATAGACTCTGAAACTGCGCATGAG    1620
MIT_Sbay_c753_19436         GATCGATATGATTTTGGTGATCTTAAGGTTTTTTGTATAGATTCGGAGACCGCACATGAG    1620
6033_MSU1                   GATAGATATGATTTTGGTGATCTTAAGGTTTTTTGTATAGATTCGGAGACTGCGCATGAG    507
                            ***.*.** ******** *******.*************** ** **.** **.******

SGD_Scer_DSS1/YMR287C       ATTGATGACGGCGTGTCGGTGAAAAACTATGGTAGAGATGGACTGTACACTTTATATATT    1680
MIT_Sbay_c753_19436         ATTGACGATGGTGTCTCCATAGAAAATCACAAAAAGGACGGATTGTACACTTTACATATT    1680
6033_MSU1                   ATTGACGATGGAATCTCGATAGAAAACCACAAAAAGGACGGATTGTACACTTTACATATC    567
                            ***** ** ** .* ** .*..****  * ..:*..** *** *********** ****

SGD_Scer_DSS1/YMR287C       CATATTGCGGATCCAACTTCTATGTTTCCGGAAAGCACTAATGTTGATATTGAGGGTATA    1740
MIT_Sbay_c753_19436         CATATTGCAGATCCAGCTTCCTTGTTTCCAGAAAGCACTGATATTGAATCTCAGGGTATT    1740
6033_MSU1                   CATATTGCAGATCCTACTTCTTTGTTTCCAGAAAGCACTAATGTTGATATTGAGGGTATA    627
                            ********.*****:.**** :*******.*********.**.****:: * *******:

SGD_Scer_DSS1/YMR287C       AGTACAGATATCCTAAATGTTGCCTTGAAAAGATCATTTACCACATATTTACCAGATACG    1800
MIT_Sbay_c753_19436         ACCACAGATGTTCTAAATGTTGCGTTGAAAAGATCGTTCACTACATATTTACCTGATTTG    1800
6033_MSU1                   AGTACAGATATCCTAAATGTTGCCTTGAAAAGATCATTTACCACATATTTACCAGATACG    687
                            *  ******.* *********** ***********.** ** ***********:***: *

SGD_Scer_DSS1/YMR287C       GTTGTTCCTATGTTACCTCAATCTATTTGTCACTTATCAGATTTGGGGAAACAAGGACAA    1860
MIT_Sbay_c753_19436         GTTGTTCCCATGCTACCTGAATCTATCTGTGACCTATCTGATTTGGGAAAGCAAGGGCAA    1860
6033_MSU1                   GTTGTTCCTATGTTACCTCAATCTATTTGTCACTTATCAGATTTGGGGAAACAAGGACAA    747
                            ******** *** ***** ******* *** ** ****:********.**.*****.***

SGD_Scer_DSS1/YMR287C       AGGACGAAAACTATATCCTTCTCTGTTGATGTTAAAATCACTTCTAAATGCAGTGGAAAA    1920
MIT_Sbay_c753_19436         AAGACAAGAACACTATCGTTTTCGGTTGACGTTAAAGTTCTCCCTAAGAGCACTGGAAAA    1920
6033_MSU1                   AGGACGAAAACTATATCCTTCTCTGTTGATGTTAAAATCACTTCTAAATGCAGTGGAAAA    807
                            *.***.*.***:.**** ** ***** ******.* .   ****.:*** *******
```

## HSP82/YPL240c, chrXVI:97019-97048

```
SGD_Scer_HSP82/YPL240C      GAAGACCCATTGTACGTTAAGCATTTCTCCGTTGAAGGTCAATTGGAATTTAGAGCTATC    960
MIT_Sbay_c60_24336          ------------------------------------------------------------    0
6033_Hsp82                  ----------------------------------AAGGTCAGTtAGAATTCAGAGCCAtc    26


SGD_Scer_HSP82/YPL240C      TTATTCATTCCAAAGAGAGCACCATTCGACTTGTTTGAGAGTAAAAAGAAGAAGAATAAT    1020
MIT_Sbay_c60_24336          ------------------------------------------------------------    0
6033_Hsp82                  tTGTACATtCCAAAGAGAnntCCATTTGantTATTcGAAAGTAAGAAGAagaagaAcAAc    86
```

```
SGD_Scer_HSP82/YPL240C      ATCAAGTTGTACGTTCGTCGTGTTTTCATCACTGATGAAGCTGAAGACTTGATTCCAGAG      1080
MIT_Sbay_c60_24336          ------------------------------------------------------------      0
6033_Hsp82                  ATCAAGTTGTATGTTCGTCGTGTTtTcntCACCGACGAAGCTGAAGACTTGATCCCAGAA      146


SGD_Scer_HSP82/YPL240C      TGGTTATCTTTCGTCAAGGGTGTTGTTGACTCTGAGGATTTACCATTGAATTTGTCCAGA      1140
MIT_Sbay_c60_24336          ---ATGTCATTCGTCAAGGGTGTTGTTGACTCCGAAGATTTACCATTGAACTTGTCCAGA      57
6033_Hsp82                  TGGATGTCCTTTGTCAAGGGTGTTGTTGACTCTGAAGACTTACCAtTGAACCTGTCCAGA      206
                               :*.** ** ******************** **.** ****** ****  ********

SGD_Scer_HSP82/YPL240C      GAAATGTTACAACAAAATAAGATCATGAAGGTTATTAGAAAGAACATTGTCAAAAAGTTG      1200
MIT_Sbay_c60_24336          GAAATGTTGCAACAAAACAAGATCATGAAGGTCATCAGAAAGAACATCGTTAAGAAGATG      117
6033_Hsp82                  GAAATGCTACAACAAAACAAGATCATGAAGGTTATCAGAAAGAACATCGTCAAGAAGGTA      266
                            ****** *.******** ************** ** *********** ** **.*** *.

SGD_Scer_HSP82/YPL240C      ATTGAAGCCTTCAACGAAATTGCTGAAGACTCTGAACAATTTGAAAAGTTCTACTCGGCT      1260
MIT_Sbay_c60_24336          ATTGAATCCTTCAACGAAATCGCTGAAGACTCTGAACAATTCGAAAAGTTCTACTCTGCC      177
6033_Hsp82                  ATTGAAGCTTTCAACGAAATTGCTGAAGACTCTGAACAATTCGAAAAGTTCTACTCTGCC      326
                            ****** * *********** ******************** ************** **

SGD_Scer_HSP82/YPL240C      TTCTCCAAAAATATCAAGTTGGGTGTACATGAAGATACCCAAAACAGGGCTGCTTTGGCT      1320
MIT_Sbay_c60_24336          TTCGCTAAGAACATCAAATTGGGTGTTCATGAAGACACTCAAAACAGAGCTGCCTTGGCC      237
6033_Hsp82                  TTCGCTAAGAACATCAAATTGGGTGTCCACGAAGACACTCAAAACAGAGCTGCCTTGGCT      386
                            *** * **.** *****.******** ** ***** ** ********.***** *****

SGD_Scer_HSP82/YPL240C      AAGTTGTTACGTTACAACTCTACCAAGTCCGTAGATGAGTTGACTTCCTTAACTGATTAC      1380
MIT_Sbay_c60_24336          AAATTGCTACGTTACAACTCCACCAAGTCCGTTGACGAATTGACTTCTTTGACTGATTAT      297
6033_Hsp82                  AAGTTACTACGTTACAACTCCACCAAGTCCGTCGACGAATTAACTTCTTTGACCGATTAC      446
                            **.**. ************* *********** ** **.**.***** **.** *****

SGD_Scer_HSP82/YPL240C      GTTACCAGAATGCCAGAACACCAAAAGAACATCTACTACATCACTGGTGAATCTCTAAAG      1440
MIT_Sbay_c60_24336          ATTACCAGAATGCCAGAACACCAAAAGAACATCTACTATATCACAGGTGAGTCTTTGAAG      357
6033_Hsp82                  ATCACCAGAATGCCAGAACACCAAAAGAACATTTACTACATCACAGGTGAATCTCTAAAG      506
                            .* ***************************** ***** *****:*****.*** *.***

SGD_Scer_HSP82/YPL240C      GCTGTCGAAAAGTCTCCATTTTTGGATGCCTTGAAGGCTAAAAACTTCGAGGTTTTGTTC      1500
MIT_Sbay_c60_24336          GCTGTTGAAAAAATCCCCATTCTTAGACGCTTTGAAAGCTAAGAACTTTGAAGTTTTGTTC      417
6033_Hsp82                  GCCGTTGAAAAAATCCCCATTCTTGGACGCTTTGAAGGCTAAGAACTTTGAAGTTTTGTTC      566
                            ** ** *****.** ***** **.** ** *****.*****.*****.** **.*********

SGD_Scer_HSP82/YPL240C      TTGACCGACCCAATTGATGAATACGCCTTCACTCAATTGAAGGAATTCGAAGGTAAAACT      1560
MIT_Sbay_c60_24336          TTGACTGATCCAATCGATGAATACGCCTTCACTCAATTAAAGGAATTCGAAGGTAAGACT      477
6033_Hsp82                  TTGACTGATCCAATCGATGAATACGCTTTCACCCAATTGAAGGAATTCGAAGGTAAGACT      626
                            ***** ** ***** *********** ***** *****.*****************.***

SGD_Scer_HSP82/YPL240C      TTGGTTGACATTACTAAAGATTTCGAATTGGAAGAAACTGACGAAGAAAAAGCTGAAAGA      1620
MIT_Sbay_c60_24336          TTAGTCGATATCACCAAGGATTTCGAGCTGGAAGAAACTGACGAAGAGAAAGCTGAAAGA      537
6033_Hsp82                  TTGGTCGATATCACCAAGGATTTCGAACTGGAAGAAACTGACGAAGAAAAAGCTGAAAGA      686
                            **.** ** ** ** **.********. *******************.************

SGD_Scer_HSP82/YPL240C      GAGAAGGAGATCAAAGAATATGAACCATTGACCAAGGCCTTGAAAGAAATTTTGGGTGAC      1680
MIT_Sbay_c60_24336          GAAAAGGAAGTTAAAGAATTCGAACCATTGACCAAGGCCTTGAAAGACATCTTGGGTGAA      597
6033_Hsp82                  GAGAAGGAGATCAAAGAATATGAACCATTGACCAAGGCCTTGAAAGAAATTTTGGGTGAC      746
                            **.*****..* *******: *************************.** ********.

SGD_Scer_HSP82/YPL240C      CAAGTGGAGAAAGTTGTTGTTTCTTACAAATTGTTGGATGCCCCAGCTGCTATCAGAACT      1740
MIT_Sbay_c60_24336          CAAGTTGAAAAGGTTGTTGTCTCTTACAAACTAGTGGATGCCCCAGCTGCCATTAGAACT      657
6033_Hsp82                  CAAGTGGAGAAAGTTGTTGTTTCTTACAAATTGTTGGATGCCCCAGCTGCTATCAGAACT      806
                            ***** **.**.******** ********* *. **************** ** ******

SGD_Scer_HSP82/YPL240C      GGTCAATTTGGTTGGTCTGCTAACATGGAAAGAATCATGAAGGCTCAAGCCTTGAGAGAC      1800
MIT_Sbay_c60_24336          GGCCAATTCGGTTGGTCCGCTAACATGGAAAGAATCATGAAGGCTCAAGCTTTGAGAGAC      717
6033_Hsp82                  GGTCAATTTGGTTGGTCTGCTAACATGGAAAGAATCATGAAGGCTCAAGCCTTGAGAGAC      866
                            ** ***** ******** ******************************** *********

SGD_Scer_HSP82/YPL240C      TCTTCCATGTCCTCCTACATGTCTTCCAAGAAGACTTTCGAAATTTCTCCAAAATCTCCA      1860
MIT_Sbay_c60_24336          TCTTCCATGTCCTCTTACATGTCCTCCAAGAAGACTTTCGAAATCTCTCCAAAATCTCCA      777
6033_Hsp82                  TCTTCCATGTCCTCCTACATGTCTTCCAAGAAGACTTTCGAAATTTCTCCAAAATCTCCA      926
                            ************** ******** ******************** ***************

SGD_Scer_HSP82/YPL240C      ATTATCAAGGAATTGAAAAAGAGAGTTGACGAAGGTGGTGCTCAAGACAAGACTGTCAAG      1920
MIT_Sbay_c60_24336          ATTATCAAGGAATTGAAAAAGAGAGTTGATGAAGGCGGTGCTCAAGATAAGACTGTCAAG      837
6033_Hsp82                  ATTATCAAGGAATTGAAAAAGAGAGTTGACGAAGGTGGTGCTCAAGACAAGACTGTcAAG      986
                            ***************************** ***** *********** ******** ***

SGD_Scer_HSP82/YPL240C      GACTTGACTAAGTTATTATATGAAACTGCTTTGTTGACTTCCGGCTTCAGTTTGGACGAA      1980
MIT_Sbay_c60_24336          GATTTGACCAACTTATTATTCGAAACCGCTCTGTTAACTTCTGGTTTCAGTCTGGAAGAA      897
```

```
6033_Hsp82                GAC---------------------------------------------------------    989
                          **
```

## PMA2/YPL036w, chrXVI:482992-483014

```
SGD_Scer_PMA2/YPL036W     ------------------------------------------------------------    0
MIT_Sbay_c676_25008       ------------------------------------------------------------    0
6033_PMA2                 GCACTGAAAAAGGGAGGATGAACGTCGACGAACGAATGTTGAAGGGGGAAATGGTTATCA    60

SGD_Scer_PMA2/YPL036W     ------------------------------------------------------------    0
MIT_Sbay_c676_25008       ------------------------------------------------------------    0
6033_PMA2                 ATGGACAAACACATAGCAATGAACACACTCTTGTCAACGTCCCCTAAACTGTGATTGGTA    120

SGD_Scer_PMA2/YPL036W     ------------------------------------------------------------    0
MIT_Sbay_c676_25008       ------------------------------------------------------------    0
6033_PMA2                 GCAGTGGTGATGGTTCGCTTACTGTACTTGCGCTTATATAAAGGCGTATCGTGTCTCCTT    180

SGD_Scer_PMA2/YPL036W     ------------------------------------------------------------    0
MIT_Sbay_c676_25008       ------------------------------------------------------------    0
6033_PMA2                 TCCTGTCCATCTCTCTTGTTCTTTCTGTATGGCGTATGCTGCTCAAGCTACTGTGCTTTC    240

SGD_Scer_PMA2/YPL036W     ------ATGTCTTCCACTGAAGCAAAGCAATACAAGGAG---AAACCCTCGAA--AGA--    47
MIT_Sbay_c676_25008       ------ATGTCTTCTAGTGGTGCAAAG---------------------------------    21
6033_PMA2                 TTTTTCATTTTTAACTGTGTAGCATTGCGATTCAAGAAAGAGAAAACATTTTTGTATTAT    300
                                ** * *:. : ** :***::*

SGD_Scer_PMA2/YPL036W     ------GTACCTCCATGCCAGTGATGGCGATGACCCTGCAAATAAT---TCTGCCGCTTC    98
MIT_Sbay_c676_25008       -------------------CCATACAGGGAGAAGCCTGCAGCGAAC---AACGGCGCTGC    59
6033_PMA2                 GTCTTCTAGTGGTGCAAAGCCATACCGAGAGAAGCCTGCAGCGAACAATGGTGCTTCGTC    360
                                             . : *  * ** .* ******.. **       *   *  *

SGD_Scer_PMA2/YPL036W     TTCGTCATCTTCGTCTTCTACATCAACTTCCGCCTCGTCATCGGCTGCAGCCGTTCCACG    158
MIT_Sbay_c676_25008       GTCTTCTTCTTCTTCCTCTTCTTCTACTTCCGCCTCGCCATCGCCTGCTGCCGCTCCACG    119
6033_PMA2                 CTCGTCTTCGTCTTCGTCTTCTTCTACATCGGCCTCGCCATCGCCTGCAGCCGCTCCACG    420
                           ** **:** ** ** ***:*:**:**:** ****** ***** ****:**** ******

SGD_Scer_PMA2/YPL036W     GAAGGCCGCAGCCGCTTCTGCCGCTGATGATTCTGACTCAGATGAAGATATAGACCAATT    218
MIT_Sbay_c676_25008       CAAGGCCGCAGCCGCTCCTGCTGCTAATGACTCCGACTCCGATGAGGATATTGATGGGCT    179
6033_PMA2                 TAAGGCCGCAGCCGCTTCTGCCGCTGATGATTCTGACTCAGATGAAGATATAGACCAATT    480
                           *************** **** ***.**** ** *****.*****.*****:**   .. *

SGD_Scer_PMA2/YPL036W     GATTGATGAACTACAATCTAACTACGGTGAGGGTGATGAATCTGGTGAAGAAGAAGTACG    278
MIT_Sbay_c676_25008       GATTGAAGAGTTGCAATCGAACTATGGGGAGAGCGAAGAGTCCAGCGAAGAAGAGAAACG    239
6033_PMA2                 GATTGATGAACTACAATCTAACTACGGTGAGGGTGATGAATCTGGTGAAGAAGAAATACG    540
                          ******:**. *.***** ***** ** ***.* **:**.** .* ********..:***

SGD_Scer_PMA2/YPL036W     TACTGATGGGGTGCACGCTGGCCAAAGGGTTGTTCCTGAAAAGGACCTTTCTACGGACCC    338
MIT_Sbay_c676_25008       CACGGACGAGGCCCACGCTGGTCAAAGGGTGATTCCCGAGAAGGACCTTTCGACAGATCC    299
6033_PMA2                 TACTGATGGGGTGCACGCTGGCCAAAGGGTTGTTCCTGAAAAGGACCTTTCTACGGacCC    600
                           ** ** *.**  ******** ******** .**** **.*********** **.*  **

SGD_Scer_PMA2/YPL036W     TGCGTATGGTTTGACTTCGGATGAAGTCGCCAGGAGAAGAAAGAAATATGGGTTAAATCA    398
MIT_Sbay_c676_25008       CGCGTACGGTTTGACTGCGGACGAAGTCACTAGGAGAAGAAAGAAGTATGGGTTGAACCA    359
6033_PMA2                 TGCGTATGGTTTGACTTCGGATGAAGTCGCCAGGAGAAGAAAGAAATATGGGTTAAATCA    660
                           ***** ********* **** ******.* **************.********.** **

SGD_Scer_PMA2/YPL036W     AATGGCTGAGGAGAATGAATCGTTGATTGTGAAGTTTTTGATGTTCTTCGTAGGGCCTAT    458
MIT_Sbay_c676_25008       AATGGCTGAAAACAATGAATCGTTAGTTGTTAAATTCATCATGTTTTTCGTTGGTCCAAT    419
6033_PMA2                 AATGGCTGAGGAGAATGAATCGTTGATTGTGAAGTTTTTGATGTTCTTCGTAGGGCCGAT    720
                          *********..* ***********..**** **.** :* ***** *****:** ** **

SGD_Scer_PMA2/YPL036W     TCAATTCGTTATGGAGGCTGCTGCTATTTTGGCTGCCGGTTTGTCTGATTGGGTTGATGT    518
MIT_Sbay_c676_25008       TCAATTCGTTATGGAAGCTGCTGCTATCTTGGCTGCCGGTTTGTCCGATTGGGTCGATTT    479
6033_PMA2                 TCAATTCGTTATGGAGGCAGCTGCTATTTTGGCTGCCGGT--------------------    760
                          ***************.**:******** ************
```

# DBVPG 6261

## UTP4/YDR324c, chrIV:1115815-1115829

```
SGD_Scer_UTP4/YDR324C      TCCATTAAATTCTGGGATTTCCAGTTTGCCACGCTAAACCAGTCATTTAAGGCGCACGAT      840
MIT_Sbay_c442_4963         TCCATAAAGTTCTGGGATTTCCAATTCGCTACATTGAACCAGTCTTTTAAAGCACATGAT      768
6261_UTP4                  ---------------------------------------------------GCACATGAT      9
                                                                              **.** ***

SGD_Scer_UTP4/YDR324C      GCAGACGTACTGTGTCTAACTACCGATACTGATAATAATTATGTTTTTAGTGCTGGTGTG      900
MIT_Sbay_c442_4963         GCAGACGTCTTATGCCTAACTACTGATATTGATAACAATTACGTGTTCAGTGCTGGTGTG      828
6261_UTP4                  GCAGacgtctTAtgccTAACTActGATATTGaTAACAATTACgtgttTAGTgatggtgTG      69
                           ****       *.     ******  **** ** *** *****       ***       **

SGD_Scer_UTP4/YDR324C      GACAGAAAAATCTTTCAATTTTCTCAAAACACTAACAAATCTCAAAAGAACAACAGATGG      960
MIT_Sbay_c442_4963         GATAGAAAAATCTTCCAATTCTCCCAAAATAGCAACAAATCTCAAAAAAACAACAGATGG      888
6261_UTP4                  GATagAAAGATCTTCCAATTTTCTCAAAATAGCAACAAATCTCAAAAGAACAACAGATGG      129
                           **   ***.***** ***** ** ***** *  **************.************

SGD_Scer_UTP4/YDR324C      GTAAATTCTTCTAATAGGTTGCTTCATGGAAACGACATTAGAGCAATATGTGCATACCAA      1020
MIT_Sbay_c442_4963         GTGAACTCTTCCAATAGGTTACTTCATGGTAACGAA------------------------      924
6261_UTP4                  gtAaaTTCTTCTAATAGGTTGCTTCATGGAAACGACATTAGAGCAATATGTGCATACCAA      189
                             .    ***** ********.********:*****.

SGD_Scer_UTP4/YDR324C      TCTAAAGGTGCAGATTTTCTAGTTTCAGGAGGTGTTGAAAAAACACTAGTCATCAACTCA      1080
MIT_Sbay_c442_4963         ------------------------------------------------------------      924
6261_UTP4                  TCTAAAGGTGCAGATTTTCTAGTTTCAGGAGGTGTTGAAAAAACACTAGTCATCAACTCA      249


SGD_Scer_UTP4/YDR324C      CTTACTTCTTTTTCTAATGGAAACTACAGGAAGATGCCAACTGTCGAACCTTATTCAAAG      1140
MIT_Sbay_c442_4963         ------------------------------------------------------------      924
6261_UTP4                  CTTACTTCTTTTtctAATGGAAACTACAGGAAGATGCCAACTGTcgAAcctTATTCAAAG      309


SGD_Scer_UTP4/YDR324C      AATGTTTTAGTTAACAAAGAGCAACGCCTTGTTGTTTCATGGAGCGAATCTACTGTTAAG      1200
MIT_Sbay_c442_4963         ------------------------------------------------------------      924
6261_UTP4                  AATGTTTTAGTTAACAAAGAGCAACGCCTTGTTGTTTCATGgagagAAtcTACTGTtaag      369


SGD_Scer_UTP4/YDR324C      ATATGGACAATGGGAACCGATTCTAGTACAGAACAGAATTATAAGCTAGTTTGCAAGTTA      1260
MIT_Sbay_c442_4963         ------------------------------------------------------------      924
6261_UTP4                  ATAtgGACAATGG-----------------------------------------------      382
```

## CHD1/YER164w, chrV:507240-507245

```
SGD_Scer_CHD1/YER164W      AGAGAAAACTCTAAGATCCTCCCACAATATTCCAGTAATTATACTTCACAGAGACCGCGT      1080
MIT_Sbay_c996_7051         AGAGAAAACTCGAAGATTCTTCCACAATATTCTAGTAACTATACTTCACAAAGACCACGT      1074
6261_CHD1                  -------------------------------------------CTTCACAAAGACCACGT      17
                                                                      *******.*****.***

SGD_Scer_CHD1/YER164W      TTTGAGAAGTTAAGCGTGCAACCTCCGTTCATTAAAGGTGGGGAATTAAGAGATTTTCAA      1140
MIT_Sbay_c996_7051         TTCGAGAAGTTAAGTGTGCAGCCACCTTTCATCAAGGGCGGAGAATTAAGAGATTTCCAA      1134
6261_CHD1                  TTCNAGAAGTTAAGNNNNCAGCCACCTTTCATAAAGGGCGGAGAACTGAGAGATTTCCAA      77
                           **  **********    . **.**:** ***** **.** **.*** *.******** ***

SGD_Scer_CHD1/YER164W      CTAACTGGTATTAATTGGATGGCATTTTTGTGGTCCAAAGGTGATAATGGTATACTGGCA      1200
MIT_Sbay_c996_7051         CTGACTGGTATCAATTGGATGGCATTTTTGTGGTCTAAAGGCGATAATGGTATCTTAGCA      1194
6261_CHD1                  CTAACCGGTACTAATTGGATGGCATTNTTATGGTCTAAAGGCGATAATGGTATCTTAGCA      137
                           **.** ****  **************.**.***** ***** ***********. *.***

SGD_Scer_CHD1/YER164W      GATGAGATGGGCCTGGGAAAAACGGTCCAGACTGTCGCCTTTATCAGTTGGCTGATATTT      1260
MIT_Sbay_c996_7051         GACGAAATGGGGCTGGGAAAAACCGTTCAGACTGTGGCATTCATCAGTTGGCTGATTTTT      1254
6261_CHD1                  GACGAAATGGGGCTAGGAAAAACCGTTCAGACCGTGGCATTTATCAGTTGGCTGATTNTT      197
                           ** **.***** **.******** ** ***** ** **.** **************:.**

SGD_Scer_CHD1/YER164W      GCTCGTAGACAAAACGGACCTCACATCATTGTCGTTCCTTTATCGACAATGCCTGCCTGG      1320
MIT_Sbay_c996_7051         GCACGTAGGCAAAACGGTCCTCACATTGTTGTGGTTCCTTTGTCTACAATGCCCGCTTGG      1314
6261_CHD1                  GCACGTAGGCAAAACGGTCCTCACATTGTTGTGGTTCCTTTGTCTACAATGCCTGCTTGG      257
                           **:*****.********:********  .**** ********.** ******** ** ***

SGD_Scer_CHD1/YER164W      TTGGATACTTTTGAGAAATGGGCGCCTGATTTGAATTGTATATGCTATATGGGCAACCAA      1380
MIT_Sbay_c996_7051         TTAGATACATTTGAAAAATGGTCACCTGATTTGAATTGTATTTGCTATATGGGTAACCAA      1374
6261_CHD1                  TTAGATACATTTGAGAGATGGTCACCTGATTTGAATTGTATTTGCTATATGGGTAACCAA      317
```

```
                             **.*****:*****.*.**** *.*****************:*********** ******

SGD_Scer_CHD1/YER164W        AAATCAAGAGATACCATTCGAGAATATGAATTTTACACCAATCCAAGGGCAAAAGGGAAA      1440
MIT_Sbay_c996_7051           AAATCAAGAGATACCCTTAGGGAATACGAATTCTACACCAATCCACAGGCAAAGGGGAAG      1434
6261_CHD1                    AAATCAAGAGATACCCTTAGGGAATACGAATTCTACACCAATCCACAGGCAAAGGGGAAG      377
                             ***************.**.*.***** ***** ************..******.*****.

SGD_Scer_CHD1/YER164W        AAAACAATGAAATTTAATGTTTTATTAACAACATACGAGTACATCTTAAAGGATCGTGCT      1500
MIT_Sbay_c996_7051           AAAACAATGAAATTTAATGTTTTATTAACAACATACGAATATATCTTAAAAGACCGTGCT      1494
6261_CHD1                    AAAACAATGAAATTTAACGTTTTATTAACAACATACGAATATATCTTAAAGGACCGTGCT      437
                             ***************** ********************.** ********.** ******

SGD_Scer_CHD1/YER164W        GAATTAGGAAGTATAAAATGGCAATTTATGGCCGTTGACGAAGCTCATAGACTAAAAAAT      1560
MIT_Sbay_c996_7051           GAATTAGGTGGCATCAAATGGCAATTTATGGCCGTTGATGAAGCCCATAGATTGAAAAAC      1554
6261_CHD1                    GAATTAGGTGGCATCAAATGGCAATTTATGGCCGTTGATGAAGCTCATAGATTAAAAAAC      497
                             ********:.* **.*********************** ***** ****** *.*****

SGD_Scer_CHD1/YER164W        GCTGAATCATCCCTTTATGAATCATTAAACAGTTTCAAGGTCGCCAACCGTATGTTAATC      1620
MIT_Sbay_c996_7051           GCGGAATCATCTCTTTATGAATCGTTGAACAGTTTCAAGGCTGCGAACCGTATGTTGATC      1614
6261_CHD1                    GCAGAATCATCTCTTTATGAATCATTGAACAGTTTCAAGGCTGCGAACCGTATGTTGATC      557
                             ** ******** ***********.**.*************  ** ***********.***

SGD_Scer_CHD1/YER164W        ACAGGCACACCTCTTCAGAATAATATTAAAGAGTTAGCTGCGTTGGTTAATTTCCTAATG      1680
MIT_Sbay_c996_7051           ACAGGTACTCCTCTTCAAAATAATATTAAAGAACTGGCTGCATTGATTAATTTCCTAATG      1674
6261_CHD1                    ACAGGTACTCCTCTTCAAAATAATATTAAAGAACTGGCTGCATTAATTAATTTCTTAATG      617
                             ***** **:********.**************. *.*****.**..******** *****

SGD_Scer_CHD1/YER164W        CCCGGAAGGTTTACGATTGATCAGGAGATTGATTTTGAAAACCAAGATGAAGAGCAAGAA      1740
MIT_Sbay_c996_7051           CCTGGAAGATTTACTATTGATCAAGAAATTGATTTTGAGAACCAAGACGCTGAACAAGAA      1734
6261_CHD1                    CCTGGAAGATTTACTATTGATCAAGAAATTGATTTTGAGAATCAAGATGCCGAACAAGAA      677
                             ** *****.***** ********.**.***********.** ***** *. **.******

SGD_Scer_CHD1/YER164W        GAATATATTCATGATTTACACCGAAGAATACAGCCTTTTATTCTTCGTCGGTTGAAGAAA      1800
MIT_Sbay_c996_7051           GAATATATTCATGATTTACATAGGAGGTTACAGCCTTTCATTCTTCGTCGTTTGAAGAAA      1794
6261_CHD1                    GAATATATCCATGATTTACATAGGAGGCTTCAGCCTTTCATTCTTCGTCGTTTAAAGAAA      737
                             ******** *********** .*.**. *:******** *********** **.******

SGD_Scer_CHD1/YER164W        GACGTAGAAAAATCACTTCCATCAAAGACAGAGCGTATTTTAAGAGTTGAATTGTCCGAC      1860
MIT_Sbay_c996_7051           GATGTAGAAAAATCATTGCCTTCAAAGACAGAACGTATTTTGAGAGTCGAACTATCCGAT      1854
6261_CHD1                    GATGTAGAAAAGTCATTGCCTTCAAAGACAGAACGTATTTTGAGAGTCGAACTATCCGAC      797
                             ** ********.*** * **:***********.********.*****.*** *.*****

SGD_Scer_CHD1/YER164W        GTACAGACTGAGTACTATAAAAATATTCTGACTAAAAACTACTCTGCTTTAACTGCTGGA      1920
MIT_Sbay_c996_7051           GTACAGACGGAGTACTATAAAAATATCTTGACCAAAAACTATTCTGCTTTAACTGCTGGT      1914
6261_CHD1                    GTACAGACTGAGTACTATAAAAATATTCTGACTAAAAACTACTCTGCTTTAACTGCTGGA      857
                             ******** *****************  **** ******** *****************:

SGD_Scer_CHD1/YER164W        GCTAAAGGGGGTCATTTCTCTTTACTGAATATTATGAACGAGTTGAAAAAGGCATCGAAC      1980
MIT_Sbay_c996_7051           CCCAAAGGTGGTCACTTTTCTTTATTGAATATCATGAATGAATTGAAAAAGGCATCAAAC      1974
6261_CHD1                    GCTAAAGGGGGTCATTTCTCTTTACTGAATATTATGAACGAGTTGAAAAAGGCATCGAAC      917
                              * ***** ***** ** ****** ******* ***** **.**************.***

SGD_Scer_CHD1/YER164W        CATCCATATCTCTTCGATAATGCTGAAGAGCGCGTCTTACAGAAATTTGGGGATGGTAAA      2040
MIT_Sbay_c996_7051           CATCCATATCTATTCGATAATGCTGAAGACCGTGTTTTACAGAAGTTTGGAGATGGTAAA      2034
6261_CHD1                    CATCCATATCTCTTCGATAATGCTGAAGAGCGCGTCTTACAGAAATTTGGGGATGGTAAA      977
                             ***********.***************** ** ** ********.*****.*********

SGD_Scer_CHD1/YER164W        ATGACTCGAGAAAACGTACTAAGAGGTTTGATCATGTCTTCGGGTAAGATGGTTCTTTTA      2100
MIT_Sbay_c996_7051           ATGACCCGTGAAAACATTTTAAGAGGGCTGATCATGTCTTCAGGTAAAATGGTTCTTTTA      2094
6261_CHD1                    ATGACTCGAGAAAACGTACTAAGAGGTTTGATCATGTCTTCGGGTAAGATGGTTCTTTTA      1037
                             ***** **:******.*: *******  **************.*****.************

SGD_Scer_CHD1/YER164W        GACCAATTATTGACCAGATTGAAGAAAGATGGGCACCGCGTGTTGATTTTTTCACAAATG      2160
MIT_Sbay_c996_7051           GATCAATTATTGACCAGATTGAAAAAAGATGGGCATCGTGTGTTAATTTTCTCTCAAATG      2154
6261_CHD1                    GACCAATTAT--------------------------------------------------      1047
                             ** *******
```

## KEM1/YGL173c, chrVII:179656-179664

```
SGD_Scer_XRN1/YGL173C        ----------------------ATGGGTATTCCAAAATTTTTCAGGTACATCTCAGAAAG      38
MIT_Sbay_c403_8392           ----------------------ATGGGTATTCCAAAATTTTTCAGATACATCTCAGAGAG      38
6261_KEM1                    AACCAACATATACTAGTACAGCATGGGTATTCCGAAATTTTTCAGGTACATCTCAGAAAG      120
                                                   ***********.***********.***********.**
```

```
SGD_Scer_XRN1/YGL173C       ATGGCCCATGATTTTACAGCTTATTGAGGGAACACAGATTCCTGAGTTTGATAACTTATA      98
MIT_Sbay_c403_8392          ATGGCCTATGATTCTACAACTTATTGAAGGGACTCAGATTCCCGAGTTTGATAACCTGTA      98
6261_KEM1                   ATGGCCCATGATTCTACAACTTATTGAAGGGACTCAGATTCCTGAGTTTGATAACCTATA      180
                            ****** ****** ****.********.**.**:******** *********** *.**

SGD_Scer_XRN1/YGL173C       CCTGGATATGAATTCGATTTTACATAATTGTACGCATGGTAACGACGATGATGTAACCAA      158
MIT_Sbay_c403_8392          TCTGGACATGAATTCGATTTTACATAACTGTACGCATGGTAACGACGATGACGTGACGAA      158
6261_KEM1                   TCTGGATATGAATTCGATTTTACATAACTGTACACATGGTAACGACGATGACGTGACGAA      240
                             ***** ******************* *****.**************** **.** **

SGD_Scer_XRN1/YGL173C       GCGATTAACTGAAGAAGAGGTTTTTGCAAAAATCTGTACGTATATCGATCACCTTTTTCA      218
MIT_Sbay_c403_8392          GCGACTAACTGAAGAAGAGGTTTTTGCCAAGATCTGCACGTATATCGACCATCTTTTTCA      218
6261_KEM1                   GCGATTAACTGAAGAAGAGGTTTTTGCAAAGATCTGCACGTATATTGATCATCTTTTTCA      300
                            **** *********************.**.***** ******** ** ** ********

SGD_Scer_XRN1/YGL173C       AACAATCAAGCCCAAGAAGATTTTCTACATGGCTATTGATGGTGTGGCCCCTCGTGCCAA      278
MIT_Sbay_c403_8392          AACCATCAAGCCCAAGCAAATTTTTTACATGGCTATCGATGGTGTGGCTCCCCGTGCGAA      278
6261_KEM1                   AACTATCAAGCCCAAGCAAATTTTTTACATGGCTATCGATGGTGTGGCCCCCCGCGCGAA      360
                            *** ************.*.***** *********** *********** ** ** ** **

SGD_Scer_XRN1/YGL173C       GATGAATCAACAAAGAGCTCGTAGATTCAGAACCGCTATGGATGCAGAAAAAGCCTTGAA      338
MIT_Sbay_c403_8392          GATGAATCAACAGAGAGCTCGTAGGTTTAGAACTGCCATGGACGCCGAAAAAGCTATGAA      338
6261_KEM1                   GATGAATCAACAGAGAGCTCGTAGATTTAGAACTGCCATGGACGCCGAAAAAGCTATGAA      420
                            ************.***********.** ***** ** ***** **.********  :****

SGD_Scer_XRN1/YGL173C       GAAGGCTATTGAGAATGGTGACGAGATTCCTAAGGGTGAGCCATTTGATTCGAATTCTAT      398
MIT_Sbay_c403_8392          AAAAGCTATTGAAAATGGTGACGAGATTCCTAAAGGTGAACCGTTCGACTCAAATTGTAT      398
6261_KEM1                   AAAAGCTATTGAAAATGGTGACGAAATTCCTAAAGGTGAGCCGTTCGACTCAAACTGTAT      480
                            .**.********.***********.********.*****.**.** ** **.** * ***

SGD_Scer_XRN1/YGL173C       TACTCCAGGTACGGAGTTTATGGCCAAATTGACCAAAAACTTACAATATTTTATTCACGA      458
MIT_Sbay_c403_8392          TACTCCAGGTACTGAATTCATGGCCAAACTGACAAAAAAATTTACAGTATTTCATTCATGA      458
6261_KEM1                   TACTCCAGGTACTGAATTTATGGCTAAACTGACAAAAAAATTTACAGTATTTCATTCATGA      540
                            ************ **.** ***** *** ****.***** *****.****** ***** **

SGD_Scer_XRN1/YGL173C       CAAGATTTCTAACGATTCCAAATGGAGGGAAGTGCAAATCATATTTTCTGGCCATGAAGT      518
MIT_Sbay_c403_8392          CAAGATCTCTAACGATTCTAAGTGGAGAGAAGTACAAATTATATTTTCTGGCCATGAAGT      518
6261_KEM1                   CAAAATTTCTAACGATTCCAAATGGAGGGAAGTGCAAATCATATTTTCTGGCCATGAAGT      600
                            ***.** *********** **.*****.*****.***** ********************

SGD_Scer_XRN1/YGL173C       TCCAGGTGAAGGTGAACACAAGATCATGAACTTTATAAGGCATTTAAAATCCCAAAAGGA      578
MIT_Sbay_c403_8392          TCCAGGTGAAGGTGAGCACAAAATCATGAATTTCATAAGGCATTTAAAATCACAGAAGGA      578
6261_KEM1                   TCCAGGTGAAGGTGAACACAAGATCATGAACTTTATAAGGCATTTAAAatcCCaAAAGGA      660
                            ***************.*****.******** ** **************   .* .*****

SGD_Scer_XRN1/YGL173C       TTTCAACCAGAATACGAGACATTGTATTTACGGTCTTGACGCAGATTTGATTATGCTGGG      638
MIT_Sbay_c403_8392          TTTTAACCAAAATACGAGACATTGTATTTATGGTCTTGATGCGGATTTGATCATGCTAGG      638
6261_KEM1                   tT----------------------------------------------------------      662
                             *
```

## TOR2/YKL203c, chrXI:60183-60193

```
SGD_Scer_TOR2/YKL203C       GCTGCTATTCAAGCTATTATGCATATTTTTCAAAACCTTGGTTTACGATGTGTCTCCTTT      2880
MIT_Sbay_c251_13959         GCGTCCATCCAAGCAATTATGCATATCTTTCAAAGCCTAGGTTTACGTTGTGTCTCATTT      2880
6261_TOR2                   ---------------ATTATGCATATTTTTCAAAACCTTGCTTTACGATGTGTCTCCTTT      45
                                           *********** *******.***:* ******:********.***

SGD_Scer_TOR2/YKL203C       TTGGATCAAATTATTCCAGGTATCATTTTAGTCATGCGTTCATGCCCGCCGTCCCAACTT      2940
MIT_Sbay_c251_13959         TTAGATCAGATAATTCCAGGTATTATCCTGGTGATGCGTTCATGTCCACCATCACAATTG      2940
6261_TOR2                   TTGGATCAAATTATTCCAGGTATCATTTTAGTCATGCGTTCGTGCCCGCTGTCCCAACTT      105
                            **.*****.**:*********** **  *.** ********.** **.* .**.*** *

SGD_Scer_TOR2/YKL203C       GACTTTTATTTTCAGCAACTGGGATCTCTCATCTCAATTGTCAAGCAACATATTAGGCCC      3000
MIT_Sbay_c251_13959         GAATTCTATTTTCAACAACTTTGTTCACTTATATTGATTGTCAAACAGCATATCAGGCCT      3000
6261_TOR2                   GACTTTTATTTTCAGCAACTGGGATCTCTCATCTCAATTGTCAAGCAACATATTAGGCCC      165
                            **.** ********.*****  *:**:** **.* .********.**.***** *****

SGD_Scer_TOR2/YKL203C       CATGTCGAGAAAATTTATGGTGTGATCAGGGAGTTTTTCCCGATCATTAAACTACAAATC      3060
MIT_Sbay_c251_13959         CATGTCGAGAAAATTTATAGTGTAATCAAGGAGTTTTTCTCAATCATCAAATTACAAATT      3060
6261_TOR2                   CATGTCGAGAAAATTTATGGTGTGATCAGGGAGTTTTTCCCGATCATTAAACTACAAATC      225
                            ******************.****.****.********** *.***** *** *******

SGD_Scer_TOR2/YKL203C       ACAATTATTTCTGTCATAGAATCGATATCTAAGGCTCTGGAAGGTGAGTTTAAAAGATTT      3120
MIT_Sbay_c251_13959         ACAATAATATCGGTTATTGAATCGATTTCTAAAGCCTTAGAAGGTGAGTTCAAAAGATTC      3120
```

```
6261_TOR2                 ACAATTATTTCTGTCATAGAATCGATATCTAAGGCTCTGGAAGGTGAGTTTAAAAGATTT   285
                          *****:**:** ** **:********:*****.**  *.*********** ********

SGD_Scer_TOR2/YKL203C     GTTCCCGAGACTCTAACCTTTTTCCTTGATATTCTTGAGAACGACCAGTCTAATAAAAGG   3180
MIT_Sbay_c251_13959       GTCCCGGAGACTTTGACATTTTTCCTCGATATCCTTGAGAACGACCAATCTAATAAAAAA   3180
6261_TOR2                 GTTCCCGAGACCCTAACCTTTTTCCTTGATATTCTTGAGAACGACCAGTCTAATAAAAAA   345
                          ** ** *****  *.**.******** ***** **************.**********..

SGD_Scer_TOR2/YKL203C     ATCGTTCCGATTCGTATATTAAAATCTTTGGTTACTTTTGGGCCGAATCTAGAAGACTAT   3240
MIT_Sbay_c251_13959       ATCGTTTCTATCCGTATACTGAAGTCTTTGGTGGCTTTTGGATCCAATTTAGAGGATTAT   3240
6261_TOR2                 ATTGTTTCTATCCGTATACTGAAGTCCTTAGTGACTTTTGGATCCAATTTAGAGGATTAC   405
                          ** *** * ** ****** *.**.** **.** .*******. * *** ****.** **

SGD_Scer_TOR2/YKL203C     TCCCATTTGATTATGCCTATCGTTGTTAGAATGACTGAGTATTCTGCTGGAAGTCTAAAG   3300
MIT_Sbay_c251_13959       TCCCATTTAATTATGCCTGTAGTCGTAAGAATGACAGAGTACTCCGCTGGAGGTTTAAAG   3300
6261_TOR2                 TCCCATTTAATTATGCCCGTAGTCGTCAGAATGACTGAATACTCCGCTGGAAGTTTAAAA   465
                          ********.******** .*.** ** ********:**.** ** ******.** ****.

SGD_Scer_TOR2/YKL203C     AAAATCTCCATTATAACTTTGGGTAGATTAGCAAAGAATATCAACCTCTCTGAAATGTCA   3360
MIT_Sbay_c251_13959       AAAATTTCAATCATTACGTTGGGGAGATTAGCAAAGAACATCAATCTCTCAGAAATGTCA   3360
6261_TOR2                 AAAaTTGCAATCATTACGTTGGGGAGATTAGCAAAGAATATCAACCTCTCAGAAATGTCA   525
                          *** *  *.** **:** ***** ************** ***** *****:*********

SGD_Scer_TOR2/YKL203C     TCAAGAATTGTTCAGGCGTTGGTAAGAATTTTGAATAATGGGGATAGAGAACTAACAAAA   3420
MIT_Sbay_c251_13959       TCAAGGATTGTCCAAGCGTTGGTGCGAATCTTGAATAATGGGGACAAAGAACTAACAAAG   3420
6261_TOR2                 TCAAGGATTGTTCAGGCACTGGTGCGAATCTTGAATAacGGGGACAGAGAACTAACAAAa   585
                          *****.***** **.**. ****..**** *******  ***** *.************

SGD_Scer_TOR2/YKL203C     GCAACCATGAATACGCTAAGTTTGCTCCTTTTACAACTAGGTACCGACTTTGTGGTCTTT   3480
MIT_Sbay_c251_13959       GCAACTATGAATACATTAAGTTTACTTCTTTTACAACTGGGTACTGACTTTGTCATATTT   3480
6261_TOR2                 gCaACCATGAATACATTGAGTTTGCTTCTttTGCAATTGGGTACCG--------------   631
                           * ** ********. *.*****.** **  *.*** *.***** *
```

## VMA5/YKL080w, chrXI:285492-285506

```
SGD_Scer_VMA5/YKL080W     TTCAAGTTAGATAAGTCTATCAAAGATTTGATAACATTGATTTCTAATGAATCTTCTCAA   420
MIT_Sbay_c262_14457       TTTAAGTTGGATAAGTCTATCAAAGATTTGATAACATTGATTTCTGATGAATCGTCTCAA   420
6261_VMA5                 ---------------------AAAGAtTtgaTAACaTTGATTTCtAATGaAtCTTCTCAA   39
                                               ***** *   **** ******** .*** * * ******

SGD_Scer_VMA5/YKL080W     TTAGACGCCGACGTCAGAGCTACTTATGCAAATTACAACAGCGCTAAAACTAACTTGGCT   480
MIT_Sbay_c262_14457       TTGGATGCTGATGTGAGAGCTACCTTTGCAAATTACAATAGTGCCAAAACTAATTTGGCT   480
6261_VMA5                 TTAGACGCCGACGTCAGAGCTACTTATGCAAATTACAACAGCGCTAAAACTAACTTGGCT   99
                          **.** ** ** ** ******** *:************ ** ** ******** ******

SGD_Scer_VMA5/YKL080W     GCTGCTGAGAGAAAGAAGACGGGTGACCTTTCTGTCAGATCCTTGCATGATATTGTCAAG   540
MIT_Sbay_c262_14457       GCTGCTGAAAGAAAGAAAACAGGCGACCTTTCTGTTAGATCCTTGCATGATATTGTCAAG   540
6261_VMA5                 GCTGCTGAGAGAAAGAAGACGGGTGACCTTTCTGTCAGATCCTTGCATGATATTGTCAAG   159
                          ********.********.**.** *********** ************************

SGD_Scer_VMA5/YKL080W     CCCGAAGACTTCGTTCTTAATTCTGAACATTTAACTACTGTTCTAGTAGCAGTTCCCAAA   600
MIT_Sbay_c262_14457       CCAGAAAATTTTGTTTTAAACTCTGAGCACTTGACCACTGTTCTTGTAGCAGTTCCAAAA   600
6261_VMA5                 CCCGAAGACTTCGTTCTTAATTCTGAACATTTAACTACTGTTCTAGTAGCAGTTCCCAAA   219
                          **.***.* ** *** *:** *****.**.** **.********:***********.***

SGD_Scer_VMA5/YKL080W     AGTTTAAAATCCGATTTCGAAAAATCGTACGAAACTTTATCCAAGAACGTTGTACCAGCA   660
MIT_Sbay_c262_14457       AGTTTGAAATCTGATTTCGAGAAAACTTATGAAACTTTGTCCAAGAACGTTGTTCCAGCA   660
6261_VMA5                 AGTTTAAAATCCGATTTCGAAAAATCGTACGAAACTTTATCCAAGAACGTTGTACCAGCA   279
                          *****.***** ********.***:* ** ********.**************:******

SGD_Scer_VMA5/YKL080W     TCTGCCAGCGTGATTGCAGAGGATGCTGAGTATGTTTTGTTCAATGTTCATTTGTTCAAG   720
MIT_Sbay_c262_14457       TCTGCAGGTGTAATTGCCGAAGATGCTGAGTATGTCTTATTCAATGTCCACTTGTTCAAG   720
6261_VMA5                 TCTGCCAGCGTGATTGCAGAGGATGCTGAGTATGTTTTGTTCAATGTTCATTTGTTCAAG   339
                          *****..* **.*****.**.************** **.******** ** *********

SGD_Scer_VMA5/YKL080W     AAAAACGTTCAAGAATTCACAACAGCTGCTAGAGAGAAGAAATTCATTCCTCGTGAATTT   780
MIT_Sbay_c262_14457       AAAAATGTTCAAGAGTTCACGGCGGCTGCAAGGGAGAAGAAATTCATCCCTCGTGAATTT   780
6261_VMA5                 AAAAACGTTCAAGAATTCACAACAGCTGCTAGAGAGAAGAAATTCATTCCTCGTGAATTT   399
                          ***** ********.*****..*.*****:**.************** ************

SGD_Scer_VMA5/YKL080W     AACTACTCGGAGGAATTAATTGACCAGTTGAAAAAAGAGCATGACTCTGCTGCCAGTTTA   840
MIT_Sbay_c262_14457       AACTACTCCGAGGAACTAATTGACCAATTGAAAAAGGAACATGACTCTGCTGCAAGTTTA   840
6261_VMA5                 AACTACTCGGAGGAATTAATTGACCAGTTGAAAAAAGAGCATGACTCTGCTGCAAGTTTA   459
                          ******** ****** **********.********.**.**************.******

SGD_Scer_VMA5/YKL080W     GAACAATCTTTGCGCGTCCAGTTGGTAAGATTGGCCAAGACAGCTTATGTCGATGTTTTT   900
```

```
MIT_Sbay_c262_14457       GAGCAATCTTTACGTGTCCAGCTAGTGAGATTGGCCAAGACAGCTTATGTTGATGTCTTC     900
6261_VMA5                 GAGCAATCTTTacgTGTCCAACTAGTTAGATTGGCCAAGACTGCTTATGTTGATGTCt--     517
                          **.*********    *****. *.** **************:******** *****
```

## GAL80/YML051w, chrXIII:172148-172153

```
SGD_Scer_GAL80/YML051W        --------------------------------ATGGACTACAACAAGAGATCTTCGGTCT     28
WashU_Sbay_Contig555.11       --------------------------------ATGGACTATAATAAGAGATCTTCCGTCT     28
6261_Gal80                    TTCGACCTCTAGTTTTCTCGTGCCTTCCAGTCATGGACTATAACAAGAGATCTTCCGTCT     120
                                                              ******** ** *********** ****

SGD_Scer_GAL80/YML051W        CAACCGTGCCTAATGCAGCTCCCATAAGAGTCGGATTCGTCGGTCTCAACGCAGCCAAAG     88
WashU_Sbay_Contig555.11       CCACAGTACCCAATGCAGCTCCCATAAGGGTCGGATTTATTGGCCTCAACGCTACCAAAG     88
6261_Gal80                    CTACAGTACCCAATGCAGCCCCCATAAGAGTCGGATTCATTGGCCTCAACGCTACCAAAG     180
                              * **.**.** ******** ********.******** .* ** ********:.******

SGD_Scer_GAL80/YML051W        GATGGGCAATCAAGACACATTACCCCGCCATACTGCAACTATCGTCACAATTTCAAATCA     148
WashU_Sbay_Contig555.11       GTTGGGCGATCAAAACTCATTATCCCGCCATACTGCAACTATCGTCCCAGTTCCAAATCA     148
6261_Gal80                    GGTGGGCGATCAAAACTCATTATCCCGCCATACTGCAGCTATCGTCCCAATTTCAAATTA     240
                              * *****.*****.**:***** **************.********.**.** ***** *

SGD_Scer_GAL80/YML051W        CTGCCTTATACAGTCCAAAAATTGAGACTTCTATTGCCACCATTCAGCGTCTAAAATTGA     208
WashU_Sbay_Contig555.11       CTGCCTTATACAGCCCAAAGATAGAGACCTCCATTTCCACTATCCAACAGCTGAAACTCA     208
6261_Gal80                    CTGCCTTATACAACCCAAAGATAGAGACCTCCATTGCCACTATCCAACAACTGAAACTTA     300
                              ************. *****.**:***** ** *** **** ** **.*. **.*** * *

SGD_Scer_GAL80/YML051W        GTAATGCCACTGCTTTTCCCACTTTAGAGTCATTTGCATCATCTTCCACTATAGATATGA     268
WashU_Sbay_Contig555.11       GTAATGCCACGGCCTTCCCCACTTTAAAATCATTTGCATCTTCTGCCACCGTGGACATGA     268
6261_Gal80                    GCAATGCCACGGCCTTCCCCACTTTAGAATCATTTGCATCTTCTGCCACCGTGGACATGA     360
                              * ******** ** ** *********.*.***********:*** ****  .*.** ****

SGD_Scer_GAL80/YML051W        TAGTGATAGCTATCCAAGTGGCCAGCCATTATGAAGTTGTTATGCCTCTCTTGGAATTCT     328
WashU_Sbay_Contig555.11       TAGTGATAACCATCCAGGTGGCCAGCCATTATGAGGTTCTGATGCCCCTCTTGAAATACT     328
6261_Gal80                    TAGTGATAACCATCCAGGTGGCCAGCCATTATGAGGTTCTAATGCCCCTGTTGAAATACT     420
                              ********.* *****.*****************.*** * ***** ** ***.***:**

SGD_Scer_GAL80/YML051W        CCAAAAATAATCCGAACCTCAAGTATCTTTTCGTAGAATGGGCCCTTGCATGTTCACTAG     388
WashU_Sbay_Contig555.11       CTCAGAACAATCCGAATCTCAAGTATCTTTTCGTAGAATGGGCCCTTGCATGTTCTCTGG     388
6261_Gal80                    CCCAAAATAATCCGAATCTCAAGTATCTTTTCGTAGAATGGGCCCTTGCATGTTCTCTGG     480
                              * .*.** ******** ***************************************:**.*

SGD_Scer_GAL80/YML051W        ATCAAGCCGAATCCATTTATAAGGCTGCTGCTGAACGTGGGGTTCAAACCATCATCTCTT     448
WashU_Sbay_Contig555.11       ATCAAGCAGAATCGATTTATAAGGCAGCTGCCGAACGTGGACTACAAACTATTATTTCTT     448
6261_Gal80                    ATCAAGCAGAATCGATTTATAAGGCAGCTGCCGAACGTGCACTACAAACTATTATTTCTT     540
                              *******.***** ***********:***** *******  . *:***** ** ** ****

SGD_Scer_GAL80/YML051W        TACAAGGTCGTAAATCACCATATATTTTGAGAGCAAAAGAATTAATATCTCAAGGCTATA     508
WashU_Sbay_Contig555.11       TACAAGGCCGTAAGTCACCATACATCTTGAGAGCCAAAGAACTAATCTCTGAAGGTTACA     508
6261_Gal80                    TACAAGGCCGTAAATCACCATACATCTTGAGAGCCAAAGAACTAATTTCTGAAGGTTACA     600
                              ******* *****.******** ** ********.****** **** *** **** ** *

SGD_Scer_GAL80/YML051W        TCGGCGACATTAATTCGATCGAGATTGCTGGAAATGGCGGTTGGTACGGCTACGAAAGGC     568
WashU_Sbay_Contig555.11       TTGGCGACATCAACTCCATAGAAATTGCAGGAAACGGCGGTTGGTATGGCTATGAAAGAC     568
6261_Gal80                    TTGGCGACATCAACTCCATAGAAATTGCAGGAAACGGCGGTTGGTATGGCTACGAAAGGC     660
                              * ******** ** ** **.**.*****:***** *********** ***** *****.*

SGD_Scer_GAL80/YML051W        CTGTTAAATCACCAAAATACATCTATGAAATCGGGAACGGTGTAGATCTGGTAACCACAA     628
WashU_Sbay_Contig555.11       CCATTAAATCGCCGAATTACATTTATGAAATTGGCAATGGTGTAGATCTGGTGACCACCA     628
6261_Gal80                    CTGTTAAATCACCAAAATACATCTATGAAATCGGGAACGGTGTAGATCTGGTAACCACAA     720
                              * .*******.**.**:***** ******** ** ** *************.*****.*

SGD_Scer_GAL80/YML051W        CATTTGGTCACACAATCGATATTTTACAATACATGACAAGTTCGTACTTTTCCAGGATAA     688
WashU_Sbay_Contig555.11       CATTTGGTCACACCATTGACCTTTTGCAATACATGACGAGCTCGTATTTTTCTAGAATAA     688
6261_Gal80                    CATTTGGTCACACAATCGATATTTTACAATACATGACAAGTTCGTACTTTTCCAGGATAA     780
                              *************.** ** .****.***********.** ***** ***** **.****

SGD_Scer_GAL80/YML051W        ATGCAATGGTTTTCAATAATATTCCAGAGCAAGAGCTGATAGATGAGCGTGGTAACCGAT     748
WashU_Sbay_Contig555.11       ACGCAATGGTGTTCAATAATATACCAGAACAAGAGCTAATAGATGAACATGGTAACCGAT     748
6261_Gal80                    ATGCAATGGTTTTCAATAATAttCCAGAGCAAGAGC------------------------     816
                              * ********* **********  *****.*******
```

## FKS3/YMR306w, chrXIII:882708-882716

```
SGD_Scer_FKS3/YMR306W     CTTGCAATCAATTTGGGACCTTCCGTGTATGTTCTGGGGTTTTTCGAATGGGATGTTCAT     1320
```

```
MIT_Sbay_c749_19333        CTCGCCATCAATCTAGGACCATCCATTTATGTTTTAGGGTTTTTCGAGTGGGATGTCTAT      1320
6261_FKS3                  ----TAATCAATTTAGGACCATCCATTTATGTTTTGGGGTTTTTCGAGTGGGATGTCCAT      56
                                .****** *.*****:***.* ****** *.***********.********  **

SGD_Scer_FKS3/YMR306W      TCAAAATCTGCGTATATCGTGTCCATCGTCCAACTAATCATTGCATTTTTAACCACTTTT      1380
MIT_Sbay_c749_19333        TCGAAATCAGCATATATCGTGTCTATCATTCAATTAATCATTGCACTTCTGACTACCCTT      1380
6261_FKS3                  tcgAAATCAGCATATATCGTGtCGATCATTCAATTAATAATTGCACTTCTAACCACCCTT      116
                              *****:**.********* * ***.* *** ****.****** ** *.** **  **

SGD_Scer_FKS3/YMR306W      TTTTTTGCTGTCAGACCTTTGGGCGGCTTATTTCGTCCATATTTGAATAAAGACAAAAAA      1440
MIT_Sbay_c749_19333        TTTTTTGCTATCAGGCCCTTGGGCGGCCTATTTCGTCCATATCTGAATAAAGACAAAAAG      1440
6261_FKS3                  TTTTTtGCTATCAGGCCCTTGGGCGGCCTATTTCGTCCATATCTGAATAAAGACAAAAGG      176
                           ***** ***.****.** ********* ************** ***************..

SGD_Scer_FKS3/YMR306W      CATCGGAGGTACATCTCATCGCAAACCTTTACCGCTTCATTTCCTAAGCTGACAGGACGC      1500
MIT_Sbay_c749_19333        CATCGAAGATACATCTCATCTCAGACTTTCACCGCTTCGTTTCCCAAGCTGGCGGGACGA      1500
6261_FKS3                  CATCGAAGATACGTCTCATCTCAGACTTTTACCGCTTCGTTTCCCAAGCTGGCAGGAAGA      236
                           *****.**.***.******* **.** ** ********.***** ******.*.***.*.

SGD_Scer_FKS3/YMR306W      AGCAAATGGTTTTCTTATGGGTTATGGGTATTTGTGTATTTGGCGAAATATATTGAGTCC      1560
MIT_Sbay_c749_19333        AGCAAATGGTTCTCTTACGGGCTATGGGTATTCGTATTTTTGGCAAAATACATTGAGTCT      1560
6261_FKS3                  AGCAAATGGTtctcTTACGGGCTATGGGTATTCGTATTTTTGGCAAAATACATTGAGTCC      296
                           **********    *** *** ********** **.*:******.***** ********

SGD_Scer_FKS3/YMR306W      TATTTTTTTTTGACCTTGTCCCTCAGGGACCCCATCAGGGTCCTCTCTATTATGGATTTG      1620
MIT_Sbay_c749_19333        TACTTCTTTTTAACCTTGTCCCTAAGGGACCCTATCAGGGTCCTATCTATAATGGATTTA      1620
6261_FKS3                  TAtTTTTTTTTGACCTTGTCCCTCAGGGACCCCATCAGGGTCCTCTCTATTATGGATTTG      356
                           ** ** *****.***********.******** ***********.*****:********.

SGD_Scer_FKS3/YMR306W      TCCAGATGTCAAGGTGAATATTTGTTGGGTCCTATTCTATGTAAATGGCAAGCCAAAATT      1680
MIT_Sbay_c749_19333        TCTAGATGTCAAGGTGAGTACCTATTGGGCCCCATACTATGTAAATGGCAGGCCAAAATT      1680
6261_FKS3                  tCCAGATGTCAAGGTGAATATTTGTTGGGTCCTATTCTATGTAAATGGCAAGCCAAAATT      416
                            * **************.**  *.***** ** **:**************.*********

SGD_Scer_FKS3/YMR306W      ACATTAGTTCTCATGCTGCTTTCTGACTTGGGCCTGTTTTTTCTCGACACTTACCTTTGG      1740
MIT_Sbay_c749_19333        ACACTGATCCTGATGTTACTTTCTGATTTGGGTCTGTTCTTTCTCGACACTTACCTTTGG      1740
6261_FKS3                  ACATTAGTTCTCATGCTGCTTTCTGACTTGGGCCTGTTTTTTCTCGACACTTACCTTTGG      476
                           *** *..* ** *** *.******** ***** ***** ********************

SGD_Scer_FKS3/YMR306W      TACATTATTTGCAACTGTATTTTTTCCATTGTACTGTCATTTTCTCTTGGTACTTCAATT      1800
MIT_Sbay_c749_19333        TACATTATTTGTAATTGTGTTTTTTCCATCATACTATCATTTTCTCTTGGTACCTCAATT      1800
6261_FKS3                  TACATTATTTGCAACTGTATTTTTTCCATTGTACTGTCATTTTCCCTTGGTACTTCAATT      536
                           *********** ** ***.********** .****.******** ******** ******

SGD_Scer_FKS3/YMR306W      CTCACGCCATGGAAGAATGTATACTCTCGATTGCCAAAAAGGATATATTCCAAAATCCTT      1860
MIT_Sbay_c749_19333        CTTACACCCTGGAAGAACGTATATTCTCGGATGCCCAAAAGGATATATTCCAAAATCCTG      1860
6261_FKS3                  CTCACGCCATGGAAGAATGTATACTCTAGATTGCCAAAAAGGATATATTCCAAAATCCTT      596
                           ** **.**.******** ***** ***.*.:****.**********************

SGD_Scer_FKS3/YMR306W      GCTACTTCAGAGATGGATGTAAAATTTAAAGCAAAAATACTGATATCGCAGGTTTGGAAT      1920
MIT_Sbay_c749_19333        GCTACCTCAGAAATGGATGTGAAATTCAAAGCAAAAATATTAATATCGCAAGTATGGAAT      1920
6261_FKS3                  GCTACTTCAGAGATGGATGTAAAATTTAAAGCAAAAATACTGATATCGCAGGTTTGGAAT      656
                           ***** *****.********.***** ************ *.********.**:******

SGD_Scer_FKS3/YMR306W      GCCATTGTTATATCAATGTATAGGGAACATCTTCTCTCCATTGAACATTTACAAAGACTC      1980
MIT_Sbay_c749_19333        GCTATCGTTATCTCAATGTATAGGGAACATCTTCTTTCCATTGAGCATTTACAGCGGCTA      1980
6261_FKS3                  GCCATTGTTATATCAATGTATAGGGAACATCTTCTCTCCATTGAGCATTTACAAAGACTC      716
                           ** ** *****.*********************** ********.********..*.**.

SGD_Scer_FKS3/YMR306W      TTGTTTCAGCAAGTTGACTCCTTAATGGGAGACACAAGAACCCTGAAATCGCCTACATTT      2040
MIT_Sbay_c749_19333        CTATTCCAACAAGTGGATTCTTTGATGGGTGATACAAGAACTTTAAAATCCCCGACATTT      2040
6261_FKS3                  TTGTTTCAGCAAGTTGACTCCTTAATGGGAGACACAAGAACCCTGAAATCGCCTACATTT      776
                            *.** **.***** ** ** **.*****:** ********  *.***** ** ******

SGD_Scer_FKS3/YMR306W      TTCGTTGCACAAGATGATTCAACGTTCAAGTCAATGGAATTTTTTCCATCAAATTCAGAG      2100
MIT_Sbay_c749_19333        TTTGTCGCTCAAGATGATTCAACTTTCAAATCTATGGAATTCTTTCCTTCCAAATCAGAA      2100
6261_FKS3                  TTCGTTGCACAAGATGATTCCACGTTCAAGTCAATGGAATTTTTTCCATCAAATT-----      831
                           ** ** **:***********.** *****.**:******** *****:**.**:*
```

## ECM3/YOR092w, chrXV:496862-496866

```
SGD_Scer_ECM3/YOR092W      AGATTGAAAATTGGAAAATTATACCCTGGTTTCTGGAAATCCGCAGTGGTATTAGTCTTT      1548
MIT_Sbay_c651_22895        AGATTGAAAATTGGGAAGTTATATCCGGGTTTCTGGAAGGCTGCTGTGGTTCTAGTGCTC      1560
6261_ECM3                  -----------------------------------------CGCAGTGGTATTAGTCTTT      19
                                                                     **:*****: ****  *
```

```
SGD_Scer_ECM3/YOR092W     CTCAGACAATGTATCATGCCGATCTTTGGTGTCTTGTGGTGTGACCGTCTAGTGAAAGCG     1608
MIT_Sbay_c651_22895       ATTAGGCAATGTATCATGCCCATTTTTGGTGTCTTGTGGTGCGATCGTCTGGTGAAAGCA     1620
6261_ECM3                 CTCAGACAATGTATCATGccnnnctTTGGTGTCTTGTGGTGTGACCGTCTAGTGAAGGCA     79
                          .* **.************       **************** ** *****.*****.**.

SGD_Scer_ECM3/YOR092W     GGATGGCTAAATTGGGAAAACGACAAGATGTTATTGTTTGTTACCGCCATTACTTGGAAC     1668
MIT_Sbay_c651_22895       GGTTGGCTAAACTGGGAAAACGACAAGATGTTACTGTTTGTCACTGCGATAGCTTGGAAT     1680
6261_ECM3                 GGATGGCTAAATTGGGAAAACGACAAGATGTTATTGTTTGTTACCGCTATTACTTGGAAC     139
                          **:******** ******************** ******* ** ** **:.*******

SGD_Scer_ECM3/YOR092W     TTACCAACAATGACCACCTTAATCTACTTCACTGCAAGTTATACCCCTGAGGACGAAACT     1728
MIT_Sbay_c651_22895       TTACCAACAATGACTACTCTAATCTATTTCACTGCGAGTTTCACTCCTGAAGATGAAACT     1740
6261_ECM3                 TTACCAACAATGACCACCTTAATCTACTTCACTGCAAGTTATACCCCTGAGGACGAAACT     199
                          ************** **  ******* ********.****: ** *****.** ******

SGD_Scer_ECM3/YOR092W     GAACCCGTTCAAATGGAATGTACCTCTTTCTTTTTGATGCTTCAATATCCTTTGATGGTC     1788
MIT_Sbay_c651_22895       GAACCCATTCAGATGGAATGCACTTCTTTCTTTTTGATGCTTCAGTATCCTTTGATGGTC     1800
6261_ECM3                 GAACCCGTTCAGATGGAATGCACTTCTTTCTTCTTGATGCTTCAGTATCCTCTGATGGTC     259
                          ******.****.******** ** ******** ***********.****** ********

SGD_Scer_ECM3/YOR092W     GTTAGTTTACCATTTTTGGTGTCTTATTTCATCAAAGTTCAAATGAAATTATAA------     1842
MIT_Sbay_c651_22895       GTTAGTTTACCATTTTTGGTCTCTTATTTCATAAAAGTTCAAATGAAATTATGA------     1854
6261_ECM3                 GTTAGTTTACCATTTTTGGTGTCTTATTTCATAAAGGTACAAATGAAACTATGATTTTCT     319
                          ******************** ***********.**.**:********* ***.*
```

## INP53/YOR109w, chrXV:526436-526442

```
SGD_Scer_INP53/YOR109W        AGAGTACAAATTACAAGATCATTTGAAGCCACCCAACCGGTATTTGACAAACATATCATG     960
WashU_Sbay_Contig620.20       AGGGTACAAATTACTAGATCATTCGAAGCTACTCAGCCAGTGTTTGACAAGCATATCATG     960
6261                          -------------------------------------CagtgTTTGACAAGCACATTATG     23
                                                                   *    ********.** ** ***

SGD_Scer_INP53/YOR109W        AAATCAGTGGAAAAGTACGGACCTGTGCATGTCGTTAATTTGTTATCAACGAAATCTTCT     1020
WashU_Sbay_Contig620.20       AAGTCAGTTGAAAAATACGGGCCCGTTCACGTTGTCAATCTGTTATCAACAAAATCTTCA     1020
6261                          AAGTCAGTAGAAAAATACGGCCCCgtTCACGTTGTCAACCTGtTATCAACAAAATCTTCA     83
                              **.***** *****.***** **    ** ** ** **  ** *******.********:

SGD_Scer_INP53/YOR109W        GAAATTGAACTTTCAAAACGATACAAGGAGCATTTAACTCATTCAAAAAAATTGAACTTC     1080
WashU_Sbay_Contig620.20       GAAATTGAACTTTCAAAGCGATACAAAGAGCATCTAACACATTCAAAAAAGTTGAATTTC     1080
6261                          GAAATAGAACTTTCAAAGCGATACAAAGAGCATCTAACGCACTCAAAAAAGTTGAATTTC     143
                              *****:***********.********.****** **** ** ********.***** ***

SGD_Scer_INP53/YOR109W        AACAAAGATATATTTTTGACAGAATTCGATTTTCATAAAGAAACTTCGCAAGAAGGCTTT     1140
WashU_Sbay_Contig620.20       AACAAAGATGTATTCTTAACAGAGTTCGATTTTCACAAAGAGACTTCTCAGGAAGGGTTT     1140
6261                          AATAAAGATGTATTCTTAACAGAGTTCGATTTTCACAAAGAGACTTCTCAGGAAGGGTTT     203
                              ** ******.**** **.*****.*********** *****.***** **.***** ***

SGD_Scer_INP53/YOR109W        TCCGGTGTCAGAAAACTCATTCCATTAATATTGGACTCTCTTTTATCTTCTGGCTATTAT     1200
WashU_Sbay_Contig620.20       TCCGGTGTCAGGAAAGTTATTCCACTCATTATGGACTCTCTTCTTTCATCTGGCTATTAC     1200
6261                          TCCGGTGTCAGAAAACTTATTCCATTAATATTGGACTCTCTTTTATCTTCTGGCTATTAT     263
                              ***********.*** * ****** *.**::*********** *:**:***********

SGD_Scer_INP53/YOR109W        TCTTACGATGTTAGAGAAAAAAAGAACATATCTGAACAACATGGCATATTTAGGACCAAC     1260
WashU_Sbay_Contig620.20       TCCTATGATGTTAAGGAAAAGAAAAATATATCCGAGCAGCACGGTATATTCAGAACTAAT     1260
6261                          TCTTACGATGTTAGAGAAAAAAAGAACATATCTGAACAACATGGCATATTTAGGACCAAC     323
                              ** ** *******..*****.**.** ***** **.**.** ** ***** **.** **

SGD_Scer_INP53/YOR109W        TGTTTAGATTGTTTGGATAGAACAAATTTAGCTCAGCAAATTATTTCTTTGGCTGCTTTT     1320
WashU_Sbay_Contig620.20       TGTTTAGATTGTTTGGATAGAACAAATTTAGCTCAGCAAGTCATTTCATTAGCTGCTCTA     1320
6261                          TGTTTAGATTGTTTGGATAGAACAAATTTAGCTCAGCAAATTATTTCTTTGGCTGCTTTT     383
                              ***************************************.* *****:**.****** *:

SGD_Scer_INP53/YOR109W        AGAACTTTTCTCGAAGATTTCCGATTGATTAGTTCAAATTCGTTCATCGACGATGATGAT     1380
WashU_Sbay_Contig620.20       AGAACCTTTTTAGAAGATTTTAGATTGATCAGTTCAAATTCTTTCATCGATGATGATGAT     1380
6261                          AGAACTTTTCTCGAAGATTTCCGATTGATTGGTTCAAATTCGTTCATCGACGATGATGAT     443
                              ***** *** *.********  .*******  .*********** ******** *********

SGD_Scer_INP53/YOR109W        TTCGTTTCTAAACATAACACCCTGTGGGCTGATCACGGTGATCAAATTTCCCAAATATAT     1440
WashU_Sbay_Contig620.20       TTTGTTTCCAAGCACAATACCCTCTGGGCTGATCACGGTGATCAGGTCTCCCAAATATAT     1440
6261                          TTCGTTTCTAAACATAACACCCTGTGGGCTGATCACGGTGATCAAATATCCCAAATATAT     503
                              ** ***** **.** ** ***** ********************..* ************

SGD_Scer_INP53/YOR109W        ACTGGTACTAATGCTTTGAAGTCCTCATTTTCAAGAAAAGGTAAAATGTCACTTGCTGGG     1500
WashU_Sbay_Contig620.20       ACCGGTACCAATGCTTTAAAGTCTTCATTTTCAAGGAAGGGTAAAATGTCATTTGCCGGG     1500
6261                          ACTGGTACTAATGCTTTGAAGTCCTCCTTTTCAAGAAAAGGTAAAATGTCACTTGCTGGG     563
```

```
                              ** ***** ********.***** **.********.**.************ **** ***

SGD_Scer_INP53/YOR109W        GCATTATCAGACGCCACAAAATCGGTCAGCAGAATATATATTAACAATTTCATGGATAAA    1560
WashU_Sbay_Contig620.20       GCATTATCAGACGCAACAAAGTCTGTCAGCAGAATATATATAAACAACTTTATGGATAAG    1560
6261                          GCATTATCAGACGCCACAAAATCGGTCAGCAGAATATATATTAACAATTTCATGGATAAA    623
                              **************.*****.** ****************:***** ** ********.

SGD_Scer_INP53/YOR109W        GAAAAGCAACAAAATATCGATACTTTGTTGGGAAGGTTACCGTATCAGAAAGCAGTGCAA    1620
WashU_Sbay_Contig620.20       GAAAAGCAGCAGAATATTGATACTTTATTGGGCAGATTGCCGTACCAGAAAGCAGTACAA    1620
6261                          GAAAAGCAACAAAATATCGATACTTTGTTGGGAAGGTTACCGTATCAGAAAGCAGTGCAa    683
                              ********.**.***** ********.*****.**.**.***** ***********.**

SGD_Scer_INP53/YOR109W        CTTTATGATCCCGTAAACGAATACGTAAGTACGAAATTACAAAGCATGTCTGATAAGTTC    1680
WashU_Sbay_Contig620.20       TTATATGATCCTGTGAACGAATATGTCAGCACTAAGCTACAAAGCATGTCTGATAAATTT    1680
6261                          CTTTATGATCCCGTAnACGAATACGTAAGTACGAnATTACAAAGCATGTCTGATAAGTTC    743
                               *:******** **. ******* **.** ** * . ******************.**

SGD_Scer_INP53/YOR109W        ACATCAACCTCCAACATTAACTTGTTAATAGGATCATTCAATGTTAATGGCGCAACCAAA    1740
WashU_Sbay_Contig620.20       ACCTCCTCCTCTAACATCAGCTTGCTGATAGGTTCGTACAATGTTAATGGGACCACCAAA    1740
6261                          ACATCAACCTCCAACATTAACTTGTTAGTAGGATCATTCAATGTTAATGGAGcnnCCAAg    803
                              **.**.:**** ***** *.**** *..****:**.*:************ .   ****

SGD_Scer_INP53/YOR109W        AAAGTTGATTTATCAAAGTGGTTATTTCCAATCGGTGAAAAATTTAAACCCGATATTGTT    1800
WashU_Sbay_Contig620.20       AAAGCAGATTTATCAAAGTGGTTATTCCCCATAGGCGAGAAATTCAAACCAGACGTCGTC    1800
6261                          nnaGTTGATTTATCnnaGTGgttATTTCCAAT----------------------------    835
                                 * :********   ***   *** **.**
```

## RGA1/YOR127w, chrXV:561420-561424

```
SGD_Scer_RGA1/YOR127W         ATGGCATCAACTGCTCCCAATGAACAATTTCCATCCTGCGTACGATGCAAAGAATTTATT    60
MIT_Sbay_c649_22742           ATGGCGTCAACCTTGCCCAACGAACAATTTCCCTCGTGTGTGCGGTGCAAAGATTCTATA    60
6261_RGA1                     ---------------------------------------------------------ATT    3
                                                                                       **:

SGD_Scer_RGA1/YOR127W         ACCACGGGGCATGCATATGAGTTGGGTTGTGATAGATGGCACACACATTGTTTCGCTTGT    120
MIT_Sbay_c649_22742           ACCACTGGGCACGCGTATGAGCTGGGTTGTGATAGATGGCACACGCATTGTTTTTCATGC    120
6261_RGA1                     ACCACGGGGCATGCATaTGAGTTGGGTTGTGATAGATGGCACAcacATTGTTTCGCTTGT    63
                              ***** ***** **.* **** ********************   *******  *:**

SGD_Scer_RGA1/YOR127W         TACAAATGTGAGAAACCATTAAGCTGCGAATCTGATTTTTTAGTCCTTGGAACAGGTGCT    180
MIT_Sbay_c649_22742           TATAAATGTGAGAAACCATTAAGTTGTGAATCTGATTTTCTAGTGCTTGGAACCGGTGCC    180
6261_RGA1                     TACAAATGTGAGAAACCATTAAGCTGCGAATCTGAtTTTTTAGTCCTTGGAACAGGTGCT    123
                              ** ******************** ** ******** *** **** ********.*****

SGD_Scer_RGA1/YOR127W         TTGATCTGCTTTGATTGTTCCGATTCTTGTAAAAAATTGCGGTAAAAAGATTGATGATTTG    240
MIT_Sbay_c649_22742           TTGATTTGTTTCGATTGTTCCGATTCATGCAAAAAATTGTGGTAAGAAAATTGATGATCTG    240
6261_RGA1                     TTGATCTGCTTTGATTGTTCCGATTCTTGTAAAAAATTGCGGTAAAAAGATTGATGATTTG    183
                              ***** ** ** **************:** ******** *****.**.********* **

SGD_Scer_RGA1/YOR127W         GCCATAATACTGTCCTCTTCAAATGAGGCCTATTGTTCAGATTGTTTTAAATGCTGTAAA    300
MIT_Sbay_c649_22742           GCCATTATATTGTCTTCGTCAAATGAAGCTTATTGTTCAGACTGTTTCAAATGTTGTAAG    300
6261_RGA1                     GCCATAATACTGTCTTCGTCAAATGAAGCTTATTGTTCAGATTGTTTCAAATGTTGCAAG    243
                              *****:*** **** ** ********.** *********** ***** ***** ** **.

SGD_Scer_RGA1/YOR127W         TGTGGTGAAAATATTGCTGACCTACGGTACGCGAAAACCAAGCGAGGTTTATTCTGTTTA    360
MIT_Sbay_c649_22742           TGTGGTGATAATATTGCTGATTTGCGGTACGCAAAGACCAAGCGGGGCCTATTCTGCTTA    360
6261_RGA1                     TGTGGTGATAATATTGCTGATTTGCGATACGCAAAGACCAAGCGGGGCTTATTCTGTTTA    303
                              ********:***********  *.**.*****.**.********.**  ******* ***

SGD_Scer_RGA1/YOR127W         AGCTGTCACGAAAAGCTATTAGCCAAACGAAAATACTACGAAGAGAAGAAAAGGCGACTC    420
MIT_Sbay_c649_22742           AATTGCCACGAGAAGTTATTGGCCAAAAGAAAATATTACGAGGAGAAAAAAAGACGACTT    420
6261_RGA1                     AATTGCCACGAGAAGCTATTAGCCAAAAGGAAATATTACGAGGAGAAAAAAAGACGACTT    363
                              *. ** *****.*** ****.******.*.***** *****.*****.*****.*****

SGD_Scer_RGA1/YOR127W         AAAAAGAATTTACCAAGTTTACCCACACCTGTGATTGATAATGGCCATACTGATGAGGTC    480
MIT_Sbay_c649_22742           AAAAAAAATTTGCCCAGTCTTCCTACTCCCGTACTTGACAGTGATTCTATCGATGCTACT    480
6261_RGA1                     aAAAAAAATTTGCCCAGTCTTCCTACTCCCGTGCTTGACAATGATTCTATTGATGTAACT    423
                               ****.*****.**.*** *:** **:** **..**** *.**.  .**  ****  .

SGD_Scer_RGA1/YOR127W         TCAGCTTCTGCAGTCCTCCCAGAAAAAACATTTAGCAGACCTGCATCACTAGTTAATGAG    540
MIT_Sbay_c649_22742           TCAATTACAGCAGTTGCACCTCAGAGGTCATCTAGTAGACCTGCCTCGCCGGTTAAGAAG    540
6261_RGA1                     TCAATTACGGCAGTTGCACCCAAAAGGTCATCTAGTAGACCTGTATCACCGGTTAAGAAA    483
                              ***. *:* *****   .**  *.*..:*** *** ******* .**.* .***** .*.

SGD_Scer_RGA1/YOR127W         ATTCCTTCAGGTTCTGAACCTTCC------AAGGACATAGAAACCAATTCGAGTGATATT    594
MIT_Sbay_c649_22742           ATATCCTTAGAGTCCGAATCTTTCAAGGATATAGCAATAGAAACCAATTCGAGTGATATT    600
```

```
6261_RGA1                    ATATCCTTAGAGTCCGAATCTATGAAAGATATAGCAATAGAAACCAACTCGAGCGATATC       543
                             **: * * **. ** *** **:        *:.*..*********** ***** *****

SGD_Scer_RGA1/YOR127W        GTTCCGCATTTTATCACTGGGTATAACGATAGCGATGACAACTCTGGAAGTTCAAAATTC       654
MIT_Sbay_c649_22742          ATTCCGCATTTCATCACTGGATATGATGATAGTGATGATAATTCTGGAAGTTCAAAATTC       660
6261_RGA1                    ATTCCGCATTTCATCACTGGGTATGATGATAGCGACGATAATTCTGGAAGTTCGAAATTC       603
                             .********** ********.***.* ***** ** ** ** ***********.******

SGD_Scer_RGA1/YOR127W        GGTTCAAATGTGTCCATAGATGTTATAGGACCGGAAGAAAATAGCACGGAGCATGTAAAT       714
MIT_Sbay_c649_22742          GGTTCTAATATATCTATAGATATTCTAGAAACACAGCAAGATAGTTCGGAGCATGCAAAA       720
6261_RGA1                    GGTTCTAATATTTCAATAGACATTATAGAACCACAGCAAGATAGCGCGGAGCATGCAAAA       663
                             *****:***.* ** ***** .**.***.*.*. *. **.****  ********* ***:

SGD_Scer_RGA1/YOR127W        GATGATGTTAAAGAGGAAGCAGAAGCACCTTCAGCGAATATGTCACTCAATGTTGCTACG       774
MIT_Sbay_c649_22742          GATGAAAAAGTCGAACAGGTGAAGGTACATTCGAGAAATGTGTCGCTTGACGTGGCGTTG       780
6261_RGA1                    GATGAGAAAGTAGAAGAGGTGAAGGTGCATTCTAGAAACGCATCGCTTGACATAGCATTG       723
                             ***** .::.:.**. *.* ..*.* .*.*** . .** . .**.** .* .* ** : *

SGD_Scer_RGA1/YOR127W        GATCCAACCC-------------------TAAGTTGTAAAGAACCTCCTAGCCATTCGAGG       816
MIT_Sbay_c649_22742          GATTCTACTCCAAATCGTAAGGTCTTGTTGGGGGACACGGAACCCCCGAATCGCTCTAAG       840
6261_RGA1                    GATGCTACTCCAAGTCATAAGGTCTTGTTGGGCAACACGGAACctCCGAGCCGCTCTAAG       783
                             *** *:** *                 *..*  . *..****  ** *. *. ** *.*

SGD_Scer_RGA1/YOR127W        AATTTGTTAAATAAAACACCGTTGAGAAATTCTTCGGGTCAGTATCTCGCAAAATCTCCA       876
MIT_Sbay_c649_22742          AATCTATTAAATAAAACCCCATTGAGGAATTCAACCGGACAATATGTCGCAAAATCGCCA       900
6261_RGA1                    ATTTTATTAAACAAAACACCATTAAGGAATTCATCCGGACAATATGTCGCAAAATCTCCA       843
                             *:* *.***** *****.**.**.**.*****::* **:**.*** ********** ***

SGD_Scer_RGA1/YOR127W        AGCTCCTATAGACAGGGAATAATTGTTAACGATAGTCTGGAAGAGAGCGATCAAATTGAT       936
MIT_Sbay_c649_22742          AGTTCTTACAGACAGGGAATAGTTGTTAATGATAGCTTTGAGGAAAATAACCAGCTTGAT       960
6261_RGA1                    AGTTCTTATAGACAGGGTATAGTTGTTAATGATAGCTT----------------------       881
                             ** ** ** ********:***.******* *****  *
```

**HSP82/YPL240c, chrXVI:97019-97037**

```
SGD_Scer_HSP82/YPL240C       TGGTTATCTTTCGTCAAGGGTGTTGTTGACTCTGAGGATTTACCATTGAATTTGTCCAGA       1140
MIT_Sbay_c60_24336           ---ATGTCATTCGTCAAGGGTGTTGTTGACTCCGAAGATTTACCATTGAACTTGTCCAGA       57
6261_HSP82                   TGGATGTCCTTTGTCAAGGGTGTTGTTGACTCTGAAGACTTACCATTGAACCTGTCCAGA       123
                                :*.** ** ******************** **.** ***********  ********

SGD_Scer_HSP82/YPL240C       GAAATGTTACAACAAAATAAGATCATGAAGGTTATTAGAAAGAACATTGTCAAAAAGTTG       1200
MIT_Sbay_c60_24336           GAAATGTTGCAACAAAACAAGATCATGAAGGTCATCAGAAAGAACATCGTTAAGAAGATG       117
6261_HSP82                   GAAATGCTACAACAAAACAAGATCATGAAGGTTATCAGAAAGAACATCGTCAAGAAGGTA       183
                             ****** *.******** ************** ** *********** ** **.*** *.

SGD_Scer_HSP82/YPL240C       ATTGAAGCCTTCAACGAAATTGCTGAAGACTCTGAACAATTTGAAAAGTTCTACTCGGCT       1260
MIT_Sbay_c60_24336           ATTGAATCCTTCAACGAAATCGCTGAAGACTCTGAACAATTCGAAAAGTTCTACTCTGCC       177
6261_HSP82                   ATTGAAGCTTTCAACGAAATTGCTGAAGACTCTGAACAATTCGAAAAGTTCTACTCTGCC       243
                             ****** * *********** ******************** ************** **

SGD_Scer_HSP82/YPL240C       TTCTCCAAAAATATCAAGTTGGGTGTACATGAAGATACCCAAAACAGGGCTGCTTTGGCT       1320
MIT_Sbay_c60_24336           TTCGCTAAGAACATCAAATTGGGTGTTCATGAAGACACTCAAAACAGAGCTGCCTTGGCC       237
6261_HSP82                   TTCGCTAAGAACATCAAATTGGGTGTCCACGAAGACACTCAAAACAGAGCTGCCTTGGCT       303
                             *** * **.** *****.******** ** ***** ** ******** *****.*****

SGD_Scer_HSP82/YPL240C       AAGTTGTTACGTTACAACTCTACCAAGTCCGTAGATGAGTTGACTTCCTTAACTGATTAC       1380
MIT_Sbay_c60_24336           AAATTGCTACGTTACAACTCCACCAAGTCCGTTGACGAATTGACTTCTTTGACTGATTAT       297
6261_HSP82                   AAGTTACTACGTTACAACTCCACCAAGTCCGTCGACGAATTAACTTCTTTGACCGATTAC       363
                             **.**. ************* *********** ** ** **.**.***** **.** *****

SGD_Scer_HSP82/YPL240C       GTTACCAGAATGCCAGAACACCAAAAGAACATCTACTACATCACTGGTGAATCTCTAAAG       1440
MIT_Sbay_c60_24336           ATTACCAGAATGCCAGAACACCAAAAGAACATCTACTATATCACAGGTGAGTCTTTGAAG       357
6261_HSP82                   ATCACCAGAATGCCAGAACACCAAAAGAACATTTACTACATCACAGGTGAATCTCTAAAG       423
                             .* ***************************** ***** *****:*****.*** *.***

SGD_Scer_HSP82/YPL240C       GCTGTCGAAAAGTCTCCATTTTTGGATGCCTTGAAGGCTAAAAACTTCGAGGTTTTGTTC       1500
MIT_Sbay_c60_24336           GCTGTTGAAAAAATCCCCATTCTTAGACGCTTTGAAAGCTAAGAACTTTGAAGTTTTGTTC       417
6261_HSP82                   GCCGTTGAAAAAATCCCCATTCTTGGACGCTTTGAAGGCTAAGAACTTTGAAGTTTTGTTC       483
                             ** ** *****.** ***** **.** ** *****.*****.*****.** **.*********

SGD_Scer_HSP82/YPL240C       TTGACCGACCCAATTGATGAATACGCCTTCACTCAATTGAAGGAATTCGAAGGTAAAACT       1560
MIT_Sbay_c60_24336           TTGACTGATCCAATCGATGAATACGCCTTCACTCAATTAAAGGAATTCGAAGGTAAGACT       477
6261_HSP82                   TTGACTGATCCAATCGATGAATACGCTTTCACCCAATTGAAGGAATTCGAAGGTAAGACT       543
                             ***** ** ***** *********** ***** *****.*****************.***

SGD_Scer_HSP82/YPL240C       TTGGTTGACATTACTAAAGATTTCGAATTGGAAGAAACTGACGAAGAAAAAGCTGAAAGA       1620
```

```
MIT_Sbay_c60_24336          TTAGTCGATATCACCAAGGATTTCGAGCTGGAAGAAACTGACGAAGAGAAAGCTGAAAGA      537
6261_HSP82                  TTGGTCGATATCACCAAGGATTTCGAACTGGAAGAAACTGACGAAGAAAAAGCTGAAAGA      603
                            **.** ** ** ** **.********. *******************.************

SGD_Scer_HSP82/YPL240C      GAGAAGGAGATCAAAGAATATGAACCATTGACCAAGGCCTTGAAAGAAATTTTGGGTGAC      1680
MIT_Sbay_c60_24336          GAAAAGGAAGTTAAAGAATTCGAACCATTGACCAAGGCCTTGAAAGACATCTTGGGTGAA      597
6261_HSP82                  GAGAAGGAGATCAAAGAATATGAACCATTGACCAAGGCCTTGAAAGAAATTTTGGGTGAC      663
                            **.*****..* *******: *************************.** ********.

SGD_Scer_HSP82/YPL240C      CAAGTGGAGAAAGTTGTTGTTTCTTACAAATTGTTGGATGCCCCAGCTGCTATCAGAACT      1740
MIT_Sbay_c60_24336          CAAGTTGAAAAGGTTGTTGTCTCTTACAAACTAGTGGATGCCCCAGCTGCCATTAGAACT      657
6261_HSP82                  CAAGTGGAGAAAGTTGTTGTTTCTTACAAATTGTTGGATGCCCCAGCTGCTATCAGAACT      723
                            ***** **.**.******** ********* *. **************** ** ******

SGD_Scer_HSP82/YPL240C      GGTCAATTTGGTTGGTCTGCTAACATGGAAAGAATCATGAAGGCTCAAGCCTTGAGAGAC      1800
MIT_Sbay_c60_24336          GGCCAATTCGGTTGGTCCGCTAACATGGAAAGAATCATGAAGGCTCAAGCTTTGAGAGAC      717
6261_HSP82                  GGTCAATTTGGTTGGTCTGCTAACATGGAAAGAATCATGAAGGCTCAAGCCTTGAGAGAC      783
                            ** ***** ******** ******************************** *********

SGD_Scer_HSP82/YPL240C      TCTTCCATGTCCTCCTACATGTCTTCCAAGAAGACTTTCGAAATTTCTCCAAAATCTCCA      1860
MIT_Sbay_c60_24336          TCTTCCATGTCCTCTTACATGTCCTCCAAGAAGACTTTCGAAATCTCTCCAAAATCTCCA      777
6261_HSP82                  TCTTCCATGTCCTCCTACATGTCTTCCAAGAAGACTTTCGAAATTTCTCCAAAATCTCCA      843
                            ************** ******** ******************** ***************

SGD_Scer_HSP82/YPL240C      ATTATCAAGGAATTGAAAAAGAGAGTTGACGAAGGTGGTGCTCAAGACAAGACTGTCAAG      1920
MIT_Sbay_c60_24336          ATTATCAAGGAATTGAAAAAGAGAGTTGATGAAGGCGGTGCTCAAGATAAGACTGTCAAG      837
6261_HSP82                  ATTATCAAGGAATTGAAAAAGAGAGTTGACGAAGGT------------------------      879
                            ***************************** *****
```

## DBVPG 6257

### YDR338c, chrIV:1148746-1148747

```
SGD_Scer_YDR338C            GATGAGGAACCTGACCTTTACTATCATGATGAAGAAGATGGCGAACTGAGCAGAACGATT      300
MIT_Sbay_c446_5035          GAAGAAGAACCTGATTTATATTACCATGACGAGGAGGATGGTGAGCTCAGTAGAACCATT      300
6257_YDR338C                 ---------cCTGATTTATATTACCACGACGAGGAGGATGGCGAGCTTAGTAGAACCATT     51
                                 ****  *:** ** ** ** **.**.***** **.** ** ***** ***

SGD_Scer_YDR338C            TCTCTACCATCAAGGGTATCAGAAACTCCTGAGCTTTCGCCACAAGATGTTGATTGGATT      360
MIT_Sbay_c446_5035          TCTTTACCTTCGAGGGTGTCAGAGACTCCTGACCTTTCTCCGCAGGACGTTGACTGGATT      360
6257_YDR338C                TCTTTACCTTCAAGAGTGTCAGAGACTCCTGAACTTTCTCCACAAGACGTTGACTGGATA      111
                            *** ****:**.**.**.*****.******** ***** **.**.** ***** *****:

SGD_Scer_YDR338C            TACATGAACATGAACGACGATACTCATCGGTATGCAACTCTGATAATGAGGAAGCAAGT       420
MIT_Sbay_c446_5035          CTACATGAACATGAACGACGATACTCATCAGTGTATAACTCTGATAATGAGGAGGAATGT      420
6257_YDR338C                CTACATGAACATGAACGACGATACTCGTCAGTGTATAACTCTGATAATGAGGAGGAGTGT      171
                            *************************.**.**.*. *****************.*..:**

SGD_Scer_YDR338C            CAAAGCAACACACCAGATAGAATACAAGAATATTCTGGAAGAGAACTAGAATATGACGAA      480
MIT_Sbay_c446_5035          CAAAGCAACACACCCGATAGAATACAAGAATATCCCGGAAGAGAACTGGAATATGACGAA      480
6257_YDR338C                CAAAGCAACATACCCGATAGAACACAAGGATATCCCGGAAGAGAACTGGAATATGACGAA      231
                            ********** ***.******* *****.**** * ***********.************

SGD_Scer_YDR338C            TTTATGAATAGACTCCAAGCTCAGAAACAGAAATTAACTCGAAGTGCGGTAACAGACGCT      540
MIT_Sbay_c446_5035          TTCATGAACAGACTTCAAGCCCAGAAACAGAAACAAAACCAAAATGTAACAATGGAAGCC      540
6257_YDR338C                TTTATGAACAGGCTTCAAGCCCAGAAACAGAAACAAAACCAAAATGCAACAATGGATGCA      291
                            ** ***** **.** ***** ************ :**. *.**.** .. ** .** **

SGD_Scer_YDR338C            AAAGGCACCTCACATCACAGAAGACCATCATTCGTATCTGTGACTAGCCGAGGCTCAGTT      600
MIT_Sbay_c446_5035          AAAGGTATCTCGCATCGCAGAAGGCCATCATTTGTATCTGTAAACAGTCGAGGTTCGGTA      600
6257_YDR338C                AAAGGTATCTCACATCGTAGAAGGCCATCATTCGTATCTGTAAACAGTCGAGGTTCGGTG      351
                            ***** * ***.****. *****.******** ********.*. ** ***** **.**

SGD_Scer_YDR338C            CCCACAATTTACCAAGAGATCGACGAGAACGATTCTGAGGCACTAGCAGAATTGGCTCAC      660
MIT_Sbay_c446_5035          CCCACAATCTACCAGGACCTAGATGAGAATGATTCAGAAGCGCTAGCTGCATTGGCTCAT      660
6257_YDR338C                CCCACGATATATCAAGACCTAGATGAGAATGATCCAGAAGCACTAGCTGCATTGGCTCAT      411
                            *****.** ** **.** .*.** ***** *** *:**.**.*****:*.*********

SGD_Scer_YDR338C            AGCCACGTCACCTTCAAATCAGAAGCAAGAGTTTTGGCATCTTATTCCTTCCCTCTAATT      720
MIT_Sbay_c446_5035           AGTAATGTGACCTTCAAATCAGAAACAAAAGTCTTAGCATCCTACTCTTTTCCCTTGATT     720
6257_YDR338C                AGTAATGTGACCTTCAAATCAGAAGCAAAAGTATTAGCATCTTACTCTTTCCCTTTAATT      471
                            ** .* ** ***************.***.*** **.***** ** ** ** **  *.***

SGD_Scer_YDR338C            TTCACATTCTTATTGGAACAGATTTTCCCTATGGTATGTTCATTAACTGTGGGCCACTTA      780
MIT_Sbay_c446_5035          TTTACATTTTTACTGGAGCAAATATTCCCTATGGTCTGTTCATTAACCGTGGGACATCTA      780
6257_YDR338C                TTCACATTCTTATTGGAACagaTTTTCCCTATGGTATGTTCATTAACTGTGGGCCActtA      531
```

```
                              ** ***** *** ****.*    *:***********.*********** *****.**   *

SGD_Scer_YDR338C              GGCAAAAATGAACTAGCAGCTGTGTCCTTAGCATCCATGACTTCTAATATAACACTAGCG      840
MIT_Sbay_c446_5035            GGTAAGAACGAGCTAGCTGCTGTATCGTTAGCTTCTATGACTTCCAATATCACATTAGCC      840
6257_YDR338C                  GgCAAAAATGAACTAGcagCTGTGTCCTTAGcatCCATGACTTCTAATATAaCactAGcg      591
                              *  **.** **.****    ****.** ****    * ******** *****. *   **

SGD_Scer_YDR338C              ATATTCGAGGGTATTGCCACTAGTCTAGATACTCTATGCCCTCAAGCGTATGGCTCTGGG      900
MIT_Sbay_c446_5035            ATATTTGAAGGTATTGCCACTAGTCTCGATACTCTTTGTCCTCAAGCATACGGTTCTGGA      900
6257_YDR338C                  aTATTcgaggGTATTGCCACTAGTCTagATACTCTATGCCCTCAAGCGTATGGCTCTGgn      651
                              ****      ****************  *******:** ********.** ** ****

SGD_Scer_YDR338C              AGATTCTACAGTGTAGGAGTTCACCTTCAACGTTGCATTGCTTTTTCATTGGTCATATAT      960
MIT_Sbay_c446_5035            AGGTTCTACAGTGTCGGGGTTCATTTTCAACGTTGTATTGCTTTTTCAATGGTCATATAT      960
6257_YDR338C                  nnntTCTACAGTGtagnagTTCacctTCAACGTTGCATTGCTTTTtcnnnggTCATATAt      711
                             *********       ***     ********* *********       *******

SGD_Scer_YDR338C              ATACCTTTTGCTGTCATGTGGTGGTATTCTGAACCTCTTCTTTCTTATATCATCCCTGAG      1020
MIT_Sbay_c446_5035            ATTCCATTTGCGTTCTTATGGTGGTATTCCGAGCCTCTTCTTTCTTATATCATTCCTGAA      1020
6257_YDR338C                  ataccTTTTGCTGTCAtGt-----------------------------------------      730
                               :*****  **: .
```

## KEM1/YGL173c, chrVII:179656-179664

```
  SGD_Scer_XRN1/YGL173C       --ATGGGTATTCCAAAATTTTTCAGGTACATCTCAGAAAGATGGCCCATGATTTTACAGC      58
  MIT_Sbay_c403_8392          --ATGGGTATTCCAAAATTTTTCAGATACATCTCAGAGAGATGGCCTATGATTCTACAAC      58
  6257_KEM1                   GCATGGGTATTCCGAAATTTTTCAGGTACATCTCAGAAAGATGGCCCATGATTCTACAAC      300
                                ***********.***********.***********.******** ****** ****.*

  SGD_Scer_XRN1/YGL173C       TTATTGAGGGAACACAGATTCCTGAGTTTGATAACTTATACCTGGATATGAATTCGATTT      118
  MIT_Sbay_c403_8392          TTATTGAAGGGACTCAGATTCCCGAGTTTGATAACCTGTATCTGGACATGAATTCGATTT      118
  6257_KEM1                   TTATTGAAGGGACTCAGATTCCTGAGTTTGATAACCTATATCTGGATATGAATTCGATTT      360
                              *******.**.**:******** *********** *.** ***** *************

  SGD_Scer_XRN1/YGL173C       TACATAATTGTACGCATGGTAACGACGATGATGTAACCAAGCGATTAACTGAAGAAGAGG      178
  MIT_Sbay_c403_8392          TACATAACTGTACGCATGGTAACGACGATGACGTGACGAAGCGACTAACTGAAGAAGAGG      178
  6257_KEM1                   TACATAACTGTACACATGGTAACGACGATGACGTGACGAAGCGATTAACTGAAGAAGAGG      420
                              ******* *****.***************** **.** ****** ***************

  SGD_Scer_XRN1/YGL173C       TTTTTGCAAAAATCTGTACGTATATCGATCACCTTTTTCAAACAATCAAGCCCAAGAAGA      238
  MIT_Sbay_c403_8392          TTTTTGCCAAGATCTGCACGTATATCGACCATCTTTTTCAAACCATCAAGCCCAAGCAAA      238
  6257_KEM1                   TTTTTGCAAAGATCTGCACGTATATTGATCATCTTTTTCAAACTATCAAGCCCAAGCAAA      480
                              *******.**.***** ******** ** ** *********** ************.*.*

  SGD_Scer_XRN1/YGL173C       TTTTCTACATGGCTATTGATGGTGTGGCCCCTCGTGCCAAGATGAATCAACAAAGAGCTC      298
  MIT_Sbay_c403_8392          TTTTTTACATGGCTATCGATGGTGTGGCTCCCCGTGCGAAGATGAATCAACAGAGAGCTC      298
  6257_KEM1                   TTTTTTACATGGCTATCGATGGTGTGGCCCCCCGCGCGAAGATGAATCAACAGAGAGCTC      540
                              **** *********** *********** ** ** ** ***************.*******

  SGD_Scer_XRN1/YGL173C       GTAGATTCAGAACCGCTATGGATGCAGAAAAAGCCTTGAAGAAGGCTATTGAGAATGGTG      358
  MIT_Sbay_c403_8392          GTAGGTTTAGAACTGCCATGGACGCCGAAAAAGCTATGAAAAAAGCTATTGAAAATGGTG      358
  6257_KEM1                   GTAGATTTAGAACTGCCATGGACGCCGAAAAAGCTATGAAAAAAGCTATTGAAAATGGTG      600
                              ****.** ***** ** ***** **.******** :****.**.********.*******

  SGD_Scer_XRN1/YGL173C       ACGAGATTCCTAAGGGTGAGCCATTTGATTCGAATTCTATTACTCCAGGTACGGAGTTTA      418
  MIT_Sbay_c403_8392          ACGAGATTCCTAAAGGTGAACCGTTCGACTCAAATTGTATTACTCCAGGTACTGAATTCA      418
  6257_KEM1                   ACGAAATTCCTAAAGGTGAGCCGTTCGACTCAAACTGTATTACTCCAGGTACTGAATTTA      660
                              ****.********.*****.**.** ** **.** * *************** **.** *

  SGD_Scer_XRN1/YGL173C       TGGCCAAATTGACCAAAAACTTACAATATTTTATTCACGACAAGATTTCTAACGATTCCA      478
  MIT_Sbay_c403_8392          TGGCCAAACTGACAAAAAAATTTACAGTATTTCATTCATGACAAGATCTCTAACGATTCTA      478
  6257_KEM1                   TGGCTAAACTGACAAAAAAATTTACAGTATTTCATTCATGACAAAATTTCTAACGATTCCA      720
                              **** *** **** ****.***** *****.***** ***** *****.** *********** *

  SGD_Scer_XRN1/YGL173C       AATGGAGGGAAGTGCAAATCATATTTTCTGGCCATGAAGTTCCAGGTGAAGGTGAACACA      538
  MIT_Sbay_c403_8392          AGTGGAGAGAAGTACAAATTATATTTTCTGGCCATGAAGTTCCAGGTGAAGGTGAGCACA      538
  6257_KEM1                   AATGGAGGGAAGTGCAAATCATATTTTCTGGCCATGAAGTTCCAGGTGAAGGTGAACACA      780
                              *.*****.*****.*****.***** ***********************************.****

  SGD_Scer_XRN1/YGL173C       AGATCATGAACTTTATAAGGCATTTAAAATCCCAAAAGGATTTCAACCAGAATACGAGAC      598
  MIT_Sbay_c403_8392          AAATCATGAATTTCATAAGGCATTTAAAATCACAGAAGGATTTAACCAAAATACGAGAC      598
  6257_KEM1                   AGATCATGAACTTTATAaGgCaTTTAAAATCCCAAAAGGATTTCAACCAGAATACGAGAC      840
                              *.******** ** *** * * *********.**.********* *****.**********
```

```
SGD_Scer_XRN1/YGL173C     ATTGTATTTACGGTCTTGACGCAGATTTGATTATGCTGGGTTTGTCTACTCATGGGCCAC      658
MIT_Sbay_c403_8392        ATTGTATTTATGGTCTTGATGCGGATTTGATCATGCTAGGTTTATCTACTCACGGGCCTC      658
6257_KEM1                 A-----------------------------------------------------------      841
                          *
```

## IRR1/YIL026c, chrIX:306349-306368

```
SGD_Scer_IRR1/YIL026C     TATCTTTTCCAGGACTATTTGACCCAACAGGCTGTTAATTTAGAAAAGAATTATTTGGCG      954
MIT_Sbay_c367_11417       TTTCTTTTCCAAGACTACTTGACCGAACAAGCTGTGAATTTGGAAAAGAATTACCTGGCC      945
6257_IRR1                 -------------------------------------------------------TAGCC      5
                                                                                 *.**

SGD_Scer_IRR1/YIL026C     AAGTTATCCAAGCAGTTATCACTAGAAGAGAAGAAAAAAAGGCCCAATAACAAGACTTTA      1014
MIT_Sbay_c367_11417       AAATTGACTAGACAATTGTCGTTAGAAGAAAAGAAAAAAAGGCCCAATAAGAAAACTGTA      1005
6257_IRR1                 AAATTGACTAGACAGTTtGTCACTAGAagAAAagAAAAAAAggCCCAatAAGAAAACTTTA     65
                          **.**.:* *..**.* .**. ****  *.*  *******  ****  ** **.*** **

SGD_Scer_IRR1/YIL026C     GAAAAGCTAGAGAGCACTATCGCCGAAACTCAAGGCAGCAAAGTCGTTATTGATAGCATT      1074
MIT_Sbay_c367_11417       GAAAAGCTAGAAAGTACTATTGCTGAAACCCAAGGTAGTAAAGTCGTCATTGAAGGTGTT      1065
6257_IRR1                 GAAAAGCtcGAaAGCACcATTGCTGAAACtCAAGGTAGTAAAGTTGTCATTGAAGGTGTT      125
                          *******  ** ** ** ** ** ***** ***** ** ***** ** *****:.* .**

SGD_Scer_IRR1/YIL026C     ATCGATAACATCGTCAAACTATGTTTTGTGCATAGGTATAAGGACGTGTCTGATTTGATT      1134
MIT_Sbay_c367_11417       ATTGACAATATTGTGAAGTTGTGTTTCGTTCACAGGTATAAGGACATCTCTGACTCAATT      1125
6257_IRR1                 ATTGATAATAttgTGAAGTTGTGTTTCGTGCACAGGTACAagGACAtntCCGATTCAATT      185
                          ** ** ** *   * **. *.***** ** ** ***** *  ***.   * ** * .***

SGD_Scer_IRR1/YIL026C     CGTTCAGAATCCATGTTGCATCTATCCATCTGGATTAAAAACTATCCCGAATATTTCCTC      1194
MIT_Sbay_c367_11417       CGTTCCGAGGCCATGCTGCACTTATCAATCTGGATTAAAAACTATCCAGAATATTTCCTC      1185
6257_IRR1                 CGTTCTGAGTCCATGTTGCACTtaTCGATCTGGATTAAaAACTATCCAGAATATTTCCTC      245
                          ***** **. ***** ****    ** *********** ********.************

SGD_Scer_IRR1/YIL026C     AAGGTTACATTTTTAAAATATTTTGGCTGGCTGCTGAGCGATAACTCTGTATCAGTCAGA      1254
MIT_Sbay_c367_11417       AAGGTCACATTTCTGAAATATTTTGGCTGGCTACTCAGCGATAACTCCGTATCTGTCAGG      1245
6257_IRR1                 AAGGTCACATnTnTGAAATATTTTGGCTGGCTACTCAGCGATAATTCCGTCTCTGTCAGG      305
                          ***** **** * *.*****************.** ******** ** **.**:*****.

SGD_Scer_IRR1/YIL026C     TTACAAGTTACGAAGATTTTACCGCATTTAATAATCCAAAATCATAACAGTAAATCCACT      1314
MIT_Sbay_c367_11417       TTACAAGTTGCCAAGATTCTTCCACATTTAATCGTCCAAAATCATAACAGCAAATCCAGT      1305
6257_IRR1                 TTACAAGTTGCCAAGATTCTTCCACATTTGATCATCCAAAATCATAACAGCAAATCCAGT      365
                          *********.* ****** *:**.*****.**..**************** ******* *

SGD_Scer_IRR1/YIL026C     GATAATTCCGCTATACGCCAAGTATTTGAACGCTTCAAAACTAAGATCCTGGAAGTGGCA      1374
MIT_Sbay_c367_11417       GATAATTCCGCCATCCGCCAAGTCTTTGAACGGTTTAAATCTAAGATCCTGGAGGTCGCC      1365
6257_IRR1                 GATAATTCCGCGATCCGCCAAGTCTTTGAAAGGTTTAAATCTAAGATCCTGGAGGTCGCC      425
                          *********** **.********.******.* ** ***:*************.** **.

SGD_Scer_IRR1/YIL026C     ATCCGTGACGTTAATCTTGATGTAAGGATCCATAGTATCCAAGTTCTAACGGAGGCGTCA      1434
MIT_Sbay_c367_11417       ATCCATGACGTTAATCTTGATGTCAGAATTCACAGTATTCAAATTCTAACCGAAGCGTCA      1425
6257_IRR1                 ATCCATGACGTCAATCTTGATGTCAGAATTCATAGTGTTCAAATTCTAACTGAAGCGTCA      485
                          ****.****** ***********.**.** ** ***.* ***.*******  **.******

SGD_Scer_IRR1/YIL026C     TCATTGGGCTATTTAGATGATTCTGAGATTCTTATCATTTCTAGTTTAATGTTCGATGAA      1494
MIT_Sbay_c367_11417       TCATTGGGCTATTTAGATGATTTTGAGATTCTAATGATTTCTAGCTTAATGTTTGATGAA      1485
6257_IRR1                 TCACTGGGCTATTTGGATGATTTTGAGATTCTAATGATTTCTAGCTTAATGTTTGATGAA      545
                          *** **********.******* *********:** ******** ******** ******

SGD_Scer_IRR1/YIL026C     GAGTTCGACCCATTTAAAACATCCTCATTCAATAAAAGATCCAAATTTCTATCCACGGTG      1554
MIT_Sbay_c367_11417       GAATTCGACCCATTCAAAACATCCTCATTTAATAAAAGGTCCAAATTTTTATCTACTGTG      1545
6257_IRR1                 GAATTCGACCCTTTCAAAACATCCTCATTCAATAAAAGGTCCAAATTTTTATCTACTGTA      605
                          **.********:** ************** ********.********* **** ** **.

SGD_Scer_IRR1/YIL026C     GCCAAATTCTTAGCAAGAGTAATAAAAGAGAAATTTGACGAATTCATCAAGACGCATGAA      1614
MIT_Sbay_c367_11417       GCCAAGTTCTTAGCAAGAGTAATAACCGAAAAAATATGAGGAATTCACCAAGACTCATGAA      1605
6257_IRR1                 GCCAAGTTCTTAGCAAGAGTAATAAAAGAGAAATTTGACGAATTCATCAAGACGCATGAA      665
                          *****.*******************..**.****:*** ******* ****** ******

SGD_Scer_IRR1/YIL026C     GACTTGCCCAAAGAAGTCGACGGGTTAGAAGTTGGACCCGTTGTTCAAGTCGGTATATTT      1674
MIT_Sbay_c367_11417       GAGTTGCCTGAAGAGGTCGATGGATTAGCTGTTGCTCCAGTCGTGCAAGTTGGGATCTTT      1665
6257_IRR1                 GACTTGCCCAAAGAAGTCGACGGGTTAGAAGTTGGACCCGTTGTTCAAGTCGGTATATTT      725
                          ** ***** .****.***** ** ****.:**** :**.** ** ***** ** **.***

SGD_Scer_IRR1/YIL026C     ATCAAGATTCTGAATGACTCTTTAATTTATCACTTGAAGGATTGCGCTGAAGTTGATTCA      1734
MIT_Sbay_c367_11417       ATCAAGATCTTGAGCAACTCCTTGATTTATCATCTAAAGGATTCCGCGGAAGTTGATTCA      1725
6257_IRR1                 ATCAAGATTCTGAATGACTCTTTAATTTATCACTTGAAGGATTGCGCTGAAGTTGATTCA      785
                          ********  ***. .**** **.********  *.******* *** ************
```

```
SGD_Scer_IRR1/YIL026C      AGGACAAAGATCCGTATGCTAACACAAGCAGCAGAGTTTTTGTCTCCTTACATTTCCACT    1794
MIT_Sbay_c367_11417        AAAACTAAGATTCGTATGCTAACGCAGGCCGCAGAGTTTTTATCTCCATACATTTCATCG    1785
6257_IRR1                  AGGACAAAGATCCGTATGCTAACACAAGCAGCAGAGTTTTTGTCTCCTTACATTTCCACT    845
                           *..**:***** ***********.**.**.***********.*****:********.:*

SGD_Scer_IRR1/YIL026C      CACTTGAAAACTATTTGCAATCTGCTGATCTCTGACACTGAATCAAATGAACTGATCCAA    1854
MIT_Sbay_c367_11417        CATTTGAAAACCATCTGCGATCTTCTGATCTCGGATACTGAATCGAATGAATTAATCCAG    1845
6257_IRR1                  CACTTGAAAACTATTTGCAATCTGCTGATCTCTGACACTGAATCAAATGAACTGATCCAA    905
                           ** ******** ** ***.**** ******** ** ********.****** *.*****.

SGD_Scer_IRR1/YIL026C      AAGTTACAAAACTCGGCCAATAATAACAGCGATGACGAGGATGTTGACGATGAGGAATTG    1914
MIT_Sbay_c367_11417        ACACTACAAAACTCGACTGACAATAACAACGATGACGACGACGAAGACGGTCAAGAACTA    1905
6257_IRR1                  AAGTTACAAAACTCGGCCAATAATAACAGCGATGACGATGATGTTGACGATGAGGAA---    962
                           *.. ***********.* .* *******.********* ** *::****.* *.***
```

## TDH2/YJR009c, chrX:453941-453961

```
SGD_Scer_TDH2/YJR009C      AAGGAATTGGACACTGCTCAAAAGCACATTGACGCTGGTGCCAAGAAGGTTGTCATCACT    360
WashU_Sbay_Contig534.6     AAGGAATTGGACACTGCTCAAAAGCACATTGACGCTGGTGCCAAGAAGGTTGTCATCACT    312
6257_TDH2                  -----------------------GCACATTGACGCTGgtGCCAAGAAGGTTGTCATCACT    37
                                                  **************  *********************

SGD_Scer_TDH2/YJR009C      GCTCCATCTTCCACCGCCCCAATGTTCGTCATGGGTGTTAACGAAGAAAAATACACTTCT    420
WashU_Sbay_Contig534.6     GCTCCATCTTCCACCGCCCCAATGTTCGTTATGGGTGTTAACGAAGACAAATACACTTCT    372
6257_TDH2                  GCTCCATCTTCCACCGccCCAATGTTCGTTATGgGTGTTAACGAAGACAAATACACTTCT    97
                           ****************  *********** *** *************.************

SGD_Scer_TDH2/YJR009C      GACTTGAAGATTGTTTCCAACGCTTCTTGTACCACCAACTGTTTGGCTCCATTGGCCAAG    480
WashU_Sbay_Contig534.6     GACTTGAAGATTGTCTCCAACGCTTCCTGTACCACTAACTGTTTGGCTCCATTGGCCAAG    432
6257_TDH2                  GACTTGAAGATTGtCTCCAACGCTTCCTGTACCACTAACTGTTtgGCTCCATTGGCCAAG    157
                           *************  *********** ******** *******  ***************

SGD_Scer_TDH2/YJR009C      GTTATCAACGATGCTTTCGGTATTGAAGAAGGTTTGATGACCACTGTTCACTCCATGACC    540
WashU_Sbay_Contig534.6     GTTATCAACGATGCTTTCGGTATTGAAGAAGGTTTGATGACCACTGTCCACTCCATGACC    492
6257_TDH2                  GTTATCAACGATGCTTTCGGTATTGAAGAAGGTTTGATGACCACTGTCCACTCCATGACC    217
                           *********************************************** ************

SGD_Scer_TDH2/YJR009C      GCCACCCAAAAGACTGTTGACGGTCCATCCCACAAGGACTGGAGAGGTGGTAGAACCGCT    600
WashU_Sbay_Contig534.6     GCCACTCAAAAGACTGTCGATGGTCCATCCCACAAGGACTGGAGAGGTGGTAGAACCGCT    552
6257_TDH2                  GCCACTCAAAAGACTGTCGATGGTCCATCCCACAAGGACTGGAGAGGTGGTAGAACCGCT    277
                           ***** *********** ** ***************************************

SGD_Scer_TDH2/YJR009C      TCCGGTAACATCATCCCATCCTCTACCGGTGCTGCTAAGGCTGTCGGTAAGGTCTTGCCA    660
WashU_Sbay_Contig534.6     TCCGGTAACATCATCCCATCCTCCACCGGTGCCGCCAAGGCTGTCGGTAAGGTCTTGCCA    612
6257_TDH2                  TCCGGTAACATCATCCCATCCTCCACCGGTGCCGCCAAGGCTGTCGGTAAGGTCTTGCCT    337
                           *********************** ******** ** ***********************:

SGD_Scer_TDH2/YJR009C      GAATTGCAAGGTAAGTTGACCGGTATGGCTTTCAGAGTCCCAACCGTCGATGTTTCCGTT    720
WashU_Sbay_Contig534.6     GAATTGCAAGGTAAGTTGACCGGTATGGCTTTCAGAGTCCCAACCGTCGATGTCTCCGTT    672
6257_TDH2                  GAATTACAAGGTAAGTTGACCGGTATGGCTTTCAGAGTCCCAACCGTCGATGTCTCCGTT    397
                           *****.*********************************************** ******

SGD_Scer_TDH2/YJR009C      GTTGACTTGACTGTCAAGTTGAACAAGGAAACCACCTACGATGAAATCAAGAAGGTTGTC    780
WashU_Sbay_Contig534.6     GTTGACTTGACTGTTAAGTTGAACAAGGAAACCACCTACGATGAAATCAAGAAGGTTGTC    732
6257_TDH2                  GTTGACTTGACTGTCAAGTTGAACAAGGAAACCACCTACGATGAAATCAAGAAGGTTGTC    457
                           ************** *********************************************

SGD_Scer_TDH2/YJR009C      AAGGCTGCCGCTGAAGGTAAGTTGAAGGGTGTCTTGGGTTACACTGAAGACGCTGTTGTC    840
WashU_Sbay_Contig534.6     AAGGCTGCCGCTGAAGGTAAGTTAAAGGGTGTTTTGGGTTACACTGAAGACGCTGTTGTC    792
6257_TDH2                  AAGGCTGCCGCTGAAGGTAAGTTGAAGGGTGTCTTGGGTTACACTGAAGACGCTGTTGTC    517
                           ***********************.******** ***************************

SGD_Scer_TDH2/YJR009C      TCCTCTGACTTCTTGGGTGACTCTAACTCTTCCATCTTCGATGCTGCCGCTGGTATCCAA    900
WashU_Sbay_Contig534.6     TCCTCTGACTTCTTGGGTGACGCTAACTCTTCCATCTTCGATGCTTCCGCTGGTATCCAA    852
6257_TDH2                  TCCTCTGACTTCTTGGGTGACTCTAACTCTTCCATCTTCGATGCTGCCGCTGGTATCCAA    577
                           ********************* *********************** **************

SGD_Scer_TDH2/YJR009C      TTGTCTCCAAAGTTCGTCAAGTTGGTTTCCTGGTACGACAACGAATACGGTTACTCTACC    960
WashU_Sbay_Contig534.6     TTGTCTCCAAAGTTCGTCAAGTTGGTCTCCTGGTACGATAACGAATACGGTTACTCTACC    912
6257_TDH2                  TTGTCTCCAAAGTTCGTCAAGTTGGTTTCCTGGTACGACAACGAATACGGTTACTCTACC    637
                           ************************** *********** *********************

SGD_Scer_TDH2/YJR009C      AGAGTTGTCGACTTGGTTGAACACGTTGCCAAGGCTTAA---------------------    999
WashU_Sbay_Contig534.6     AGAGTTGTCGACTTGGTTGAACACGTTGCCAAGGCTTAA---------------------    951
6257_TDH2                  AGAGTTGTCGACTTGGTTGAACACGTTGCCAAGGCTTAAATTTAACTCCTTAAGTTACTT    697
```

```
                                    ***************************************
```

## PRI2/YKL045w, chrXI:354021-354023

```
SGD_Scer_PRI2/YKL045W        CAGTTTATCTCAAATGAAGAAAAGGCCGAATTATCTCATCAGTTGTATCAAACAGTTTCT     600
WashU_Sbay_Contig652.30      CAGTTTATTTCCAACGAAGAAAAGGCGGAATTGTCGCACCAGTTGTACCAAACCGTTTCA     600
6257_PRI2                    --------------------------------------------------------TTCC     4
                                                                                     ***

SGD_Scer_PRI2/YKL045W        GCGTCTCTACAGTTCCAATTGAATTTAAACGAGGAACATCAAAGAAAACAGTATTTCCAA     660
WashU_Sbay_Contig652.30      GCTTCTTTACAGTTTCAATTGAATCTTAACGAGGAACACCAAAGAAGACAGTACTTCCAA     660
6257_PRI2                    GCGTCTCTACAGTTCCAATTGAATTTAAACGAGGAACATCAAAGAAAACAGTATTTTCAA     64
                             ** *** ******* ********* *:*********** *******.****** ** ***

SGD_Scer_PRI2/YKL045W        CAGGAAAAATTCATAAAATTACCTTTCGAAAATGTGATAGAACTGGTAGGAAACCGTTTA     720
WashU_Sbay_Contig652.30      CAAGAGAAGTTCATAAAGCTACCGTTCGAAAACGTGATAGAGCTAGTGGGTAACCGTCTA     720
6257_PRI2                    CAGGAAAAATTCATAAAATTACCTTTCGAAAATGTGATAGAACTGGTAGGAAACCGTTTA     124
                             **.**.**.********. **** ******** ********.**.**.**:****** **

SGD_Scer_PRI2/YKL045W        GTGTTTTTGAAGGACGGGTACGCATATTTACCACAATTCCAGCAATTGAATTTACTTTCT     780
WashU_Sbay_Contig652.30      GTGTTTTTGAAAAACGGCTATGCTTATTTGCCACAGTTCCAACAACTGAATCTGCTGTCT     780
6257_PRI2                    GTGTTTTTGAAGGACGGGTACGCATATTTACCACAATTCCAGCAATTGAATTTACTTTCT     184
                             ***********..**** ** **:*****.*****.*****.***.***** *.** ***

SGD_Scer_PRI2/YKL045W        AATGAGTTTGCTAGCAAATTGAACCAGGAGTTAATAAAAACGTACCAGTACTTACCAAGA     840
WashU_Sbay_Contig652.30      AATGAGTTTGCTAGTAAGTTGAACGAGGAGCTATTGAAGACCTACCAGTATCTTCCAAGA     840
6257_PRI2                    AATGAGTTTGCTAGCAAATTAAACCAGGaGTTAATAAAAACGTACCAGTACTTACCAAGA     244
                             ************** **.**.*** *** * **:*.**.** ********  *:******

SGD_Scer_PRI2/YKL045W        CTGAATGAGGATGACCGGTTGTTACCAATTCTAAATCATCTTTCGTCGGGGTACACTATC     900
WashU_Sbay_Contig652.30      TTGAACGAGGATGACAGGTTGCTGCCGATTTTAAACCATCTTTCTTCAGGGTACACAATT     900
6257_PRI2                    CTGAATGAGGATGACCGGTTGTTACCAATTCTGAATCATCTTTCGTCAGGATACACAATT     304
                              **** *********.***** *.**.*** *.** ******** **.**.*****:**

SGD_Scer_PRI2/YKL045W        GCGGATTTCAACCAGCAAAAGGCAAACCAATTCAGTGAGAACGTAGATGATGAGATTAAT     960
WashU_Sbay_Contig652.30      GCGGATTTTAACCAGCAAAAGGCAAACCAATTCGGTGAAAATGTAGACGATGAAATAAAT     960
6257_PRI2                    GCAGACTTTAACCAGCAAAAGGCAAACCAATTCGGTGAAAATGTAGACGATGAAATAAAT     364
                             **.** ** ************************.****.** ***** *****.**:***

SGD_Scer_PRI2/YKL045W        GCGCAAAGTGTCTGGTCTGAAGAGATTAGCTCAAACTATCCGCTATGTATCAAAAACCTG     1020
WashU_Sbay_Contig652.30      GCACAAAGTGTGTGGTCCGAGGAGATTAGCTCAAACTATCCTTTGAGTATCAAAAACTTG     1020
6257_PRI2                    GCACAAAGTGTGTGGTCCGAGgAgATCAGCTCGAATTATCCATTGAGTATCAAAAACTTA     424
                             **.******** ***** **. * ** *****.** *****  *.:*********** *.

SGD_Scer_PRI2/YKL045W        ATGGAGGGTCTTAAAAAGAACCATCATTTGAGGTATTATGGGAGACAACAACTGAGTCTG     1080
WashU_Sbay_Contig652.30      ATGGAGGGTTTGAAGAAAAACCATCATTTGAGATATTACGGTAGACAACAACTAAGCTTG     1080
6257_PRI2                    ATGGAGGGTTTGAAGAAAAACCATCATTTGAGATATTACGGTAGGCAACAACTAAGCCTG     484
                             ********* * **.**.**************.***** ** **.********.**  **

SGD_Scer_PRI2/YKL045W        TTTTTGAAAGGTATCGGCCTGAGCGCTGATGAAGCGTTAAAATTTTGGTCTGAGGCATTT     1140
WashU_Sbay_Contig652.30      TTTCTAAAGGGAATTGGATTGAGTGCCGATGAAGCCTTGAAATTTTGGTCAGAAGCATTC     1140
6257_PRI2                    TTTCTAAAGGGAATTGGGTTGAGCGCCGACGAAGCTTTGAAATTTTGGTCAGAAGCTTTC     544
                             *** *.**.**:** **  **** ** ** ***** **.***********:**.**:**

SGD_Scer_PRI2/YKL045W        ACAAGAAATGGGAACATGACAATGGAGAAGTTCAATAAAGAATACCGTTACAGCTTCAGG     1200
WashU_Sbay_Contig652.30      ACAAGAAACGGCAACATGACGATGGAAAAGTTCAATAAAGAGTACCGCTACAGTTTCAGA     1200
6257_PRI2                    ACAAGAAACGGCAACATGACGATGGAAAAGTTCAATAAAGAGTACCGCTACAGTTTTAGA     604
                             ******** ** ********.*****.**************.***** ***** ** **.

SGD_Scer_PRI2/YKL045W        CATAATTACGGTCTTGAAGGTAACAGAATCAACTACAAACCATGGGACTGTCACACTATC     1260
WashU_Sbay_Contig652.30      CACAACTACGGTCTCGAAGGTAACAGAATTAACTACAAGCCATGGGACTGTCACACCATT     1260
6257_PRI2                    CACAACTACGGCCTCGAAGGTAACAGAATCAACTACAAACCGTGGGACTGTCACACCATT     664
                             ** ** ***** ** ************** ********.**.************** **

SGD_Scer_PRI2/YKL045W        CTTTCCAAGCCCAGACCTGGGCGCGGAGATTATCATGGATGCCCTTTCCGTGATTGGAGT     1320
WashU_Sbay_Contig652.30      CTCTCCAAGCCCAGACCCGGCCGTGGCGACTATCACGGATGCCCATTCCGTGACTGGAGC     1320
6257_PRI2                    CTATCTAAGCCCAGACCCGGCCGCGGCGATTATCACGGATGCcCATTCCGTGACTGGAGC     724
                             ** ** *********** ** ** ** **.** ***** ****** *:******** *****

SGD_Scer_PRI2/YKL045W        CACGAGAGACTATCTGCAGAATTGCGCTCTATGAAGTTGACCCAAGCGCAGATCATCAGT     1380
WashU_Sbay_Contig652.30      CACGACAGGCTATCTGCAGAACTGCGTTCCATGAAACTCACCCAAGCACAAATCATAAGC     1380
6257_PRI2                    CACGACAGACTATCTGCAGAACTGCGTTCCATGAAACTCACcCAAGCACAAATCATAAGT     784
                             ***** **.************ **** ** *****. * ** *****.**.*****.**
```

```
SGD_Scer_PRI2/YKL045W      GTTCTGGATTCCTGCCAGAAAGGTGAATACACAATTGCTTGCACTAAAGTGTTTGAAATG     1440
WashU_Sbay_Contig652.30    GTCCTAGATTCGTGCCAAAAGGGCGAATACACAATCGCTTGCACTAAAGTGTTCGAAATG     1440
6257_PRI2                  GTCCTagaTTCGTGCCAAAnagGCGagTACACAATCGCTTGCACCAAAGTATTCGAAATA     844
                           ** **   *** *****.*   * *  ******** ******** *****.** *****.

SGD_Scer_PRI2/YKL045W      ACACACAACTCTGCATCAGCGGATTTGGAAATTGGCGAGCAAACGCATATCGCGCATCCT     1500
WashU_Sbay_Contig652.30    ACACACAATTCCGCCCCAGCCGACTTGGAGATCGGCGAACAAACTCATATCGCACATCCT     1500
6257_PRI2                  ACACAC------------------------------------------------------     850
                           ******
```

## ALD2/YMR170c, chrXIII:602993-602998

```
SGD_Scer_ALD2/YMR170C      ------------------------------------------------------------     0
MIT_Sbay_c933_18910        ------------------------------------------------------------     0
6257_ALD2                  TTTGGAAAATAGCttatcttnttcATCTATCATCTATCATTTTTTtctCGCTTTTGTTtC     60

SGD_Scer_ALD2/YMR170C      ------------------------------------------------------------     0
MIT_Sbay_c933_18910        ------------------------------------------------------------     0
6257_ALD2                  GAGTGACCCGACCCTTAtctAATTTGCCGCCAGAGGCAAGCAGATCGCTATCGCCGTGCT     120

SGD_Scer_ALD2/YMR170C      ------------------------------------------------------------     0
MIT_Sbay_c933_18910        ------------------------------------------------------------     0
6257_ALD2                  TTGCGtctTACTtCTCCACGCCCTTTGAAGCACGGCGGCCTTTAGATACTAATAATATAC     180

SGD_Scer_ALD2/YMR170C      ------------------------------------------------------------     0
MIT_Sbay_c933_18910        ------------------------------------------------------------     0
6257_ALD2                  AATCTAAAACCCTnTACTATACACCCCTCtCCAATGTGCTATCTCCAAGTCGTTTAGGGG     240

SGD_Scer_ALD2/YMR170C      ------------------------------------------------------------     0
MIT_Sbay_c933_18910        ------------------------------------------------------------     0
6257_ALD2                  TTGGGGACGTGATTATGCTTAATGATCAATGCTCACGTAATTGAAGCACAGTCAATAGGA     300

SGD_Scer_ALD2/YMR170C      ------------------------------------------------------------     0
MIT_Sbay_c933_18910        ------------------------------------------------------------     0
6257_ALD2                  CTTATATAAAAGCGCGAGGCCAGTGAAAATAGTTTCAAATTACTCTCCTTTCTTTCGCCT     360

SGD_Scer_ALD2/YMR170C      ----------------------------------------ATGCCTACCTTGTATACTGA     20
MIT_Sbay_c933_18910        ----------------------------------------ATGCCTAATTTATATACAGA     20
6257_ALD2                  CACTTTTATCAACAAAATCCACAAACAAAACTAAAGCGCCATGCCTGATTTATACACAGA     420
                                                                   ******.. **.** **:**

SGD_Scer_ALD2/YMR170C      TATCGAAATCCCACAATTGAAAATCTCTTTAAAGCAACCGCTAGGGTTGTTTATCAACAA     80
MIT_Sbay_c933_18910        CCTCGAAATCCCACAACTGAAAATTTCTGTTAGACAACCACTTGGGTTGTTCATCAACAA     80
6257_ALD2                  CCTCAAAATCCCACAATTGAAAATCTCTGTTAAACAACCACTTGGGTTGTTCATCAACAA     480
                            .**.*********** ******* *** *:*..*****.**:******** ********

SGD_Scer_ALD2/YMR170C      TGAGTTTTGTCCATCATCAGATGGAAAGACCATCGAAACTGTGAACCCAGCTACTGGCGA     140
MIT_Sbay_c933_18910        CGAATTTTGCCCTTCATCAGACGGCAAAACTATCGAAACTGTAAACCCAAGTACCGGTGA     140
6257_ALD2                  CGAATTTTGTCCATCATCAGATGGAAAAAACCATCGAAACTGTGAACCCAGCTACTGGCGA     540
                            **.***** **:******** **.**.** ************.******. *** ** **

SGD_Scer_ALD2/YMR170C      ACCGATAACATCCTTCCAAGCAGCTAACGAAAAGGATGTAGACAAAGCTGTGAAAGCTGC     200
MIT_Sbay_c933_18910        GGCTATAACCTCTTTCCAAGCCGCTAGCGAAAAGGATGTTGATAAGGCGGTCAAAGCAGC     200
6257_ALD2                  ACCGATAACATCCTTCCAAGCAGCTAACGAAAAGGATGTAGACAAAGCTGTGAAAGCTGC     600
                           . * *****.** ********.****.************:** **.** ** *****:**

SGD_Scer_ALD2/YMR170C      CAGGGCTGCTTTTGATAACGTTTGGTCGAAGACATCTTCTGAGCAACGTGGTATTTATCT     260
MIT_Sbay_c933_18910        CAGAGATGCTTTTGAGAATGTCTGGTCGAAGACATCTGCTGAGCAACGTGGCATATATCT     260
6257_ALD2                  CAGGGCTGCTTTTGATAACGTTTGGTCGAAGACATctTCTGAGcaaCGTGGTATTTATCT     660
                           ***.*.********* ** ** *************   *****   ***** **:*****

SGD_Scer_ALD2/YMR170C      TTCAAACTTATTAAAACTTATTGAGGAGGAGCAAGACACACTTGCCGCATTAGAGACTTT     320
MIT_Sbay_c933_18910        CTCAAACCTACTGAAACTCATCGAAGAAGAACAGGAAACGCTAGCCGCCCTGGAGACTTT     320
6257_ALD2                  TTCAAACTTATTAAAACTTATTGAGGAGGAGcAaGACACACTTGCCGCGTTAGAGACTTT     720
                            ****** ** *.***** ** **.**.**. * **.**.**:*****  *.********
```

```
SGD_Scer_ALD2/YMR170C      AGACGCTGGAAAGCCTTACCATTCAAATGCCAAAGGTGATTTGGCACAAATTTTACAGCT      380
MIT_Sbay_c933_18910        AGACGCTGGTAAACCTTTCCATTCTAATGCTATGGGAGATTTAGCTCAAATCATGCAACT      380
6257_ALD2                  AGACGCTGGAAAGCCTTATCATTCAAATGCCAAAGGTGATTTGGCACAAATTTTacAGCT      780
                           *********:**.****: *****:***** *:.**:*****.**:***** :*  *.**

SGD_Scer_ALD2/YMR170C      TACCAGATATTTTGCTGGGTCCGCTGATAAGTTTGACAAAGGTGCAACCATACCATTGAC      440
MIT_Sbay_c933_18910        TACAAGATATTTTGCCGGATCTGCCGATAAGTATAACAAGGGTGATACTATTCCATTATC      440
6257_ALD2                  TACCAGATATTTTGCTGGGTCCGCTGATaAGTTTGAcaaGGGTGCAACCATACCATTGAC      840
                           ***.*********** **.** ** *** ***:*.*   .****.:** **:*****.:*

SGD_Scer_ALD2/YMR170C      TTTTAACAAGTTTGCATATACTCTAAAAGTTCCTTTTGGCGTTGTTGCTCAAATCGTTCC      500
MIT_Sbay_c933_18910        TTCTGAAAAGTTTGCGTACACTTTGAAGGTTCCATTTGGTGTTGTTGCGCAAATCATTCC      500
6257_ALD2                  TTTTAaCaaGTT------------------------------------------------      852
                           ** *. .  ***
```

## YMR196w, chrXIII:657849-657854

```
SGD_Scer_YMR196W        GACTTCAAAGTAGAGTGTCCCGTAGGTTCAGGAGATTATTTGAATCTTGCTGAAGTTGCC      2610
MIT_Sbay_c938_19031     GATTTTAAGGTCGAGTGTCCGGTAGGTTCGGGTGATTATTTAAATCTAGCAGAGGTTGCT      2640
6257_YMR196W            ---------------------------------------------------------TTG      3


SGD_Scer_YMR196W        GAAGAACTTGGATATCGTATGATTCACTTATTTGTTCCAGACGAAAACGGGGAGCGCGCC      2670
MIT_Sbay_c938_19031     GAAGAACTTGGGTACCGTATGATTCACTTGTTTGTACCAGATGAAAATGGTGAGCGTGCC      2700
6257_YMR196W            CCGANACTTGGGTATCGTATGATTCACTTGTTTGTACCAGATGAAAATGNNNAGCGTGCC      63
                         ... ******.** **************.*****:***** ***** *   **** ***

SGD_Scer_YMR196W        ATTCATTATGGTGATCACTCTAAGTTTCTGTCTTCTGATCCATATTTCAGGGATTATGTG      2730
MIT_Sbay_c938_19031     GTTCATTATGGCGATCATTCTAAATTTTTATCTACTGATCCATATTTCAAGGACTATGTA      2760
6257_YMR196W            GTTCATTATGGTGATCACTCTAAATTCTTATCCACTGATCCATATTTCAAGGATTATGTA      123
                        .********** ***** *****.**  *.** :***************.*** *****.

SGD_Scer_YMR196W        CCATTTTTTGAATACTTTGACGGTGATACTGGAAGAGGGCTTGGCGCTTCACACCAATGT      2790
MIT_Sbay_c938_19031     CCATTCTTTGAATACTTCGACGGTGACACAGGAAGAGGCCTTGGTGCTTCTCACCAGTGC      2820
6257_YMR196W            CCATTCTTTGAATACTTCGATGGTGACTCAGGAAGAGGGCTTGGTGCTTCACACCAATGT      183
                        ***** *********** ** ***** :*:******** ***** *****:*****.**

SGD_Scer_YMR196W        GGTTGGACTGCTCTTGTGGCCAAATGGATAAGTGATGTAGGTATATCCTGTGTAAGACTA      2850
MIT_Sbay_c938_19031     GGTTGGACCGCTCTTGTGGCTAAATGGATTAGTGATGTAGGAATATCCTGTGTGAGGTTG      2880
6257_YMR196W            GGTTGGACTGCTCTTGTGGCCAAATGGATAAGTGATGTAGGTATATCCTGTGTAAGACTA      243
                        ******** *********** ********:***********:***********.**. *.

SGD_Scer_YMR196W        CCTCGTACGCCAAGATCATCTGTGGCAACGACCGCTTCAACAGAGAGCTCTGAGCAAGGT      2910
MIT_Sbay_c938_19031     CCTCGTACTCCAAGATCATCTGTAGCTACCACTGCGTCCACAGAAAGCTCTGGACAAGAA      2940
6257_YMR196W            CCTCGTACGCCAAGATCATCTGTGGCAACGACCGCTTCAACAGAGAGCTCTGAGCAAGGT      303
                        ******** *************.**:** ** ** **.*****.*******..****.:

SGD_Scer_YMR196W        CCCAAAATGAAGAGAATGGCAAGACGTAAGAGTGCAAAGTCTTTGGTAAACTACACTGCC      2970
MIT_Sbay_c938_19031     CACAAAATGAAGAGAATGGCAAGACGTAAGAGCGCAAAGTCCTTAGTGAACTACACCGCC      3000
6257_YMR196W            CCCAAAATGAAGAGAATGGCAAGACGTAAGAGTGCAAAGTCTTTGGTAAACTACACTGCC      363
                        *.****************************** ******** **.**.******** ***

SGD_Scer_YMR196W        ACCATTTTGGACTTAACCGAAGAAGAAAAGCGCCATCATAGGATAGGGGGCACCCATTCT      3030
MIT_Sbay_c938_19031     ACTATTTTGGACTTAACTGAAGAAGAAAAGCGTCATCATAGAATAGGAGGAACCCATTCT      3060
6257_YMR196W            ACCATTTTGGACTTAACCGAAGAAGAAAAGCGCCATCATAGGATAGGGGGCACCCATTCT      423
                        ** ************** ************** ********.*****.**.*********

SGD_Scer_YMR196W        GGGTTGACACCACAAAGCAGCATTTCAAGTGACAAGGCTAGACATTTGATGGAGGAAATG      3090
MIT_Sbay_c938_19031     GGGTTGACGCCACAAAGCAGTATATCAAGTGATAAGGCAAAAAACCTCATGGAGGAAATG      3120
6257_YMR196W            GGGTTGACACCACAAAGCAGCAATTCAAGTGACAAGGCTAGACATTTGATGGAGGAAATG      483
                        ********.*********** *::******** *****:*.*.*  * ************

SGD_Scer_YMR196W        AATGAAGAGGAAGGTATTCACGAAACTGTGGTACCTGAAGATCGTCACAACTTTGAAACC      3150
MIT_Sbay_c938_19031     AATGAAGAGGAAGGTATTCATGAAGCCGTGATACCAGAAGATCGTCATAATTTCGAAACA      3180
6257_YMR196W            AATGAAGAGGAAGGTATTCACGAAACTGTGGTACCTGAAGATCGTCACAACTTTGAAACC      543
                        ******************* ***.* ***.****:*********** ** ** *****.

SGD_Scer_YMR196W        AAGCTTATAGGCAAGCTAAAAGATAAGGTGAAAAATATGAAAGTAACTGACAAGGCTAAA      3210
MIT_Sbay_c938_19031     AAATTGATAAGCAAACTAAAGGATAAAGTGAAAAATATGAAAGTAAGCGACAAGACTAAA      3240
6257_YMR196W            AAGCTTATAGGCAAGCTAAAAGATAAGGTGAAAAATATGAAAGTAACTGACAAGGCTAAA      603
                        **. * ***.****.*****.*****.******************  ******.*****
```

```
SGD_Scer_YMR196W          GATGAGGATATAGACCCAATGGACCCAATGAGTCCGTTGAATAAG-GATGTGTCTTGA--      3267
MIT_Sbay_c938_19031       GACGAAGATATAGATCCAATGGATCCAATGTGTCCATTGAATAAGGATTTATTTTGA---      3297
6257_YMR196W              GATGAGGCTATAGACCCAATGGACCCGATGAGTCCGTTTGAATAANGATGTGTCTTGATT      663
                          ** **.*.****** ******** **.***:****.** .*::*. .:* : * * .

SGD_Scer_YMR196W          ------------------------      3267
MIT_Sbay_c938_19031       ------------------------      3297
6257_YMR196W              AACTGCGGTGAACCTTTCAATATC      687
```

## HSP82/YPL240c, chrXVI:97019-97048

```
SGD_Scer_HSP82/YPL240C    GAAGACCCATTGTACGTTAAGCATTTCTCCGTTGAAGGTCAATTGGAATTTAGAGCTATC      960
MIT_Sbay_c60_24336        ------------------------------------------------------------      0
6033_Hsp82                ----------------------------------AAGGTCAGTtAGAATTCAGAGCCAtc      26


SGD_Scer_HSP82/YPL240C    TTATTCATTCCAAAGAGAGCACCATTCGACTTGTTTGAGAGTAAAAAGAAGAAGAATAAT      1020
MIT_Sbay_c60_24336        ------------------------------------------------------------      0
6033_Hsp82                tTGTACATtCCAAAGAGAnntCCATTTGantTATTcGAAAGTAAGAAGAagaagaAcAAc      86


SGD_Scer_HSP82/YPL240C    ATCAAGTTGTACGTTCGTCGTGTTTTCATCACTGATGAAGCTGAAGACTTGATTCCAGAG      1080
MIT_Sbay_c60_24336        ------------------------------------------------------------      0
6033_Hsp82                ATCAAGTTGTATGTTCGTCGTGTTtTcntCACCGACGAAGCTGAAGACTTGATCCCAGAA      146


SGD_Scer_HSP82/YPL240C    TGGTTATCTTTCGTCAAGGGTGTTGTTGACTCTGAGGATTTACCATTGAATTTGTCCAGA      1140
MIT_Sbay_c60_24336        ---ATGTCATTCGTCAAGGGTGTTGTTGACTCCGAAGATTTACCATTGAACTTGTCCAGA      57
6033_Hsp82                TGGATGTCCTTTGTCAAGGGTGTTGTTGACTCTGAAGACTTACCAtTGAACCTGTCCAGA      206
                             :*.** ** ******************** **.** ****** ****  ********

SGD_Scer_HSP82/YPL240C    GAAATGTTACAACAAAATAAGATCATGAAGGTTATTAGAAAGAACATTGTCAAAAAGTTG      1200
MIT_Sbay_c60_24336        GAAATGTTGCAACAAAACAAGATCATGAAGGTCATCAGAAAGAACATCGTTAAGAAGATG      117
6033_Hsp82                GAAATGCTACAACAAAACAAGATCATGAAGGTTATCAGAAAGAACATCGTCAAGAAGGTA      266
                          ****** *.******** ************** ** *********** ** **.*** *.

SGD_Scer_HSP82/YPL240C    ATTGAAGCCTTCAACGAAATTGCTGAAGACTCTGAACAATTTGAAAAGTTCTACTCGGCT      1260
MIT_Sbay_c60_24336        ATTGAATCCTTCAACGAAATCGCTGAAGACTCTGAACAATTCGAAAAGTTCTACTCTGCC      177
6033_Hsp82                ATTGAAGCTTTCAACGAAATTGCTGAAGACTCTGAACAATTCGAAAAGTTCTACTCTGCC      326
                          ****** * *********** ******************** ************** **

SGD_Scer_HSP82/YPL240C    TTCTCCAAAAATATCAAGTTGGGTGTACATGAAGATACCCAAAACAGGGCTGCTTTGGCT      1320
MIT_Sbay_c60_24336        TTCGCTAAGAACATCAAATTGGGTGTTCATGAAGACACTCAAAACAGAGCTGCCTTGGCC      237
6033_Hsp82                TTCGCTAAGAACATCAAATTGGGTGTCCACGAAGACACTCAAAACAGAGCTGCCTTGGCT      386
                          *** * **.** *****.******** ** ***** ** ********.***** *****

SGD_Scer_HSP82/YPL240C    AAGTTGTTACGTTACAACTCTACCAAGTCCGTAGATGAGTTGACTTCCTTAACTGATTAC      1380
MIT_Sbay_c60_24336        AAATTGCTACGTTACAACTCCACCAAGTCCGTTGACGAATTGACTTCTTTGACTGATTAT      297
6033_Hsp82                AAGTTACTACGTTACAACTCCACCAAGTCCGTCGACGAATTAACTTCTTTGACCGATTAC      446
                          **.**. ************* *********** ** **.**.***** **.** *****

SGD_Scer_HSP82/YPL240C    GTTACCAGAATGCCAGAACACCAAAAGAACATCTACTACATCACTGGTGAATCTCTAAAG      1440
MIT_Sbay_c60_24336        ATTACCAGAATGCCAGAACACCAAAAGAACATCTACTATATCACAGGTGAGTCTTTGAAG      357
6033_Hsp82                ATCACCAGAATGCCAGAACACCAAAAGAACATTTACTACATCACAGGTGAATCTCTAAAG      506
                          .* **************************** ***** *****:*****.*** *.***

SGD_Scer_HSP82/YPL240C    GCTGTCGAAAAGTCTCCATTTTTGGATGCCTTGAAGGCTAAAAACTTCGAGGTTTTGTTC      1500
MIT_Sbay_c60_24336        GCTGTTGAAAAAATCCCCATTCTTAGACGCTTTGAAAGCTAAGAACTTTGAAGTTTTGTTC      417
6033_Hsp82                GCCGTTGAAAAAATCCCCATTCTTGGACGCTTTGAAGGCTAAGAACTTTGAAGTTTTGTTC      566
                          ** ** *****.** ***** **.** ** *****.*****.***** **.*********

SGD_Scer_HSP82/YPL240C    TTGACCGACCCAATTGATGAATACGCCTTCACTCAATTGAAGGAATTCGAAGGTAAAACT      1560
MIT_Sbay_c60_24336        TTGACTGATCCAATCGATGAATACGCCTTCACTCAATTAAAGGAATTCGAAGGTAAGACT      477
6033_Hsp82                TTGACTGATCCAATCGATGAATACGCTTTCACCCAATTGAAGGAATTCGAAGGTAAGACT      626
                          ***** ** ***** *********** ***** *****.*****************.***

SGD_Scer_HSP82/YPL240C    TTGGTTGACATTACTAAAGATTTCGAATTGGAAGAAACTGACGAAGAAAAAGCTGAAAGA      1620
MIT_Sbay_c60_24336        TTAGTCGATATCACCAAGGATTTCGAGCTGGAAGAAACTGACGAAGAGAAAGCTGAAAGA      537
6033_Hsp82                TTGGTCGATATCACCAAGGATTTCGAACTGGAAGAAACTGACGAAGAAAAAGCTGAAAGA      686
                          **.** ** ** ** ** **.********. ******************.***********

SGD_Scer_HSP82/YPL240C    GAGAAGGAGATCAAAGAATATGAACCATTGACCAAGGCCTTGAAAGAAATTTTGGGTGAC      1680
MIT_Sbay_c60_24336        GAAAAGGAAGTTAAAGAATTCGAACCATTGACCAAGGCCTTGAAAGACATCTTGGGTGAA      597
6033_Hsp82                GAGAAGGAGATCAAAGAATATGAACCATTGACCAAGGCCTTGAAAGAAATTTTGGGTGAC      746
```

```
                           **.*****..* *******: **************************.** ********.

SGD_Scer_HSP82/YPL240C     CAAGTGGAGAAAGTTGTTGTTTCTTACAAATTGTTGGATGCCCCAGCTGCTATCAGAACT     1740
MIT_Sbay_c60_24336         CAAGTTGAAAAGGTTGTTGTCTCTTACAAACTAGTGGATGCCCCAGCTGCCATTAGAACT     657
6033_Hsp82                 CAAGTGGAGAAAGTTGTTGTTTCTTACAAATTGTTGGATGCCCCAGCTGCTATCAGAACT     806
                           ***** **.**.******** ********* *. **************** ** ******

SGD_Scer_HSP82/YPL240C     GGTCAATTTGGTTGGTCTGCTAACATGGAAAGAATCATGAAGGCTCAAGCCTTGAGAGAC     1800
MIT_Sbay_c60_24336         GGCCAATTCGGTTGGTCCGCTAACATGGAAAGAATCATGAAGGCTCAAGCTTTGAGAGAC     717
6033_Hsp82                 GGTCAATTTGGTTGGTCTGCTAACATGGAAAGAATCATGAAGGCTCAAGCCTTGAGAGAC     866
                           ** ***** ******** ******************************** *********

SGD_Scer_HSP82/YPL240C     TCTTCCATGTCCTCCTACATGTCTTCCAAGAAGACTTTCGAAATTTCTCCAAAATCTCCA     1860
MIT_Sbay_c60_24336         TCTTCCATGTCCTCTTACATGTCCTCCAAGAAGACTTTCGAAATCTCTCCAAAATCTCCA     777
6033_Hsp82                 TCTTCCATGTCCTCCTACATGTCTTCCAAGAAGACTTTCGAAATTTCTCCAAAATCTCCA     926
                           ************** ******** ******************** ***************

SGD_Scer_HSP82/YPL240C     ATTATCAAGGAATTGAAAAAGAGAGTTGACGAAGGTGGTGCTCAAGACAAGACTGTCAAG     1920
MIT_Sbay_c60_24336         ATTATCAAGGAATTGAAAAAGAGAGTTGATGAAGGCGGTGCTCAAGATAAGACTGTCAAG     837
6033_Hsp82                 ATTATCAAGGAATTGAAAAAGAGAGTTGACGAAGGTGGTGCTCAAGACAAGACTGTcAAG     986
                           ***************************** ***** *********** ******** ***

SGD_Scer_HSP82/YPL240C     GACTTGACTAAGTTATTATATGAAACTGCTTTGTTGACTTCCGGCTTCAGTTTGGACGAA     1980
MIT_Sbay_c60_24336         GATTTGACCAACTTATTATTCGAAACCGCTCTGTTAACTTCTGGTTTCAGTCTGGAAGAA     897
6033_Hsp82                 GAC---------------------------------------------------------     989
                           **
```

## GPH1/YPR160w, chrXVI:862792-862803

```
SGD_Scer_GPH1/YPR160W      TTGTATCCAAACGATAACTTTGCTCAAGGTAAGGAGTTGAGGTTGAAACAGCAGTACTTC     1071
MIT_Sbay_c607_26280        CTGTATCCAAACGACAACTTTGCCCAAGGTAAAGAACTGAGATTGAAACAGCAATACTTC     1080
6257_GPH1                  ------------------------------------GTTGAGGTTGAAACAGCAGTACTTC     25
                                                               . ****.***********.******

SGD_Scer_GPH1/YPR160W      TGGTGTGCTGCATCCTTACACGACATCTTAAGAAGATTCAAAAAATCCAAGAGGCCATGG     1131
MIT_Sbay_c607_26280        TGGTGTGCTGCATCCTTGCACGATATCCTAAGAAGATTCAAAAAGTCTAAGAGATCATGG     1140
6257_GPH1                  TGGTGTGCTGCATcCttACACGACATCTTAAGAAGATTCAAAAAATCCAAGAGGCCATGG     85
                           ************* *  .***** *** ****************.** *****. *****

SGD_Scer_GPH1/YPR160W      ACTGAATTTCCTGACCAAGTGGCTATTCAGTTGAATGATACTCATCCAACTTTAGCCATC     1191
MIT_Sbay_c607_26280        ACCGAGTTCCCTGAGCAAGTGGCTATTCAATTGAATGATACCCATCCAACGTTAGCTATC     1200
6257_GPH1                  ACTGAATTTCCTGACCAAGTGGCTATTCAGTTGAATGATACTCATCCAACTTTAGCCATC     145
                           ** **.** ***** ***************.*********** ******** ***** ***

SGD_Scer_GPH1/YPR160W      GTTGAATTACAGAGAGTTTTGGTCGATCTAGAAAAACTAGATTGGCACGAGGCTTGGGAC     1251
MIT_Sbay_c607_26280        GTTGAATTGCAAAGAGTCTTGGTCGATTTGGAGAAATTAGATTGGCACGAAGCTTGGGAC     1260
6257_GPH1                  GTTGAATTACAGAGAGTTTTGGTCGATCTAGAAAAACTAGATTGGCACGAGGCTTGGGAC     205
                           ********.**.***** ********* *.**.*** *************.*********

SGD_Scer_GPH1/YPR160W      ATCGTGACCAAGACTTTTGCTTATACTAACCACACTGTTATGCAAGAGGCCCTGGAAAAA     1311
MIT_Sbay_c607_26280        ATTGTCACCAAGACCTTTGCTTACACCAACCACCGTTATGCAAGAAGCCTTGGAAAAA     1320
6257_GPH1                  ATCGTGACCAAGACTTTTGCTTATACTAACCACACTGTTATGCAAGAGGCCCTGGAAAAA     265
                           ** ** ******** ******** ** ******** ***********.*** ********

SGD_Scer_GPH1/YPR160W      TGGCCCGTCGGCCTCTTTGGCCATTTGCTACCCAGACATTTGGAAATTATATATGATATC     1371
MIT_Sbay_c607_26280        TGGCCCGTTGGCTTATTCGGCCATTTGTTGCCCAGACATCTGGAAATTATCTACGATATT     1380
6257_GPH1                  TGGCCCGTCGGCCTCTTTGGCCATTTGCTACCCAGACATTTGGAAATTATATATGATATC     325
                           ******** *** *.** ********* *.********* **********.** *****

SGD_Scer_GPH1/YPR160W      AACTGGTTCTTCTTGCAAGATGTGGCCAAAAAAATTCCCCAAGGATGTTGATCTTTTGTCT     1431
MIT_Sbay_c607_26280        AATTGGTTCTTCTTACAGGATGTTGCAAAGAAATTCCCCAAGGATGTTGATCTTTTGTCT     1440
6257_GPH1                  AACTGGTTCTTCTTGCAAGATGTGGCCAAAAAAATTCCCCAAGGATGTTGATCTTTTGTCT     385
                           ** ***********.**.***** **.**.*****************************

SGD_Scer_GPH1/YPR160W      CGTATATCCATCATCGAAGAAAACTCTCCAGAAAGACAGATCAGAATGGCCTTTTTGGCT     1491
MIT_Sbay_c607_26280        CGTATATCCATCATCGAGGAAAACTCTCCAGAAAGACAGATCAGAATGGCCTTTTTGGCC     1500
6257_GPH1                  CGTATATCCATCATCGAAGAAAACTCTCCAGAGAGACAGATCAGAATGGCCTTTTTGGCT     445
                           *****************.**************.**************************

SGD_Scer_GPH1/YPR160W      ATTGTTGGTTCACACAAGGTTAATGGTGTTGCTGAATTGCACTCTGAATTAATCAAAACG     1551
MIT_Sbay_c607_26280        ATTGTTGGTTCTCATAAGGTCAACGGTGTTGCTGAATTGCACTCTGAATTAATTAAGACC     1560
6257_GPH1                  ATTGTTGGTTCTCATAAAGTCAACGGTGTTGCGGAATTGCACTCTGAATTAATTAAGACC     505
                           ***********:** ** .** ** ******** ******************** **.**

SGD_Scer_GPH1/YPR160W      ACCATATTTAAAGATTTTGTCAAGTTCTATGGTCCATCAAAGTTTGTCAATGTCACTAAC     1611
MIT_Sbay_c607_26280        ACCATTTTCAAAGATTTCGTCAAATTCTACGGTGCATCAAAGTTCGTAAATGTTACTAAC     1620
6257_GPH1                  ACCATCTTCAAAGATTTCGTCAAATTCTACGGTGCATCAAAGTTTGTCAACGTTACTAAC     565
                           ***** ** ******** *****.***** *** ********** **.** ** ******
```

```
SGD_Scer_GPH1/YPR160W      GGTATCACACCAAGGAGATGGTTGAAGCAAGCTAACCCTTCATTGGCTAAACTGATCAGT      1671
MIT_Sbay_c607_26280        GGTATCACACCAAGAAGATGGTTAAAGCAAGCCAACCCTGCCTTGGCTAGGTTGATTAGT      1680
6257_GPH1                  GGTATCACACCAAGAAGATGGTTGAAGCAAGCTAACCCTAACTTGGCTAGATTGATTAGC      625
                           **************.*********.******** ****** ..*******.. **** **

SGD_Scer_GPH1/YPR160W      GAAACCCTTAACGATCCAACAGAGGAGTATTTGTTGGACATGGCCAAACTGACCCAGTTG      1731
MIT_Sbay_c607_26280        GAAACTCTTAACGATCCTACAGAGGAGTATCTACTAGATATGACAAAGTTAACTCAATTG      1740
6257_GPH1                  AAAACTCTTAACGATCCTACAGAGGACTATCTACTAGACATGACAAAGTTAACTCAACTG      685
                           .**** ***********:******** *** *. *.** ***.*.**. *.** **. **

SGD_Scer_GPH1/YPR160W      GGAAAATATGTTGAAGATAAGGAGTTTTTGAAAAAAATGGAACCAAGTCAAGCTTAATAAT      1791
MIT_Sbay_c607_26280        GCAAAACACATTGAGGATAAGAAGTTTTTGAAGGAATGGAACCAAGTCAAGCTCAACAAT      1800
6257_GPH1                  GCAAAGCACCTTGAGGATAAGAAGTTTTTGAAAGAGTGGAATCAAGTCAAACTCAATAAT      745
                           * ***. *  ****.******.**********..*.***** ********.** ** ***

SGD_Scer_GPH1/YPR160W      AAGATCAGATTAGTAGATTTAATCAAAAAGGAAAATGATGGAGTAGACATCATTAACAGA      1851
MIT_Sbay_c607_26280        AAGATCAGATTAGTGGATCTAATCAAAAAAGAAAATGACGGTGAAGACATCATTAACAGA      1860
6257_GPH1                  AAGATCAGATTGGTGGACCTAATCAAAAAaGAAAATGGTGGTGAAGACATCATTAACAGA      805
                           ***********.**.**  ********** *******. **:*:****************

SGD_Scer_GPH1/YPR160W      GAGTATTTGGACGACACCTTGTTTGATATGCAAGTTAAACGTATTCATGAATATAAGCGT      1911
MIT_Sbay_c607_26280        AAGTATCTAGATGACACTTTGTTTGATATGCAAGTTAAACGTATTCATGAGTATAAGCGT      1920
6257_GPH1                  GAGTATCTAGACGATACTtTGTTTGATATGCAAGTTAAACGTATTCACGAGTATAAACGT      865
                           .***** *.** ** **  **************************** **.*****.***

SGD_Scer_GPH1/YPR160W      CAACAGCTAAACGTCTTTGGTATTATATACCGTTACCTGGCAATGAAGAATATGCTGAAG      1971
MIT_Sbay_c607_26280        CAACAACTAAACGTCTTTGGTATTATTTACCGTTACTTGGCAATGAAAAATATGCTAGAG      1980
6257_GPH1                  CAACAACTAAACGTCTTTGGTATTATTTACCGTTACTTAGCAATGAAAAATATGCTAgag      925
                           *****.********************:********* *.********.********.

SGD_Scer_GPH1/YPR160W      AACGGTGCTTCGATCGAAGAAGTTGCCAAGAAATATCCACGCAAGGTTTCAATCTTTGGT      2031
MIT_Sbay_c607_26280        AACGGTGCTTCTATCGAAGAAGTGGCCAAGAAATATCCACGTAAGGTTTCTATCTTCGGT      2040
6257_GPH1                  aACGGTGCTTCCATCGAGGAAGTGGCCAanaAATATCCACGTAAGG--------------      971
                            ********** *****.***** ****   ********** ****
```

## QCR2/YPR191w, chrXVI:919950-919951

```
SGD_Scer_QCR2/YPR191W      TATGCAACCAAGGATGGTGTGGCCCATCTTTTAAACAGATTCAACTTTCAAAACACGAAC      180
MIT_Sbay_c606_26157        TATGCAACCAAGGACGGTGTAGCTCATCTCTTGAACAGGTTCAACTTCCAAAACACAAAT      180
6257_QCR2                  -------------------------ATCTTTTGAACAGGTTCAACTTCCaGaACACAAAT      35
                                                    **** **.*****.******** * . ****.**

SGD_Scer_QCR2/YPR191W      ACTAGATCAGCTTTGAAATTAGTCAGAGAATCCGAATTATTAGGGGGAACTTTTAAGTCT      240
MIT_Sbay_c606_26157        GCCAGGTCTGCGTTGAGATTGGTCAGAGAATCCGAATTATTAGGGGGAATTTTTAAGTCC      240
6257_QCR2                  GCTAGGTCTGCGTTGAGATTGGTCAGAGAATCCGAATTATTAGGGGGAAATTTTAAGTCC      95
                           .* **.**:** ****.***.*********************** *********

SGD_Scer_QCR2/YPR191W      ACCTTGGATAGGGAATACATCACTTTGAAAGCTACCTTTTTGAAGGACGACCTTCCCTAC      300
MIT_Sbay_c606_26157        ACTTTGGACAGGGAATATATCACTCTAAAAGCCACCTTCTTGAGGGATGACCTTCCTTAC      300
6257_QCR2                  ACTTTGGATAGGGAATACATCACTCTAAAAGCTACATTCTTGAGGGACGACCTTCCTTAC      155
                           ** ***** ******** ****** *.***** **.** ****.*** ******** ***

SGD_Scer_QCR2/YPR191W      TACGTCAATGCCCTAGCAGACGTGCTATACAAGACTGCCTTCAAACCTCACGAGCTCACC      360
MIT_Sbay_c606_26157        TACGTCAATGCCCTGGCAGATGTGTTGTACAAGACTGCCTTCAAGCCCCACGAGCTACCT      360
6257_QCR2                  TACGTCAATGCCTTGGCAGATGTGTTGTATAAGACTGCCTTCAAACCCCACGAGCTGTCT      215
                           ************ *.***** *** *.** **************.** ********  *

SGD_Scer_QCR2/YPR191W      GAATCTGTTTTGCCTGCTGCTAGATACGATTATGCGGTCGCTGAACAATGTCCGGTAAAG      420
MIT_Sbay_c606_26157        GAATCTGTTTTGCCTGCTGCCAGATACGATTATGCTGTTGCTGAGCAGTGCCCCGAAAAA      420
6257_QCR2                  GAATCTGTTTTGCCTGCTGCCAGATACGATTATGCGGTTGCTGAGCAGTGCCCCGTAAAA      275
                           ******************** ************** ** *****.**.** ** *:***.

SGD_Scer_QCR2/YPR191W      AGCGCCGAAGACCAATTGTATGCCATTACATTCAGAAAGGGTTTAGGAAACCCATTGTTA      480
MIT_Sbay_c606_26157        AATGCAGAGGAACAGTTGTTCGCTATTACATTCAGAAAGGGCCTGGGAAACCCATTGCTC      480
6257_QCR2                  AGCGCAGAAGAACAGTTATTCGCTATTACATTTAGAAAGGGCTTGGGAAATCCATTGTAT      335
                           *. **.**.**.**.**.*: ** ******** ********  *.***** ****** :

SGD_Scer_QCR2/YPR191W      TACGATGGTGTGGAAAGAGTCAGTTTGCAAGATATCAAGGACTTTGCGGACAAAGTCTAT      540
MIT_Sbay_c606_26157        TACGACGGCGTGGAAAAGGTCAGCTTGCAAGATATCAAGGATTACGCTGACAAAGTCTAC      540
6257_QCR2                  TACGACGGTGTGGAAAAAGTCAGCTTGCAAGATATCAAGGATTACGCTGaCAAAGTCTAC      395
                           ***** ** *******..***** **************** *: ** * *********
```

```
SGD_Scer_QCR2/YPR191W     ACCAAGGAGAACCTTGAAGTTAGCGGTGAAAATGTTGTTGAGGCCGATTTGAAAAGATTT     600
MIT_Sbay_c606_26157       ACTAAAGAGAATCTTGAAATCACAGGTGAAAATATCGTTGAGGCCGATTTGAAAAGATTC     600
6257_QCR2                 ACTAAAGAGAATCTTGAAATTACAGGTGAAAATATTGTTGAGGCCGATTTGAAAAGATTT     455
                          ** **.***** ******.* * .*********.* ***********************

SGD_Scer_QCR2/YPR191W     GTTGACGAGTCACTGTTAAGCACTTTGCCTGCAGGTAAGTCATTGGTGAGTAAATCCGAA     660
MIT_Sbay_c606_26157       GTTGACGATTCCTTGTTGGCCACTTTACCCACAGGCAAATCACTGGTAAGTAAATCCGAA     660
6257_QCR2                 GTTGACGAGTCACTGTTAAGCACTTTGCCTGCAGGTAAGTCATTGGTGAGTAAATCCGAA     515
                          ******** **. ****.. ******.** .**** **.*** ****.************

SGD_Scer_QCR2/YPR191W     CCAAAATTCTTTTTGGGTGAAGAAAACAGGGTAAGGTTTATCGGTGACTCCGTTGCCGCC     720
MIT_Sbay_c606_26157       CCAAAATCCTTCTTAGGTGAAGAAAACAGACTAAGATTCCTCGGTGATTCCGTTGCTGCC     720
6257_QCR2                 CCAAAATCCTTTTTGGGTGAAGAAAACAGGGTAAGGTTTATCGGTGACTCCGTTGCCGCC     575
                          ******* *** **.**************. ****.** .******* ******** ***

SGD_Scer_QCR2/YPR191W     ATTGGTATCCCGGTAAACAAAGCCTCCCTAGCTCAATATGAAGTATTGGCCAACTATTTG     780
MIT_Sbay_c606_26157       ATCAGTATCCCTGTGAACAAAGCATCTCTAGCCCAATACGAAGTCTTGGCCAGCTATTTG     780
6257_QCR2                 ATTGGTATCCCGGTAAACAAAGCCTCCCTAGCTCAATATGAAGTATTGGCCAACTATTTG     635
                          ** .******* **.********.** ***** ***** *****.*******.*******

SGD_Scer_QCR2/YPR191W     ACCTCTGCCCTATCCGAGCTTTCCGGTTTAATCAGCTCGGCTAAACTTGATAAATTCACT     840
MIT_Sbay_c606_26157       ACCTCCGCGCTTTCCGATCTCTCCGGCTTGGTCAACACTGCCAAACTACAAAAATTCAGT     840
6257_QCR2                 ACCTCTGCCCTATCCGACCTTTCCGGCTTAGTCAGCTCGGCTAAACTTGATAAATTCACT     695
                          ***** ** **:***** ** ***** **..***.*:* ** *****: *:******* *

SGD_Scer_QCR2/YPR191W     GACGGCGGCCTATTTACTCTGTTTGTAAGAGACCAGGACAGCGCCGTGGTATCTTCCAAC     900
MIT_Sbay_c606_26157       GACGGCGGTCTCTTCACTTTGACCGTCAGAGATCAAGACAGCTCTGTGGTGTCCGCAAAC     900
6257_QCR2                 GACGGCGGCCTGTTTACTCTGTTTGTAAGAgACCAAGACAGCGCCGTGGTATCTTCCAAC     755
                          ******** ** ** *** **:  **.*** * **.****** * *****.**  *.***

SGD_Scer_QCR2/YPR191W     ATCAAGAAAATTGTTGCGGATTTGAAGAAGGGCAAGGACTTATCCCCTGCAATAAATTAC     960
MIT_Sbay_c606_26157       ATCAAGAAGATTGTTGCAGACTTGAAGAAGGGTAAGGATTTATCACCTGCGGTAAATTAC     960
6257_QCR2                 ATCAAGAAAATTGTTGCGGATTTGAA----------------------------------     781
                          ********.********.** *****
```